\documentclass[journal]{IEEEtran}

\usepackage{booktabs} % For formal tables
\usepackage{graphicx}
\graphicspath{{./graphics/}}
\usepackage[utf8]{inputenc}

\usepackage{amsmath}
\usepackage{amssymb}
\usepackage{microtype}
\usepackage{amsmath}
\usepackage{hyperref}
\usepackage{cite}
\usepackage{url}
\usepackage[process=auto]{pstool}
\usepackage{makecell}
\usepackage{longtable}
\usepackage{supertabular,booktabs}

\usepackage{tabularx,booktabs, multirow, longtable, paralist}
\usepackage{placeins}

\newcolumntype{C}[1]{>{\raggedright}p{#1}}
\newcolumntype{Z}{>{\raggedright\arraybackslash}X}
\newcommand{\textbfsf}[1]{\textbf{\textsf{#1}}}
\newtheorem{remark}{Remark}

\definecolor{MyRed}{RGB}{0,0,250}

\begin{document}
\title{A Survey of Biological Building Blocks for Synthetic Molecular Communication Systems}	
	\author{Christian A. S\"{o}ldner, Eileen Socher, Vahid Jamali, Wayan Wicke,\\  Arman Ahmadzadeh, Hans-Georg Breitinger, Andreas Burkovski,\\ Kathrin Castiglione, Robert Schober, and Heinrich Sticht
	\thanks{This work was supported in part by the German Research Foundation under Projects SCHO 831/7-1 and SCHO 831/9-1, in part by the Friedrich-Alexander
		University Erlangen-N{\"u}rnberg under the Emerging Fields Initiative, and in part by the STAEDTLER Foundation.}
	\thanks{Christian A. S\"{o}ldner and Heinrich Sticht are with the Division of Bioinformatics, Institute of Biochemistry, Friedrich-Alexander-Universität Erlangen-Nürnberg (FAU), Fahrstr. 17, 91054 Erlangen, Germany. (email: \{christian.soeldner, heinrich.sticht\}@fau.de)}
	\thanks{Eileen Socher is with the Institute of Biochemistry and the Institute of Anatomy, Friedrich-Alexander-Universität Erlangen-Nürnberg (FAU), 91054 Erlangen, Germany. (email: eileen.socher@fau.de)}
\thanks{ Vahid Jamali, Wayan Wicke, Arman Ahmadzadeh, and Robert Schober are with the Institute for Digital Communications, Department of Electrical, Electronics, and Communication Engineering (EEI), Friedrich-Alexander-Universität Erlangen-Nürnberg (FAU), Cauerstr. 7, 91058 Erlangen, Germany. (email: \{vahid.jamali, wayan.wicke, arman.ahmadzadeh, robert.schober\}@fau.de)}
\thanks{Hans-Georg Breitinger is with the Department of Biochemistry, Faculty of Pharmacy and Biotechnology, German University in Cairo (GUC), New Cairo 11835, Egypt. (email: hans.breitinger@guc.edu.eg)}
\thanks{Andreas Burkovski is with the Division of Microbiology, Department of Biology, Friedrich-Alexander-Universität Erlangen-Nürnberg (FAU), Staudtstr. 5, 91058 Erlangen, Germany. (email: andreas.burkovski@fau.de)}

\thanks{Kathrin Castiglione is with the Institute of Bioprocess Engineering, Department of Chemical and Bioengineering, Friedrich-Alexander-Universität Erlangen-Nürnberg (FAU), Paul-Gordanstr. 3,	91052 Erlangen,
	Germany. (email: kathrin.castiglione@fau.de)}
}

\markboth{IEEE COMMUNICATIONS SURVEYS \& TUTORIALS}{}

\maketitle

\begin{abstract} 
Synthetic molecular communication (MC) is a new	communication engineering paradigm which is expected to enable revolutionary applications such as smart drug delivery and real-time health monitoring. The design and implementation of synthetic MC systems (MCSs) at nano- and microscale is very challenging. This is particularly true for synthetic MCSs employing biological components as transmitters and receivers or as interfaces with natural biological MCSs. Nevertheless, since such biological components have been optimized by nature over billions of years, using them in synthetic MCSs is highly promising. This paper provides a survey of biological components that can potentially serve as the main building blocks, i.e., transmitter, receiver, and signaling particles, for the design and implementation of synthetic MCSs. Nature uses a large variety of signaling particles of different sizes and with vastly different properties for communication among biological entities. Here, we focus on three important classes of signaling particles: cations (specifically protons and calcium ions), neurotransmitters (specifically acetylcholine, dopamine, and serotonin), 
and phosphopeptides. These three classes have unique and distinct features such as their large diffusion coefficients, their specificity, and/or their uniqueness of signaling that make them suitable candidates for signaling particles in synthetic MCSs. 
For each of these candidate signaling particles, we present several specific transmitter and receiver structures mainly built upon proteins that are capable of performing the distinct physiological functionalities required from the transmitters and receivers of MCSs. Moreover, we present options for both microscale implementation of MCSs as well as the micro-to-macroscale interfaces needed for experimental evaluation of  MCSs. One of the main advantages of employing proteins for signal emission and detection is that they can be modified with tools from synthetic biology and be tailored to a wide range of application needs. We discuss the properties, limitations, and applications of the proposed biological building blocks for synthetic MCSs in detail. Furthermore, we outline new research directions for the implementation and the theoretical design and analysis of the proposed transmitter and receiver architectures.% and options for intersymbol interference mitigation for the considered signaling particles. 
 
 \end{abstract}

\begin{IEEEkeywords}
Molecular communications, transmitter and receiver architecture, signaling particles, synthetic biology, and test-bed implementation.
\end{IEEEkeywords}

%--------------Figure---------------
\begin{figure*}%[t]
	\centering
	\includegraphics[width=0.7\textwidth]{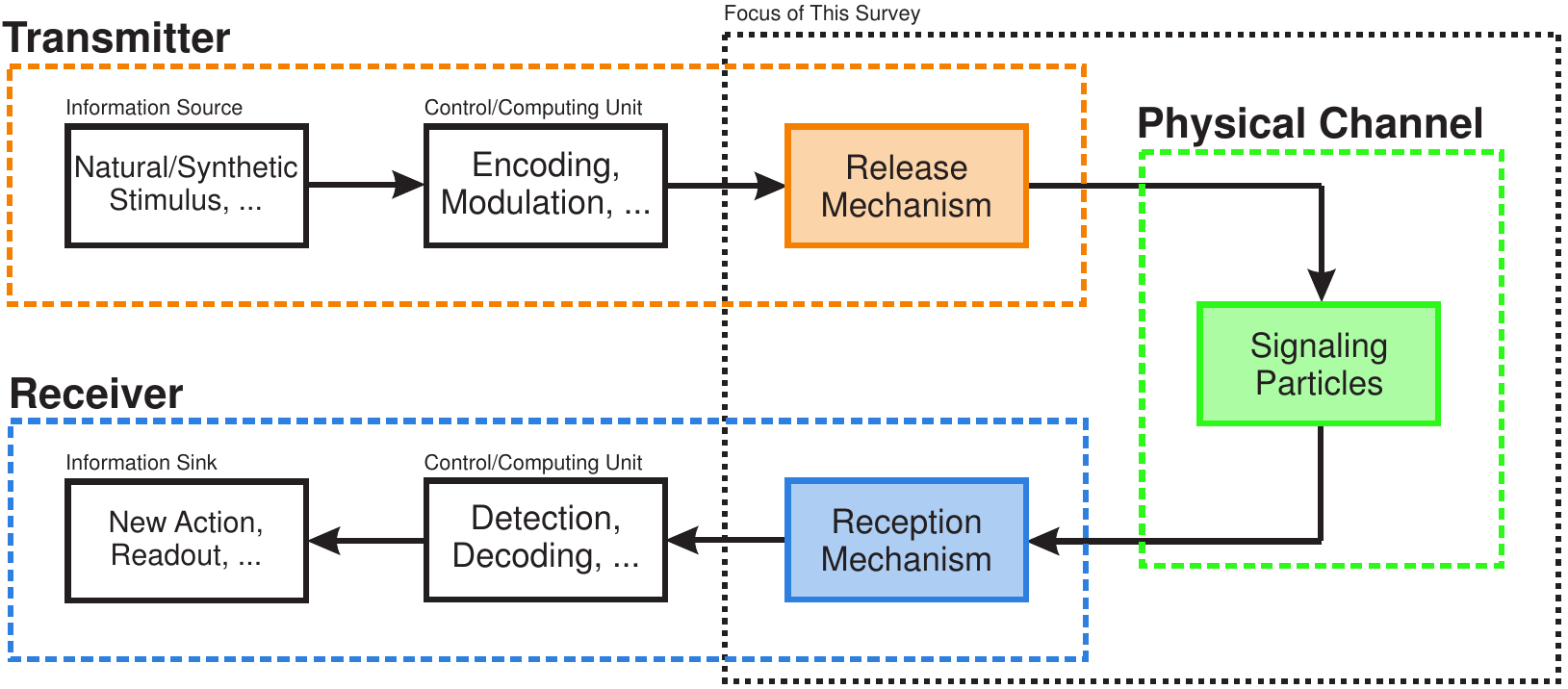}
	\caption{General block diagram of an MCS. This paper surveys suitable biological building blocks for implementation of the release and reception mechanisms for several classes of signaling particles. The color code used to represent transmitter, channel, and receiver will be applied throughout the paper.} %A transmitter encodes a signal by releasing the signaling particle upon a certain input of information which may be artificial or biological. The signaling particles (ions, neurotransmitters, phosphopeptides) propagate through the channel and are then to be detected by a receiver which decodes the information and can either be read out artificially or induce a biological reaction. The main difference compared to conventional communication systems is the use of biological or chemical components for transmitter, signaling particle, and receiver.
	\label{fig:mcs_einfuehrung}
\end{figure*}
%-----------------------------------
	
%%%%%%%%%%%%%%%%%%%%%%%%%
\section{Introduction}
%%%%%%%%%%%%%%%%%%%%%%%%%
%Nanomachines
The development of nanomachines for medical applications such as  real-time health monitoring and targeted drug delivery is a focus area of current nanotechnology research \cite{Hess2004, Gao2014, Langer2007}. In order to realize the full potential of such applications, it is necessary that the nanomachines be able to efficiently communicate with each other \cite{Moritani2006,Akyildiz2008,dressler2010survey,Nakano2012}. In particular, it is envisioned that a network of
	 communicating nanomachines can help realize the concept of the Internet of Bio-NanoThings which is expected to enable nanomachines to perform complex tasks \cite{Akyildiz2015Internet,Abbasi2017Insulin}. For instance, a group of nanomachines may detect a metabolic condition and communicate this observation to another nanomachine which is then responsible for triggering the release of a drug into the body.   Since conventional communication techniques are not well suited for communication at nano- and microscale, especially in liquid media, molecular communication (MC), where molecules are used as information carriers, has been proposed as a promising bio-inspired mechanism for enabling communication among nanomachines  \cite{Akyildiz2008, Moritani2006}. 

The general structure of a (synthetic) MC system (MCS) is depicted in Fig.~\ref{fig:mcs_einfuehrung}.  In response to a certain input signal, which may be artificial (e.g. a light impulse or an electrical stimulation) or biological (e.g. a nerve signal), the transmitter releases a pattern of signaling particles\footnote{Throughout this paper we use the terms molecules and particles interchangeably, although the latter term is broader as not all particles are molecules.}, which represents the information to be conveyed.  Depending on how sophisticated the transmitter is, it may also apply advanced encoding and modulation techniques for efficient representation of the data before releasing the corresponding signaling particles into the channel. The signaling particles propagate through the channel, e.g. via free diffusion where the propagation may be further  accelerated by advection \cite{Jamali2019Channel}. The receiver observes the signaling particles and recovers the data by applying suitable demodulation and decoding techniques. Thereby, the data may either be read out using an artificial mechanism (e.g. via a light emission or an electrical current) or trigger a biological process (e.g. a nerve signal).

%%%%%%%%%%%%%%%%%%%%%%%%%
\subsection{Motivation and Scope}\label{Sec:Scope}		
%Goal
Although synthetic MC has received considerable interest from the research community over the past decade, the research area is still in its infancy. 
 In particular, the design, analysis, and implementation of microscale biological MCSs require inherently a multidisciplinary approach with contributions from different engineering disciplines, including electrical, biological, and chemical engineering, and different branches of science, including 
biology, chemistry, physics, and medicine. Particularly, the field of synthetic biology is expected to play a crucial role in the fabrication and implementation of the main components of future synthetic MCSs, i.e., the 
transmitter, receiver, and signaling particles.

\begin{figure}
	\centering
	\includegraphics[width=0.8\linewidth]{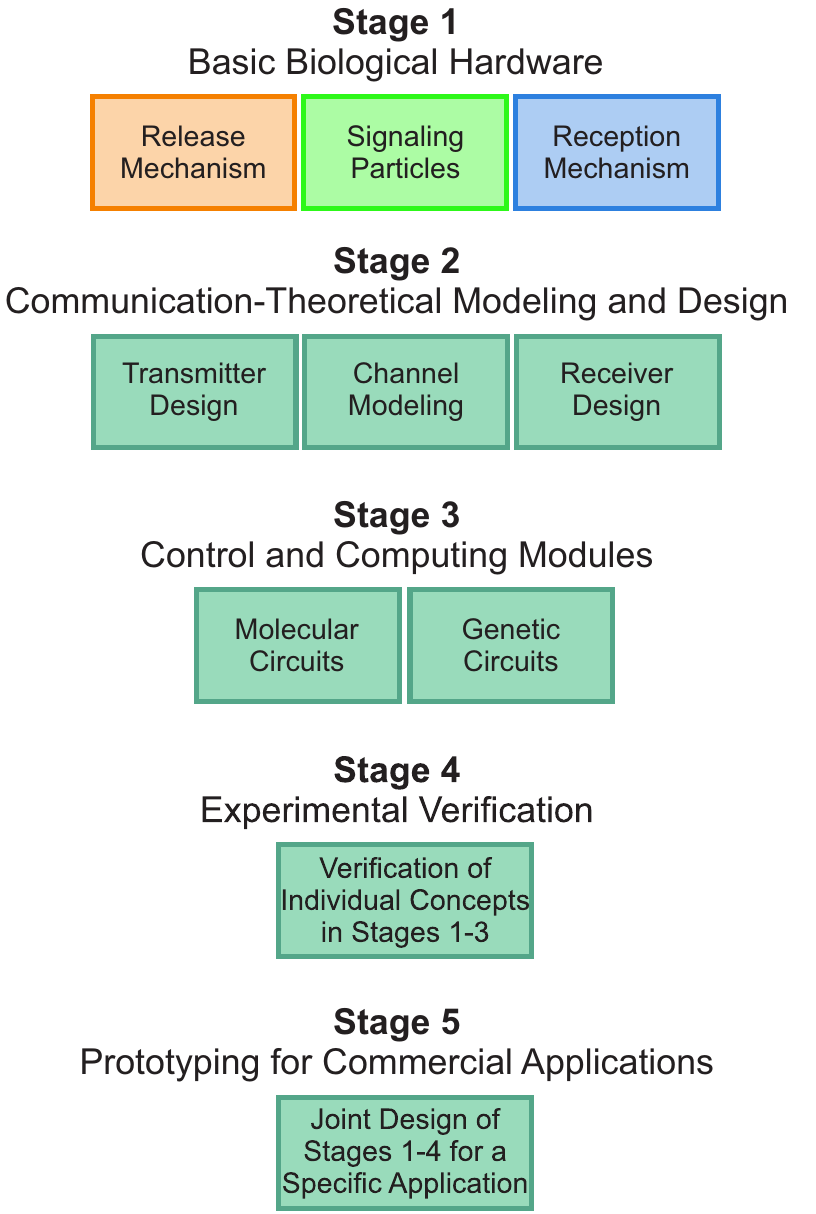}
	\caption{This figure illustrates a roadmap for the development of synthetic biological MCSs towards commercial applications. The focus of this survey is on Stage~1, where we present different biological  options for realizing the release mechanism at the transmitter and the reception mechanism at the receiver for several different types of signaling particles. The biological building blocks proposed in this paper may help theoreticians to develop models and communication-theoretical system designs that account for the relevant biological constraints and features imposed by the proposed building blocks  (in Stage~2). Moreover, we propose  architectures suitable for implementation of both microscale MCSs as well as the micro-to-macroscale interfaces required for experimental evaluation of synthetic MCSs (in Stage~4).}
	\label{fig:MCS_Roadmap}
\end{figure}

In this paper, we review biological components suitable for implementation of MCSs. In order to define the scope of this survey paper and to facilitate the classification of the different research directions in the field of synthetic MC, we present a roadmap for the development of synthetic biological MCSs from the basic biological building blocks to commercial applications, see Fig.~\ref{fig:MCS_Roadmap}. 	
\begin{itemize}
	\item \textbf{Stage 1 -- Enabling Basic Biological Hardware:} The fundamental feature of MCSs is that signaling particles are employed as information carriers \cite{MC_Book}. Therefore, in its most basic form, an MCS consists of a transmitter that is able to release signaling particles into the channel and a receiver which is able to detect the presence of the signaling particles. The primary focus of this survey is the compilation of various biological options for realizing the release mechanism at the transmitter and the reception mechanism at the receiver for several different types of signaling particles.
	\item \textbf{Stage 2 -- Communication-Theoretical Modeling and Design:}  The next step needed for the design of an MCS is the development of communication-theoretical models for the release, propagation, and reception of the signaling particles that account for the features and constraints of the adopted biological building blocks  \cite{Farsad_Channel_2014,Yilmaz_Poiss,Arman_ReactReciever,Bicen2016Channel,Damrath2017Equivalent,Jamali2018Reactive}. Based on these models, the basic functionalities of MCSs such as channel coding \cite{Shih2013Codes,Jamali2018Codes}, modulation \cite{Gursoy2019Modulation,Kuran2011Modulation}, detection \cite{Kuran2011Modulation,Kilinc2013,Adam_OptReciever}, decoding \cite{Chou2015MAP,Jamali2018Codes}, synchronization \cite{Lin2015Synchronization,TNBC_Sync}, and estimation \cite{TCOM_MC_CSI,Noel2015Estimation} can be developed and their performance can be analyzed. 
	\item \textbf{Stage 3 -- Control and Computing Modules:} The implementation of the communication-theoretical concepts developed in Stage 2 depends on where the corresponding operations are to be performed. For instance, for health monitoring applications where the observations can be collected and accessed from outside the MC environment, a personal computer may be responsible for part of the processing. For other applications, such as targeted drug delivery, sophisticated nano-transmitters and nano-receivers may have to process the data themselves. Various options have been proposed for realizing control/computing units at nano- and microscale for biological transmitters/receivers including molecular circuits (i.e., cascaded networks of chemical reactions) \cite{Chou2015MAP,Awan2019Circuit,Chou2019Circuits} and genetic circuits \cite{Marcone2018LDPC,Madsen2012Genetic,Ramakrishnan2009Proteins}.   
	\item \textbf{Stage 4 -- Experimental Verification:} The concepts and designs developed in Stages~1-3 have to be verified via laboratory experiments \cite{Nakano_Microplatform_2008,Akyildiz_testbed_2015,Modulator_Expriment}. Stage~4 is challenging due to the fact that controlling an MCS at microscale
	is difficult. Therefore, in addition to the development of microscale MCSs, it is advantageous to also develop micro-to-macroscale interfaces that enable their observation and test. The role of this stage in the development of MCSs is analogous to employing spectrum analyzers, channel sounders, and other measurement devices to test and analyze the components of wireless communication systems \cite{carvalho2013microwave}. Therefore, this paper does not only survey options for the implementation of MCSs but also the interfaces required for their experimental evaluation.   
	\item \textbf{Stage 5 -- Prototyping for Commercial  Applications:} Depending on the application, suitable building blocks from Stages~1-3 (that have also been experimentally verified in Stage~4) are chosen to develop first-order prototypes and ensure that these blocks successfully work together. At this level, there are numerous applications for MCSs including smart drug delivery, health monitoring, and even the realization of the Internet of Bio-NanoThings \cite{Akyildiz2015Internet,Abbasi2017Insulin,Mosayebi2019Cancer,cao2019diffusive}.   
\end{itemize}

% Figure --------------------
% ---------------------------

So far, the main focus of the MC literature has been on Stages~2 and 3 assuming often quite abstract and simple models for the underlying biological building blocks in Stage~1. In addition, as a proof of concept, MCSs have been demonstrated at macroscale \cite{Farsad_Tabletop_2013,Farsad_Novel_2017,Koo_Molecular_2016,Farsad_Channel_2014,TestBed_Harold,schlechtweg2019magnetic} and at microscale \cite{Nakano_Microplatform_2008, Krishnaswamy_Time_2013, Nakano_Interface_2014, Felicetti_Modeling_2014,Akyildiz_testbed_2015,Modulator_Expriment,Martins2018Bacterial,Nakano2018Leader} (i.e., Stage~4).  The latter systems employ biological components as transmitter and/or receiver but were either demonstrated only for single pulse transmission or offer very low data rates on the order of one symbol per hour. However, for nano- and microscale MC to become practical, continuous transmission at much higher data rates 
is needed. Both the design and the implementation of such systems require a sound understanding of the biological building blocks that can be used to construct them. 
This paper surveys  candidates for signaling molecules, release mechanisms, and reception mechanisms needed in Stage 1 as well as the micro-to-macroscale interfaces needed for their experimental evaluation in Stage~4.

Several survey and tutorial papers focusing on different aspects of MC have been published over the past few years \cite{Akyildiz2008, dressler2010survey,Nakano2012, darchini2013molecular, 120-Nakano2014, Farsad2016,felicetti2016applications,chahibi2017molecular, Okonkwo2017,Wang2017survey,Chen2018Protein,Akyildiz2019Information,Rose2019Capacity,Jamali2019Channel,Kuscu2019Design,Nakano2019methods,Nguyen2019Genetic,McBride2019Loads,Veletic2019Synaptic,Kim2019Redox,Jornet2019Interfaces,Qadri2020},  see Table~\ref{tab:survey} for a brief summary of these survey and tutorial papers. In particular, the authors of  \cite{Akyildiz2008,dressler2010survey,Nakano2012,Wang2017survey} provide general overviews of the field of MC, its future applications, and related challenges. In \cite{Nakano2012,darchini2013molecular,120-Nakano2014,Okonkwo2017}, networking aspects of MCSs are discussed and potential network layer architectures for MCSs are proposed. A general survey on synthetic MCSs, including aspects such as particle transport, communication engineering aspects, testbeds, and applications, is presented in \cite{Farsad2016}. The theoretical aspects of MC are surveyed in detail from the perspectives of information theory in \cite{Akyildiz2019Information,Rose2019Capacity}, physical-layer channel modeling in \cite{Jamali2019Channel}, and transmitter and receiver design in \cite{Kuscu2019Design}. The design of genetic circuits is surveyed in \cite{Nguyen2019Genetic} and the mutual impact of connected biological components is discussed in \cite{McBride2019Loads}.  Potential medical applications of synthetic MCSs are surveyed in \cite{felicetti2016applications,Qadri2020}. Drug delivery applications are discussed in \cite{Okonkwo2017} and applications of mobile MCSs are presented in \cite{Nakano2019methods}. In addition, the authors of \cite{Veletic2019Synaptic} survey research works with a particular focus on MCs in the synaptic cleft and \cite{Chen2018Protein} surveys tools from bioinformatics for the analysis of protein-protein interactions. Finally, the authors of \cite{Kim2019Redox} propose electrochemical methods to interface with biological MCSs whereas \cite{Jornet2019Interfaces} studies optogenomic interfaces for controlling genes and their interactions in the cell nucleus.  These survey papers are either general overviews (e.g., \cite{Akyildiz2008,dressler2010survey,Nakano2012,Wang2017survey,Farsad2016}) or  focus mainly on Stage~2 (e.g., \cite{Akyildiz2019Information,Rose2019Capacity,Jamali2019Channel,Kuscu2019Design,Nakano2019methods}) and Stage~3 (e.g.,  \cite{Nguyen2019Genetic,McBride2019Loads}) of the development roadmap illustrated in Fig.~\ref{fig:MCS_Roadmap}, and although most of them also consider nanomachines as components of MCSs, the corresponding survey of the related literature is very brief. A comprehensive survey of potential biological building blocks of MCSs (in Stage~1) and their micro-to-macroscale interfaces (in Stage~4) is not available in the literature, yet. 

%--------------Table---------------	
\begin{table*}
	\footnotesize
	\raggedleft
	\caption{Summary of survey and tutorial papers on MCSs. Most of the papers listed in this table  provide an overview of MCSs and contain content related to all development stages shown in Fig.~\ref{fig:MCS_Roadmap}. For clarity, the main focus of each survey/tutorial paper is reported in the fourth column of this table.}
	\label{tab:survey}
	\begin{tabularx}{\linewidth}{p{2.8cm}p{0.7cm}p{0.8cm}p{1.8cm}Z}
		\toprule
		\textbfsf{Reference} & \textbfsf{Year} & \textbfsf{Type}  & \textbfsf{Main Focus}  & \textbfsf{More Detailed Notes}\\
		\midrule
		Akyildiz et al. \cite{Akyildiz2008} &  2008 &  Survey & Overview&
		Nanonetworks, architectural aspects, and expected features \\ 
		Dressler et al. \cite{dressler2010survey} &  2010 &  Survey & Overview&
		Bio-inspired networking approaches \\ 
		Nakano et al. \cite{Nakano2012} &  2012 &  Survey & Stage 2&
		Physical and network layers of MCSs \\ 
		Darchini et al. \cite{darchini2013molecular} &  2013 &  Survey & Stage 2&
		MCSs via microtubules and physical contact \\
		Nakano et al. \cite{120-Nakano2014} &  2014 &  Survey & Stage 2&
		Layered architecture of MCSs \\
		Farsad et al. \cite{Farsad2016} &  2016 &  Survey & Overview/ Stage~2&
		Underlying physical principles of MCSs, communication engineering aspects, and simulation tools   \\
		Felicetti et al. \cite{felicetti2016applications} &  2016 &  Survey & Overview&
		Medical applications of MCSs, diagnostic and treatment applications, and implementation interfaces    \\
		Chahibi et al. \cite{chahibi2017molecular} &  2017 &  Survey & Overview&
		Targeted drug delivery, component and modeling approaches    \\
		Okonkwo et al. \cite{Okonkwo2017} &  2017 &  Survey & Stages 1, 2, $\,\,\,\,\,$ and 4&
		Targeted drug delivery, application concepts, propagation channel modeling, and system design   \\
		Wang et al. \cite{Wang2017survey} &  2017 &  Survey & Overview&
		Diffusive MCSs, communication theoretical designs, and cooperative relay-based networks   \\
		Jamali et al. \cite{Jamali2019Channel} &  2019 &  Tutorial & Stage 2&
		Channel modeling of diffusive MCSs, physical principles,  communication theoretical,  simulation-based, and data-driven models, and model derivation methodologies\\ 
		Akyildiz et al. \cite{Akyildiz2019Information} &  2019 &  Tutorial & Stage 2&
		MC theory and models for functional blocks of MCSs based on chemical kinetics and statistical mechanics\\ 
		Rose et al. \cite{Rose2019Capacity} &  2019 &  Tutorial & Stage 2&
		Capacity of point-to-point MCSs\\ 
		Kuscu et al. \cite{Kuscu2019Design} &  2019 &  Survey & Stage 2&
		Transmitter and receiver architectures of MCSs, modulation, coding, and
		detection techniques\\ 
		Nakano et al. \cite{Nakano2019methods} &  2019 &  Survey & Stage 2&
		Mobile MCSs, modeling approaches, and networking\\ 
		Nguyen et al. \cite{Nguyen2019Genetic} &  2019 &  Tutorial & Stage 3&
		Asynchronous genetic circuits\\ 
		McBride et al. \cite{McBride2019Loads} &  2019 &  Tutorial & Stage 3&
		Synthetic biomolecular circuits\\ 
		Veleti\'{c} et al. \cite{Veletic2019Synaptic} &  2019 &  Survey & Stage 2&
		Synaptic communication engineering and brain–machine interface\\ 
		Kim et al. \cite{Kim2019Redox} &  2019 &  Survey & Overview&
		Reduction/oxidation (redox) reactions\\ 
		Jornet et al. \cite{Jornet2019Interfaces} &  2019 &  Survey & Overview&
		Optogenomic interfaces \\ 
		Qadri et al. \cite{Qadri2020} &  2020 &  Survey & Stage~2&
		Internet of Nano-Things for healthcare applications\\ 
		{Söldner et al.}  {(this paper)} &  -- &  Survey & Stages~1 and 4&
		Potential biological building blocks of MCSs, microscale implementations, and macroscale interfaces \\ 
		\bottomrule	
	\end{tabularx}
\end{table*}
%-----------------------------------	

While biological building blocks of MCSs have received little attention in the MC literature, in the field of synthetic biology, there is a vast body of literature  on biological systems that can potentially be used as components
of MCSs. However, for researchers not well versed in synthetic biology, it can be challenging to find the relevant literature and to relate it to MCS design. Therefore, in this paper, we provide a comprehensive survey of biological building blocks 
that can potentially be engineered to serve as components of nano- and microscale synthetic MCSs operating in aqueous
	environments. Since, unlike what is often assumed in the MC literature, biological systems are very specific, the signaling particles, transmitters, and receivers have to be
carefully matched to each other. In particular, the design of the transmitter and receiver in MCSs crucially depends on the adopted signaling particles. Hence, in this survey, we adopt a signaling particle centric approach and first 
present several candidate signaling particles for synthetic MCSs. Then, for each of the considered signaling particles, we provide several candidate transmitter and receiver structures. We believe that this survey is useful to both 
theoreticians and experimentalists. For researchers working on the theoretical aspects of MCS design, taking into account the specific properties of the underlying biological building blocks, which are reviewed here and described in 
detail in the provided references from synthetic biology, will allow them to develop more realistic communication-theoretical models and designs of MCSs. For researchers interested in developing MC testbeds and experiments, the survey outlines 
the advantages and disadvantages of potential design choices and the provided references contain the detailed information needed for implementation. In the following subsections, we first explain some basic biological concepts and components. Then, we provide a brief overview of the considered signaling particles and matching biological components, which can be used to construct transmitters and receivers.                                                                                                                            
%%%%%%%%%%%%%%%%%%%%%%%%%
\subsection{Some Important Basic Biological Concepts and Components} 
In the following, to assist readers that do not have a background in biology, we explain some essential basic biological concepts and components used throughout this article. A summary of additional biological concepts and terminology appearing in the text is provided in Appendix~\ref{Sec;Glossary}. 
\begin{itemize}
	
\item \textbf{Vesicles:} A vesicle is a small, round or oval-shaped container whose wall consists of a lipid bilayer membrane which encloses a liquid substance \cite{Rosoff1996}. Being much smaller than a cell, natural vesicles are formed by deformation and subsequent budding from the cell membrane or the membrane of cellular organelles such as the Golgi apparatus or the endoplasmic reticulum. They are used for transport purposes, e.g. in the context of secretion, for the storage of certain biomolecules, and as compartments with particular reaction conditions. Different proteins may be embedded in their lipid membrane which may facilitate for instance transmembrane transport. Moreover, vesicles with transmembrane proteins can be created artificially using biochemistry and molecular biology tools. Such artificially generated lipid vesicles are called liposomes.

%Such artificial vesicles are called liposomes.

\item \textbf{Ion channels:} An ion channel is a special type of transmembrane protein with a pore that becomes permeable for specific types of ions under certain circumstances \cite{Zheng2016}. Ion channels only allow passive transport which means that they merely facilitate diffusion along an already existing concentration gradient. Ion channels may be subdivided into different groups based on the conditions that lead to channel opening (``gating''). For instance, there are voltage-gated (a certain transmembrane potential is required), ligand-gated (a certain molecule has to bind from the outside), and mechanosensitive (stretching of/pressure on the membrane is required) ion channels \cite{Zheng2016}.  

\item \textbf{Carriers:} Another kind of protein involved in the movement of ions or small molecules across membranes are carriers \cite{Wolfersberger1994}. The transported particles are also referred to as substrate in this context. So called uniporters transport only one specific type of substrate. Upon stimulation, they undergo a conformational change whereby the particle is carried through the membrane to be released at the other side. 
%Like ion channels, this is a form of passive transport that only works in the direction of a concentration or charge gradient \cite{Wolfersberger1994}. 
As other secondary carriers (see below) they are able to accumulate their substrate against a concentration gradient \cite{Wolfersberger1994}. However, there are carriers which do not transport one type of substrate alone, but two or more different types of substrates (e.g. particles A and B) simultaneously, either both in the same direction (symporter)  or in opposite directions (antiporter). This principle, which is called secondary active transport, allows to transport substrate A against its concentration gradient if there is a sufficiently high concentration or charge gradient for substrate B that can be used as a source of energy for the transport process. Often, the concentration gradient for substrate B is actively maintained by use of an ion pump \cite{Wolfersberger1994}.    

\item \textbf{Ion pumps:} Similar to ion channels and carriers, ion pumps are transmembrane proteins which can transport ions across a membrane. In contrast to ion channels, ion pumps use an external source of energy such as light or adenosine triphosphate (ATP) to facilitate an active transport which also works against the concentration gradient of the respective ion \cite{Gadsby2009}. This type of transport is also called primary active transport.

\item \textbf{Voltage-clamp method:} The voltage-clamp technique is a method where a microelectrode is placed inside a vesicle or a cell to manipulate or measure the current across the vesicle membrane at a certain voltage \cite{Rubaiy2017}. Thereby, changes in membrane potential can be induced, e.g. in order to open a voltage-gated ion channel. Moreover, the two-electrode voltage clamp technique can be used to measure the transmembrane current that arises when ion channels are opened. The two-electrode voltage clamp technique allows adjustment of the transmembrane potential and recording of currents through separate electrodes \cite{Rubaiy2017}, and is mostly used for measurements on oocytes or very large vesicles/cells ($>$ 1-2~mm in diameter)  \cite{Guan2013}.

\item \textbf{Reversibility:} In this article, we refer to transmitters as ``reversible'' if the signaling particles are recycled after their release so that they can be used repeatedly. For example, a simple vesicle-based transmitter may eventually get exhausted over time having released all signaling particles that were stored inside at the beginning. In contrast, a ``reversible'' transmitter is able to regenerate its content, e.g. by pumping the signaling particles back inside. In addition, reversibility will require that the vesicle exhibits a high stability and remains intact during multiple regeneration cycles of the transmitter. 
    
\end{itemize}

\subsection{Signaling Particles}\label{subsec:signaling_particles}
\begin{figure*}
\centering
\includegraphics[width=0.7\textwidth]{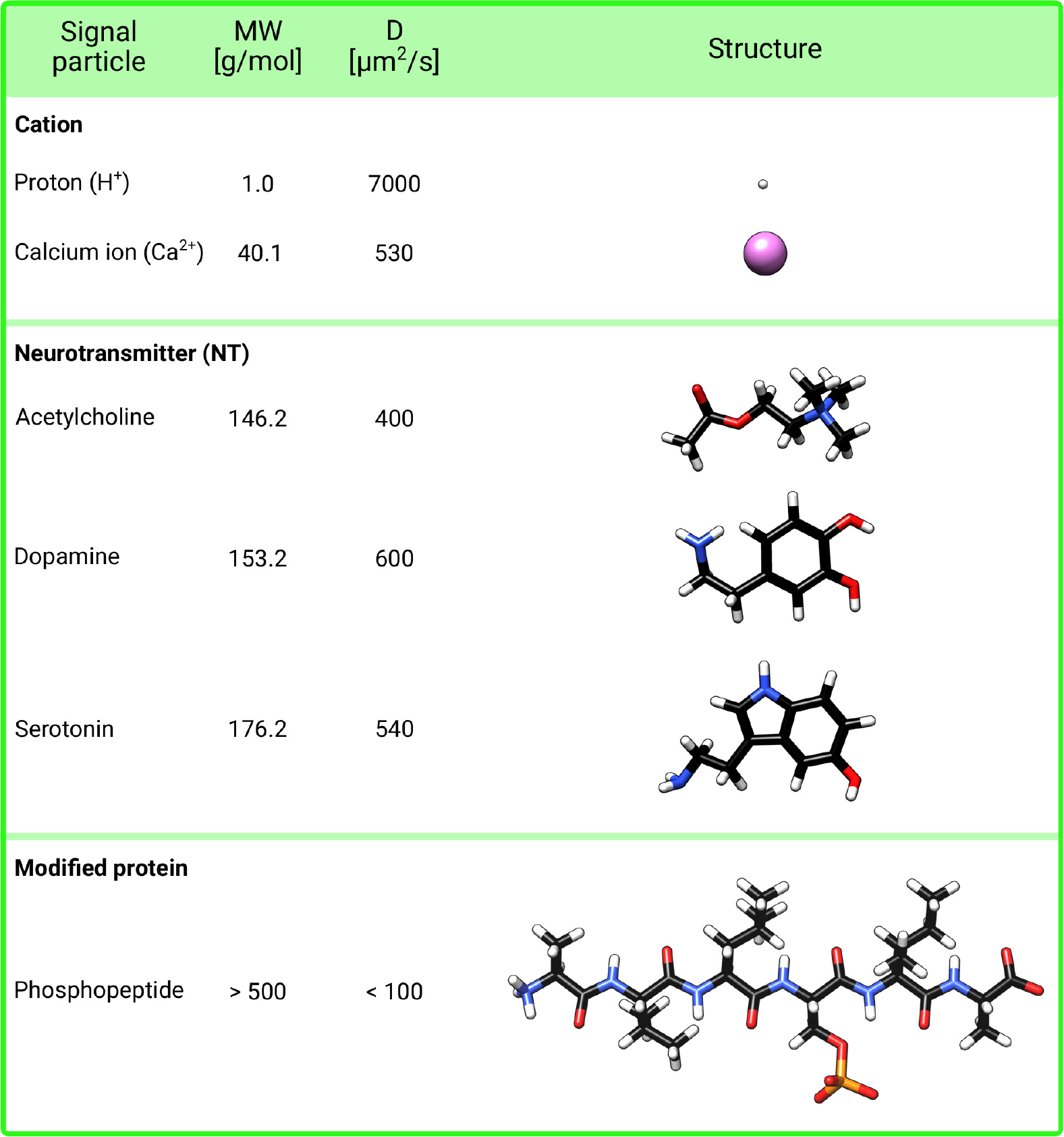}
\caption{Overview of the considered signaling particles, their molecular weight (MW), diffusion coefficient (D), and chemical structure. Carbon atoms are colored in black, hydrogen atoms in white, oxygen atoms in red, nitrogen atoms in blue, and phosphorous atoms in orange. Diffusion coefficients were taken from \cite{BioNum} (protons, acetylcholine), \cite{Donahue1987} (calcium ions), and \cite{Gerhardt1982} (dopamine, serotonin). The structure images were created using UCSF Chimera \cite{Pettersen2004}.}
\label{fig:signal_particles}
\end{figure*}

%%%%%%%%%%%%%%%%%%%%%%%%%
Nature uses a vast number of different molecules for information exchange between different entities. For concreteness, in this survey, we focus on three important types of signaling particles, namely cations, neurotransmitters (NTs), and phosphopeptides as a representative class of modified proteins, see Fig.~\ref{fig:signal_particles}. These classes of signaling particles are attractive for use in synthetic MCSs as they allow the design of simple transmitter and receiver structures employing only a small number of protein components. Furthermore, as will be explained in the following, the considered classes of signaling particles differ substantially in their behavior and properties, such that they are collectively suitable for a wide class of different MCSs. 
	
%Cations
\textbf{Cations:} Cations\footnote{In a similar manner, anions (i.e., small negatively charged ions) can be used for signaling in MCSs.} are small positively charged ions that have the advantage of fast diffusion \cite{BioNum, Donahue1987}. In addition, protons, as a specific ion, can jump from one water molecule to the next through  the formation and concomitant cleavage of covalent bonds, the so-called Grotthuss mechanism \cite{DeGrotthuss1805, Cukierman2006}, which makes them appear to move even faster than they would already be by classic diffusion due to their small size. Since ions interact with a plethora of proteins in organisms \cite{Matthew1985, Riccardi2000, Brown1995}, there exists a large set of possible biological structures that can be used as components of transmitters and receivers. Although many types of cations and anions may be appropriate for MCSs, the present article focuses on protons due to their unique speed of diffusion, and on calcium ions, because they play an important role as messengers in cellular signaling \cite{Clapham2007}. For instance, calcium ions are involved in the process of apoptosis (programmed cell death of damaged cells) \cite{Pinton2008} as well as in the coupling between the electrical excitation and the consecutive contraction of muscle fibers \cite{Caldern2014}.
	
%Neurotransmitters
\textbf{Neurotransmitters:} As an alternative class of particles for MCSs, we consider NTs. NTs are bigger than cations and therefore diffuse more slowly. On the other hand, they are more specific and are therefore more likely to avoid unintended signal interference from and to different natural processes e.g. in the human body. The class of NTs comprises several distinct molecules (e.g. acetylcholine, dopamine, and serotonin) that all function in a similar fashion in signal transduction. This has the general advantage that similar building blocks can be used for the construction of transmitters and receivers for different NTs. In nature, NTs are used to trigger or suppress neuronal firing (the so-called action potentials) between neurons or from a neuron to a muscle fiber \cite{Wecker2010}. Thus, NTs are particularly interesting because they could be used to directly interact with neurons \cite{Nakano2012}, e.g. in the context of nerve lesions which need to be bridged, the control of a prosthesis, and the treatment of neurological diseases. One possible application would be targeted drug delivery. For example, Parkinson's disease which is caused by a selective loss of dopamine producing neurons in a specific region of the brain, the so-called \textit{substantia nigra}, is nowadays treated by a systemic oral administration of levodopa, a precursor of dopamine synthesis \cite{Dhall2016}. In this context, a direct delivery of dopamine via MC to the region where it is required would greatly enhance the treatment effectiveness of Parkinson's disease.       
	 
%Peptides
\textbf{Modified proteins:} The final class of signaling particles discussed in this article are proteins that become signaling particles by post-translational modification (PTM). PTM describes a covalent enzymatic modification of proteins that occurs after protein biosynthesis. There are different types of PTMs (e.g. phosphorylation, methylation, acetylation, and ubiquitination), which differ in their functional groups, their size, and the way they are linked to the protein \cite{Seet2006}. Protein phosphorylation, one of the most frequent types of PTM, is an important cellular regulatory mechanism as many proteins (e.g. enzymes, receptors) are activated/deactivated by phosphorylation and dephosphorylation events. More than two-thirds of the 21,000 proteins encoded by the human genome were shown to be phosphorylated \cite{Ardito2017} indicating that phosphorylation plays a role in almost every physiological process. Abnormal protein phosphorylation may lead to severe diseases including Alzheimer's disease, Parkinson's disease, and cancer. From a communication-theoretic perspective, the protein moiety is always present in the channel but it is only  activated for signaling if it carries a phosphate group (orange phosphorous with red oxygen atoms in Fig.~\ref{fig:signal_particles}) that is transferred by a kinase  \cite{Miller2018}, a special type of enzyme. We consider phosphorylations because of their versatility due to the large number of different signaling particles that can be generated by variation of the protein sequence. In addition, phosphorylated proteins may be miniaturized to generate smaller signaling particles (henceforth termed `phosphopeptides') that have the advantage of faster diffusion due to their smaller size compared to proteins.

%--------------Figure---------------	
%-----------------------------------	

%Generality
 Fig.~\ref{fig:signal_particles} provides an overview of the properties of the different signaling particles that we survey for their applicability in MCSs. From  top to  bottom, they are ordered according to their molecular weights (MWs). As mentioned before, one common  property of all three considered classes of signaling particles is that their respective transmitters and receivers require only a few biological components which makes them attractive for application in synthetic MCSs. In addition, they are intrinsically rather homogeneous groups of signaling particles, which facilitates the design of transmitter and receiver structures as well as communication protocols that can be applied to multiple members of these classes.
	
%Scope
\begin{remark}
We refrained  from considering hormones as signaling particles in this survey, because they are very divergent as far as their size (molecular weights from about 150 g/mol up to insulin which is a protein with 51 amino acids and a molecular weight of 5800 g/mol), the involved receptors (transmembrane, intracellular, and nuclear), and also the effect that they have on their target cell are concerned \cite{Norman2014}. Therefore, it is not possible to develop a general MCS architecture valid for the entire group of hormones. We also do not consider complex biological communication systems such as chemotaxis or bacterial quorum sensing, which would require the challenging implementation of the signaling cascades involved \cite{Witzany2016, Miller2001}. Quorum sensing, for example, depends on the conditional activation of gene transcription \cite{Miller2001} and would thus only be possible if complete bacterial cells were used as transmitters/receivers. 
In this paper, we focus on the design of simple transmitters and receivers consisting of only a few protein components. 
\end{remark}

\begin{remark} 
In the MC literature, there are several works that have analyzed the performance and the design of MCSs for the signaling particles introduced above. For example, the use of calcium/proton ions as signaling particles in MCSs is considered in \cite{Modulator_Expriment, Nakano_C1, Barros_C2, Barros_C3, Bicen_C4, Barros_C5, Barros_C6, Nakano_C7}. Furthermore, several theoretical works investigate the design and analysis of neuronal MCSs employing NTs as information particles, see e.g. \cite{Maham_N1, Veletic_N2, He_N3, Malak_N4, Ramezani_N5, Ramezani_N6}. However, these existing theoretical works either considered abstract models for the transmitter and receiver or focused on transmitter and/or receiver structures of \emph{natural} MCSs. In this paper, unlike the existing works, we investigate different transmitter and/or receiver structures for each of the considered signaling particles with the design of synthetic MCSs in mind. Furthermore, the proposed transmitter and receiver structures are not limited to those existing in nature. 
\end{remark}

\subsection{Biological Components of Transmitter and Receiver}

	%--------------Figure---------------
\begin{figure*} 
	\centering
	\includegraphics[width=0.7\textwidth]{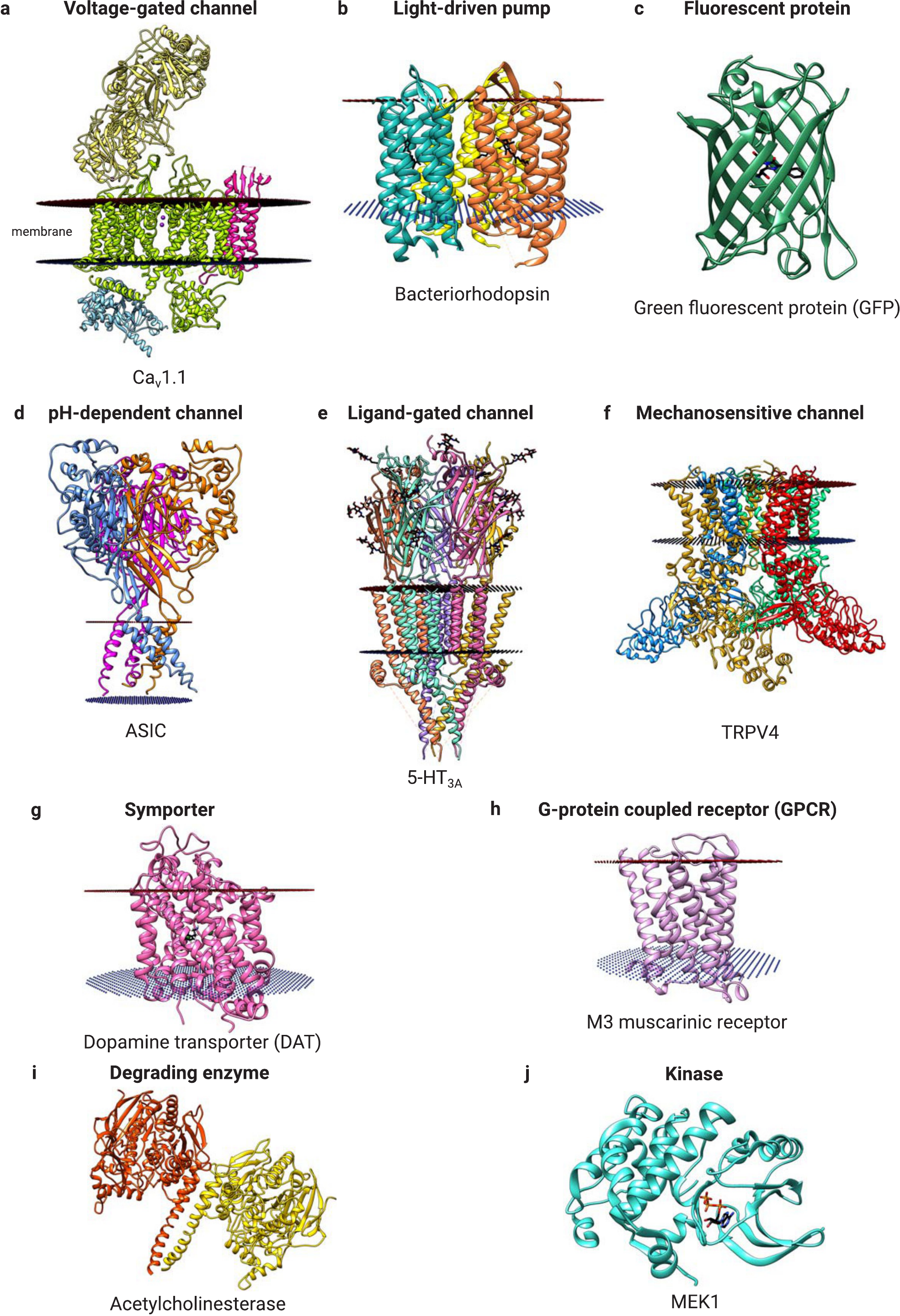}
	\caption{Exemplary protein structures of potential MCSs building blocks. For transmembrane proteins, the position of the membrane is highlighted by two planes. All structures are listed with a unique identifier (PDB code) referring to the protein data bank (\url{www.rcsb.org}). (\textbf{a}) Voltage-gated channel: Ca$_v$1.1 (PDB code \texttt{5GJV} \cite{Wu2016}). (\textbf{b}) Light-driven pump: Bacteriorhodopsin (PDB code \texttt{1M0K} \cite{Schobert2002}). (\textbf{c}) Fluorescent protein: Green fluorescent protein (GFP; PDB code \texttt{1GFL} \cite{Yang1996}). (\textbf{d}) pH-dependent channel: Acid-sensing ion channel (ASIC; PDB code \texttt{4FZ0} \cite{Baconguis2012}). (\textbf{e}) Ligand-gated ion channel: Serotonin receptor subtype 5-HT3A (PDB code \texttt{4PIR} \cite{Hassaine2014}). (\textbf{f}) Mechanosensitive ion channel: Transient receptor potential vanilloid 4 (TRPV4; PDB code \texttt{6C8F} \cite{Deng2018}). (\textbf{g}) NT symporter: Dopamine transporter (DAT; PDB code \texttt{4XP1} \cite{Wang2015}). (\textbf{h}) G-protein coupled receptor (GPCR): M3 muscarinic receptor (PDB code \texttt{4DAJ} \cite{Kruse2012}). (\textbf{i}) NT degrading enzyme: Acetylcholinesterase (PDB code \texttt{4BDT} \cite{Nachon2013}). (\textbf{j}) Kinase: Mitogen-activated protein (MAP) kinase 1 (MEK1; PDB code \texttt{1S9J} \cite{Ohren2004}). Structure images were created using UCSF Chimera \cite{Pettersen2004}.}
	\label{fig:strukturabbildungen}
\end{figure*}
%-----------------------------------

	The signaling particles described in Section~\ref{subsec:signaling_particles} are compatible with several biomolecules that can be used to construct transmitters and receivers for MCSs.  An overview of some relevant classes of such biomolecules is given in Fig.~\ref{fig:strukturabbildungen}.  One relevant protein family are channels that allow ions to pass through a membrane via a channel pore in response to an external stimulus (e.g. voltage, pH, ligands, or mechanical stress) (Figs. \ref{fig:strukturabbildungen}a, d, e, f). In contrast to such ion channels, which only allow diffusion of ions along a concentration gradient, pumps (Fig.~\ref{fig:strukturabbildungen}b) and symporters (Fig.~\ref{fig:strukturabbildungen}g) 
can move molecules against a concentration gradient at the expense of energy consumption. G-protein coupled receptors (Fig.~\ref{fig:strukturabbildungen}h) detect ligands outside the cell and convert the corresponding signal to intracellular responses. These biomolecules can be integrated into the membrane of cells or 
artificial vesicles using the tools of synthetic biology \cite{Beales2017, Garni2017, Skrzypek2018, Kuhn2017, Jorgensen2016}.
In addition to these membrane-bound components, there exist also soluble proteins with properties relevant for MCSs. For instance, kinases (Fig.~\ref{fig:strukturabbildungen}j) can convert peptides into signaling particles and enzymes 
(Fig.~\ref{fig:strukturabbildungen}i) allow the degradation of signaling particles.

\begin{figure*}
	\centering
	\includegraphics[width=1.25\textwidth,angle =90]{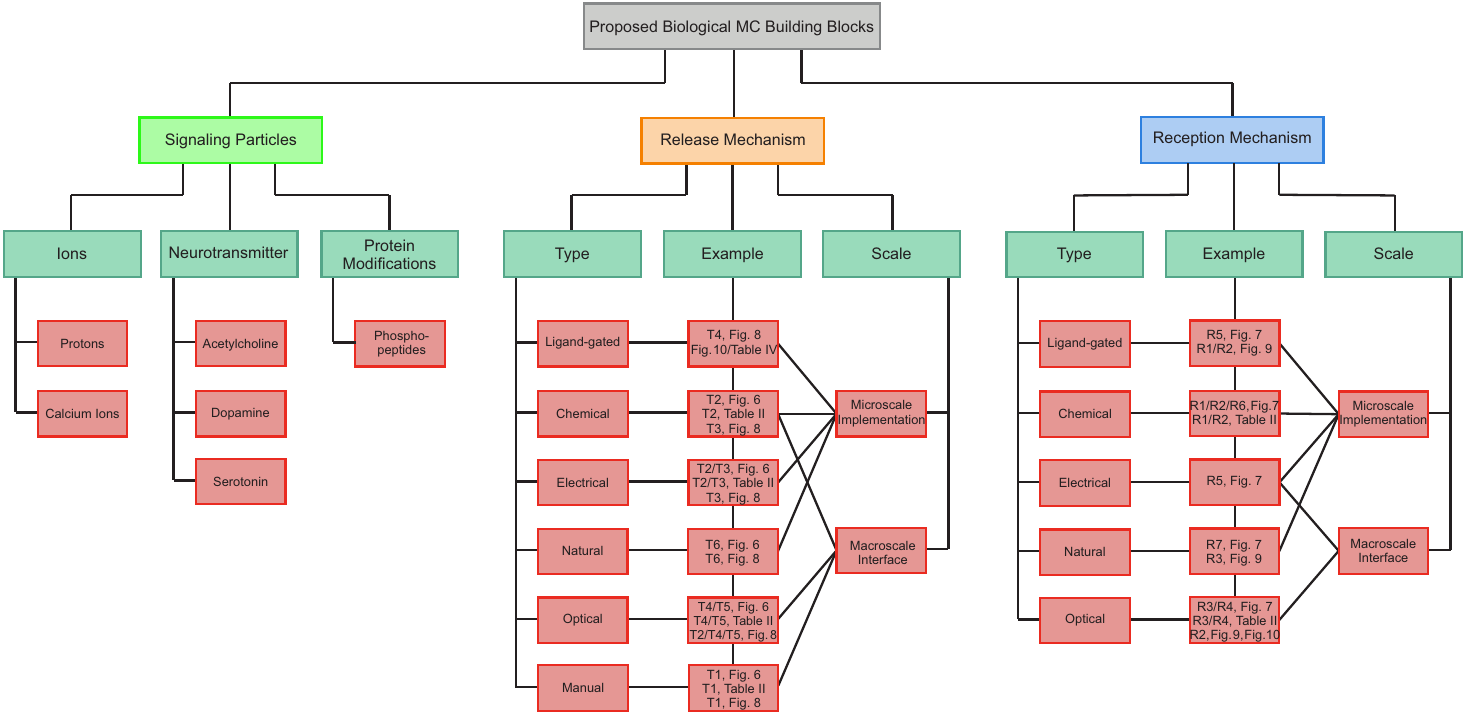}
	\caption{Classification of the biological building blocks for the MCSs presented in this paper.}
	\label{fig:taxonomy}
\end{figure*}

{Fig.~\ref{fig:taxonomy}} presents different classifications of the biological transmitter and receiver architectures discussed in this paper. As mentioned in Section~\ref{Sec:Scope}, experimental verification of the concepts conceived in Stages~1-3 of the proposed roadmap for the development of MCSs is in general difficult since controlling an MCS at microscale is a challenging task. Therefore, for each class of signaling particles, we present options for both microscale implementation of the MCSs and microscale-to-macroscale interfaces that can be used for their experimental evaluation (i.e., in Stage~4). For instance, light-driven ion-pumps can be inserted into the membrane of the transmitting cell to enable the release of ions (as signaling particles) in response to an external light stimulus which is easy to control in laboratory experiments. For practical microscale MCSs, other stimuli may be preferred such as other molecules (originating potentially from other synthetic or natural sources) which open corresponding ligand-gated channels to release the signaling particles.

Conceptually, transmitter architectures can be also categorized based on whether the particle release is induced manually (e.g., via a pipette in a laboratory experiment); initiated by binding other molecules to  ligand-gated channels on the transmitter surface; or triggered by an electrical, optical, or chemical stimulus. Similarly, receiver architectures can be classified based on whether the signaling particles bind to  receptors on the receiver surface and trigger another secondary signal inside the receiver or the reception process causes an electrically, optically, or chemically detectable signal. In addition to design principles for synthetic transmitters and receivers, we also discuss natural mechanisms in the human body for the release and detection of the considered signaling particles. These mechanisms can be interpreted as natural transmitters and receivers or natural sources of interference.

%--------------Figure---------------

\subsection{Organization of the Paper} 
In the following, we outline how the remainder of the paper and the individual sections are organized. In particular, in Sections~\ref{sec:protons} and \ref{sec:Calcium}, we discuss transmitter and receiver mechanisms for protons and calcium ions, respectively. In Section~\ref{sec:neurotransmitter}, transmitter and receiver structures for NT signaling particles are presented. Section~\ref{sec:peptides} introduces transmitters that use phosphorylation as a mechanism for controlling the release of peptide particles as well as different receivers that can detect phosphorylated peptides. Several extensions of the proposed transmitter/receiver structures to more complex architectures, capable of performing more elaborate processing tasks, are also provided. In Sections~\ref{sec:protons}-\ref{sec:peptides}, the respective different synthetic transmitter and/or receiver architectures are presented in order of
increasing \emph{sophistication} and  \emph{complexity} required for integrating the necessary components into the transmitter and receiver. Comparatively simple building blocks, which require external devices or the use of detergents, are rather intended for \textit{in
	vitro} purposes such as testbeds or benchmarking experiments. They could be used for an initial
	proof of concept whereas more complex cell- or vesicle-based systems may be better suited for
	\textit{in vivo} applications.  We note that there is no one-to-one mapping between the orders in which the transmitters and the receivers are presented. In other words, depending on the specific application, the proposed transmitter and receiver structures can be flexibly combined. Moreover, in Section~\ref{sec:GC}, we compare the different signaling particles studied in this paper, discuss potential medical applications of the considered MCSs, and present several practical considerations for the implementation of these MCSs including relevant \emph{biological} mechanisms for inter-symbol interference (ISI) mitigation  and bottlenecks for the achievable data rates. Furthermore, in Section~\ref{sec:futurework}, we outline possible future research directions including the new research problems that should be tackled for the modeling and design of MCSs based on the proposed biological building blocks as well as challenges that need to be addressed for the implementations of such MCSs. Finally, the conclusions of the paper are drawn in Section~\ref{sec:conclusions}.
		
%%%%%%%%%%%%%%%%%%%%%%%%%%%%%%%%%%%%
\section{Protons as Signaling Particles\label{sec:protons}}
\begin{figure*}
		\centering
		\includegraphics[width=0.7\textwidth]{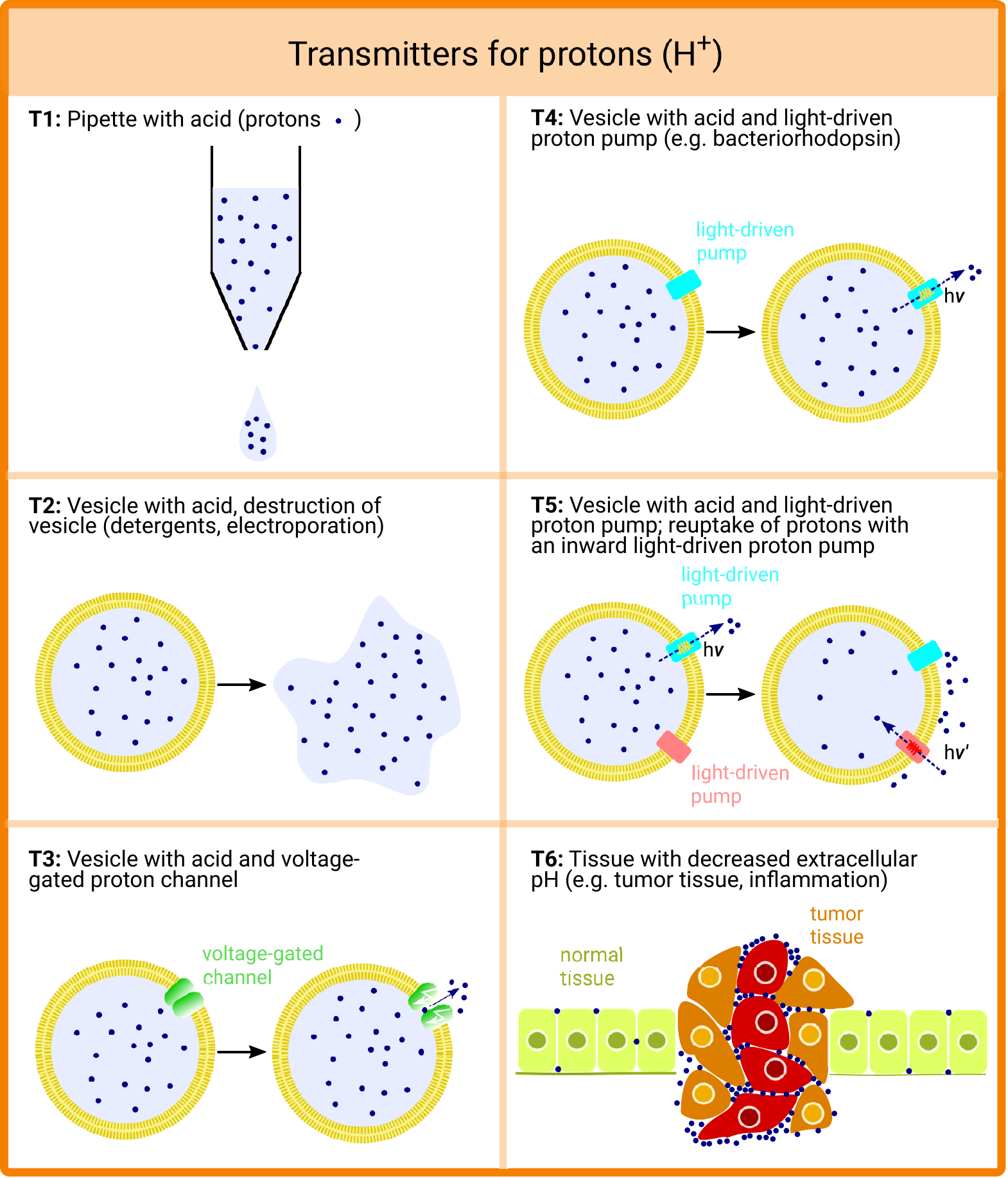}
		\caption{Transmitters for protons as signaling particles. The suggested building blocks for different transmitter systems are presented in order of increasing complexity and include designed artificial systems and physiological emitters, which may naturally occur in the human body.}
		\label{fig:proton_transmitters}
	\end{figure*}

%%%%%%%%%%%%%%%%%%%%%%%%%%%%%%%%%%%%			
Protons (represented by the symbol H$^+$) are the first type of signaling particles we  consider for MCSs. The corresponding options for transmitter systems are depicted in Fig.~\ref{fig:proton_transmitters}, those for receiver systems are presented in Fig.~\ref{fig:proton_receivers}. They are described in detail in the following.
%%%%%%%%%%%%%%%%%%%%%%%%%%%%%%%%%%%%	
\subsection{Transmitters}
%%%%%%%%%%%%%%%%%%%%%%%%%%%%%%%%%%%
We consider six different transmitter structures (Fig.~\ref{fig:proton_transmitters}, T1-T6). While the first  transmitter (T1) is quite simple (for laboratory experiment purposes), transmitters T2-T5 are more complex and consist of a vesicle containing an acidic solution and employ different mechanisms for the controlled release and/or reuptake of protons. Finally, we also briefly consider proton emitters (T6) which may naturally occur in the human body.\\
%Pipette or destroying vesicle
\textbf{T1 (Pipette):} The simplest transmitter for protons is a pipette by which an acidic solution is released dropwise into the channel (Fig.~\ref{fig:proton_transmitters}, T1). This transmitter may be suitable for 
controlling the release of protons at macroscale.\\
% Vesicle
\textbf{T2 (Degenerating vesicle):} As second and more complex transmitter system for protons, we propose the usage of vesicles that contain an acidic solution (Fig.~\ref{fig:proton_transmitters}, T2). In the simplest case, the release of the vesicle content could be triggered by adding a detergent destroying the membrane or by means of electroporation \cite{Weaver1996} where the membrane permeability is increased due to an externally applied electric field. Both approaches are very effective; however, they also destroy the entire vesicle and thus release the entire content at once. In order to transport only partial quantities of the signaling particles from the interior of the vesicle to the outside, different transporter proteins can 
be incorporated into the vesicle membrane as will be explained in the following.\\
%Ion channels
\textbf{T3 (Ion channels):}  For the third transmitter model, we propose to use ion channels (e.g. voltage-gated proton channels) for a controlled release of the signaling particles from the vesicle (Fig.~\ref{fig:proton_transmitters}, T3). 
Voltage-gated ion channels are transmembrane proteins (Fig.~\ref{fig:strukturabbildungen}a), which undergo conformational rearrangements due to changes in the electrical membrane potential near the channel. Such changes of membrane potential can be induced 
artificially, e.g. by the 
voltage-clamp technique \cite{Rubaiy2017}. The channel opens through these conformational changes and ions can leave the vesicle. Voltage-gated proton channels exhibit a high selectivity allowing only protons to leave the vesicle \cite{DeCoursey2018}, which makes them perfect candidates for outward transportation. \\
% ion pumps
\textbf{T4 (Ion pumps):} Alternatively, for the fourth transmitter model, light-driven or ATP-driven pumps are exploited for the proton transmitter system (Fig.~\ref{fig:proton_transmitters}, T4). While voltage-gated ion channels conduct cations or anions in a 
passive manner, ion pumps need an external source of energy and function as active transporters, which allows them to build up a proton concentration gradient \cite{Inoue2016}.  As light-driven outward proton pump, for instance, bacteriorhodopsin \cite{Arjmandi2016} may be used (Fig.~\ref{fig:strukturabbildungen}b), which can
be embedded in the membrane of proton containing vesicles. In particular, there are several known variants of bacteriorhodopsin which differ in the wavelength required  for activation \cite{Soppa1989}. Recently, an experimental testbed employing bacteriorhodopsin has been reported in \cite{Modulator_Expriment}. As an ATP-driven pump, V-ATPases or P-ATPases \cite{Anandakrishnan2017} could be used and activated by addition of ATP in the vicinity of the vesicle.\\
            %Inward pump
\textbf{T5 (Inward ion pumps):} The previous two proposed transmitter mechanisms (Fig.~\ref{fig:proton_transmitters}, T3 and T4) enable the controlled release of signaling particle; however, they lack reversibility, i.e., the recycling of the 
signaling particles by the transmitter. By recycling the signaling particles, the transmitter can harvest some of the previously released protons for future releases. Moreover, when transmitter and receiver are placed close to each other, as e.g. in the synaptic cleft, recycling the signaling particles aids in clearing the channel for future releases, and thereby, reducing ISI. Reversibility can be achieved by adding a second biomolecule that transports the signaling particle back into the vesicle after signal detection. As inward transporter, light-driven inward proton pumps can be additionally integrated into the vesicle \cite{Inoue2016}. These inward pumps are able to transport protons from the surrounding solution into the interior of the vesicle (Fig.~\ref{fig:proton_transmitters}, T5). In case that light-driven outward proton pumps are used for signaling particle release (Fig.~\ref{fig:proton_transmitters}, T4), it is important to choose
outward and inward proton pumps that are activated by different wavelengths in order to avoid the simultaneous operation of both proton pumps.
\begin{remark}	
In Fig.~\ref{fig:proton_transmitters}, light-driven pumps are labelled by $h\nu$ because the energy of the photon used as a driving force may be calculated as $E=h\nu$, where $h$ denotes the Planck constant and $\nu$ is the frequency of the light \cite{Tipler2015}. In T5, the inward light driven pump is referred to as $h\nu^\prime$, indicating the fact that a different activation frequency/wavelength is needed compared to the outward pump.   
\end{remark}

%Proton signals also generated naturally
\textbf{T6 (Natural transmitters):} Besides these designed synthetic transmitter systems (T1-T5), there are also some processes in the human body that may be interpreted as proton emission. For example, some tissue has a higher proton concentration (lower pH value) than other tissue (Fig.~\ref{fig:proton_transmitters}, T6). Lower pH values can be observed e.g. in inflamed tissue but also the extracellular pH of tumors can be heterogeneous and acidic \cite{Van1999}, which may be used for detection of cancer cells.  
%%%%%%%%%%%%%%%%%%%%%%%%%%%%%%%%%%%
\subsection{Receivers}
%%%%%%%%%%%%%%%%%%%%%%%%%%%%%%%%%%%
Protons as signaling particles lead to an increase of the proton concentration at the receiver or equivalently a reduction of the solution pH. We discuss five possible synthetic receiver architectures (R1-R5) for protons and two natural processes
in the human body (R6, R7) that are triggered by changes in the proton concentration.\\
%%%%%%%%%%%%%%%%%%%%%%%%
	\begin{figure*} 
		\centering
		\includegraphics[width=0.7\textwidth]{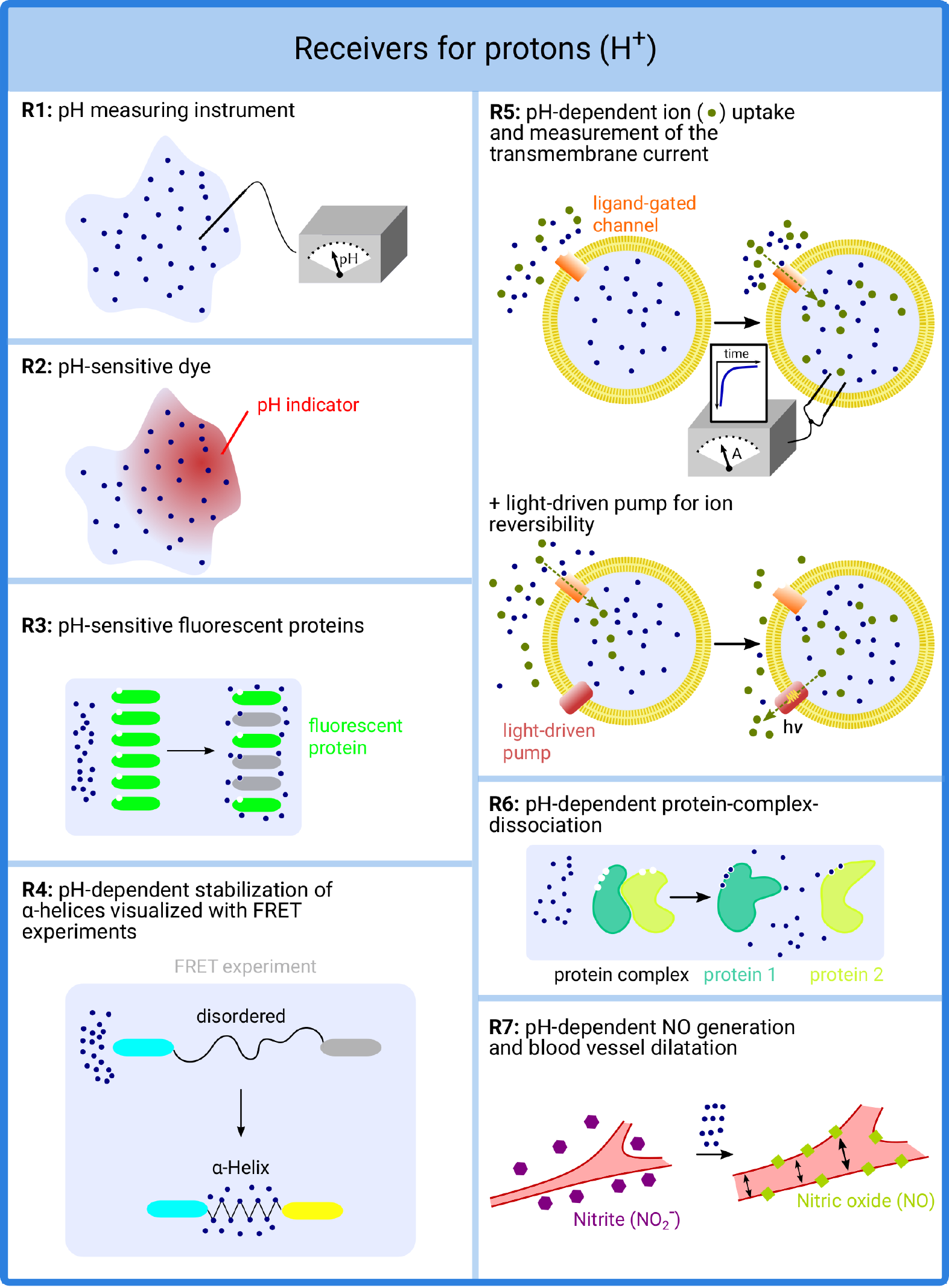}
		\caption{Receivers for protons as signaling particles. The suggested building blocks for different receiver systems are presented in order of increasing complexity and include designed artificial systems and physiological receivers, which naturally occur in the human body. Protons are indicated as blue dots.}
		\label{fig:proton_receivers}
	\end{figure*}
%%%%%%%%%%%%	
\textbf{R1 and R2 (pH sensors):} The simplest approaches to detect a change in the proton concentration of a solution are pH-measuring instruments (Fig.~\ref{fig:proton_receivers}, R1) and pH-sensitive dyes (pH indicators) (Fig.~\ref{fig:proton_receivers}, R2) as they are routinely used in chemical and biological laboratories. The changes of dye colors can for example be detected by photometry in a second step. \\
%GFP
\textbf{R3 (Fluorescence proteins):} In biological experiments, pH-sensitive fluorescence proteins are also often used as pH indicators and they are good candidates for use in nanoscale MCSs. One example for such a fluorescent protein is the green fluorescent protein (GFP) (Fig.~\ref{fig:proton_receivers}, R3) 
with its characteristic $\beta$-barrel structure and the fluorescent chromophore (fluorophore) in the center (Fig.~\ref{fig:strukturabbildungen}c). The fluorescence is dependent on the protonation of the fluorophore and in the last years many different variants were developed, which possess an altered pH-sensitive excitation spectrum. For some of these variants, a response time of less than 20 ms could be demonstrated \cite{Miesenbck1998}. Besides GFP variants with disappearing fluorescence through decreasing pH values, also GFP variants which emit light of different color at different pH values were reported \cite{Day2009}. For example, for excitation at 388 nm, the GFP from the sea cactus \textit{Cavernularia obesa} has a blue fluorescence at pH 5 and below, whereas it shows a green fluorescence at pH 7 and above \cite{Ogoh2013}. In case of long-term fluorescent measurements, the light-induced bleaching of the fluorophores has to be considered. Photobleaching results in a decreasing signal that is independent from pH variations. Here, ratiometric pH-sensitive GFP variants, such as pHluorin2, have a big advantage since the pH is not estimated from simple changes in the fluorescence level but from the ratio of the fluorescence intensities at two different excitation wavelengths \cite{mahon2011phluorin2}. Thus, photobleaching will not affect the correctness of the pH measurement, but only the lifetime of the receiver.\\
%Detect conformational change by some light excitement experiment
\textbf{R4 (FRET):} Another possibility for a receiver system is the usage of biomolecules, which undergo large conformational changes upon a decrease in the surrounding pH value. One example are peptides consisting of polyglutamate, which form $\alpha$-helices in acidic solution while being disordered at intermediate or high pH (Fig.~\ref{fig:proton_receivers}, R4) \cite{Finke2007}. This conformational rearrangement of the structure can be visualized and made detectable with a coupled FRET (F\"{o}rster resonance energy transfer or fluorescence resonance energy transfer) experiment. The underlying mechanism is an energy transfer between two light-sensitive molecules (chromophores). A donor chromophore, which is excited by irradiation with light of a certain wavelength, transfers energy to an acceptor chromophore. The shorter the distance between the two chromophores, the more energy can be transferred from the donor to the acceptor chromophore.  Hence, FRET is very sensitive to small changes in distance between the chromophores. This effect can be read out by monitoring the ratio of the respective light intensities emitted by the donor and the acceptor at different wavelengths. Alternatively, the intensity of the donor can be compared in the presence and absence of the acceptor \cite{Clegg2009}.\footnote{We note that the underlying FRET mechanism as an option for communication at nanoscale has been theoretically investigated in \cite{Kuscu_F1, Kuscu_F2, Kuscu_F3, Wojcik_F4, Solarczyk_F5}.} We propose the construction of a FRET-based receiver employing a fusion protein with two FRET partners at the termini and a polyglutamate sequence in between as described in \cite{Finke2007}. Upon helix formation in acidic pH, the distance between the two ends decreases. In biological experiments, a common donor-acceptor pair is the cyan fluorescent protein (CFP) combined with the yellow fluorescent protein (YFP) \cite{Bajar2016, Day2009} which are both color variants of GFP (Fig.~\ref{fig:strukturabbildungen}c).\\
            %Let proton open ion channel (for other ion) and detect that; also reversible
\textbf{R5 (Proton-gated channels):} As a further option for a pH-dependent receiver system, we propose the usage of proton-gated channels, like acid-sensing ion channels (ASICs, Fig.~\ref{fig:strukturabbildungen}d) \cite{Grnder2015, Hanukoglu2016} or proton-gated proton channels (e.g. the viral p7 protein) \cite{Breitinger2016} embedded in a vesicle (Fig.~\ref{fig:proton_receivers}, R5). These ligand-gated ion channels are permeable for certain types of cations after activation by high proton concentration at their extracellular side \cite{Grnder2015, Hanukoglu2016}. The cations can flow via the channel pore in the vesicle membrane into the interior of the vesicle and increase the concentration of cations in the vesicle, which can be detected by measuring the transmembrane current by methods such as the two-electrode voltage clamp method \cite{Guan2013}. For reversibility, an additional light-driven outward pump (e.g. KR2 of the marine bacterium \textit{Krokinobacter eikastus} \cite{Inoue2013} for sodium ions or bacteriorhodopsin for protons) can be embedded into the vesicle membrane. With such light-driven pumps, it would be possible to pump the cations from the interior of the vesicle back into the surrounding solution.\\
            %Examples of proton-sensitive biological processes
 \textbf{R6 (Proton-triggered protein dissociation):} In nature, there are numerous examples where a protein complex dissociates in response to decreasing pH values (Fig.~\ref{fig:proton_receivers}, R6). One such example is described for the periplasmic protein HdeA, which forms a well-folded homodimer at neutral pH. Under acidic conditions, HdeA unfolds and exhibits an enhanced tendency to dissociate into monomers \cite{Salmon2018, Socher2016}. Another example is the human Hsp47 protein, which is a collagen-specific molecular chaperone and indispensable for molecular maturation of collagen \cite{Ito2017}. Hsp47 transiently binds to procollagen in the endoplasmic reticulum (ER, neutral pH) and dissociates from it in a pH-dependent manner once this complex is transported to a compartment with lower pH (e.g. the \textit{cis}-Golgi or the ER-Golgi intermediate compartment) \cite{Ito2017}. These examples demonstrate that various physiological processes may be triggered by modulating proton concentrations.\\
            %Application: Let blood vessel shrink and expand
\textbf{R7 (Proton-modulated ``acidic-metabolic'' vasodilatation):} In the human body, decreasing pH values during hypoxia/ischaemia can also lead to a physiological mechanism known as ``acidic-metabolic'' vasodilatation \cite{Modin2001} (Fig.~\ref{fig:proton_receivers}, R7), which improves blood flow and oxygen supply. Vasodilator nitric oxide (NO) is generated through a non-enzymatic reduction of inorganic nitrite (NO$_2^-$) to NO, a reaction that takes place predominantly 
during acidic/reducing conditions \cite{Modin2001}. Thus, ``acidic-metabolic'' vasodilatation represents another physiological process that may be modulated by proton-based synthetic MCSs. 
            	
	\section{Calcium Ions as Signaling Particles\label{sec:Calcium}}
		Besides protons, other types of ions may be suitable for synthetic MCSs as well. In many cases, there exist building blocks similar to those presented in Section~\ref{sec:protons}. Due to their particular role as second messengers\footnote{Intracellular chemical substance whose concentration changes upon a primary signal, e.g. a ligand binding to a transmembrane receptor \cite{Krauss2008}.} in the human body, we illustrate in this section how the concepts proposed for protons may be transferred to calcium ions (represented by the symbol Ca$^{2+}$). Table~\ref{tab:proton_calcium_vergleich} summarizes the proposed building blocks and underlines the high similarity of the underlying components for both types of signaling particles.
		 
		 		\begin{table*}[!htb]
		\footnotesize
		\raggedleft
		\caption{Transferability of the building blocks suggested for protons (Figs. \ref{fig:proton_transmitters} and \ref{fig:proton_receivers}) to MCSs using Calcium ions as signaling particle. Comparison of possible transmitters and receivers.}
		\label{tab:proton_calcium_vergleich}
		\begin{tabularx}{\textwidth}{lXX}
			\toprule
			\textbfsf{Component} & \textbfsf{Protons (H$^+$)} & \textbfsf{Calcium ions (Ca$^{2+}$)} \\ \addlinespace
			\midrule
			\textbfsf{T1 (Pipette)} & Pipette with acidic solution & Pipette with solution containing calcium salt (e.g. CaCl$_2$) \\ 
			\textbfsf{T2 (Degenerating carriers)} & Vesicle with acidic solution & Vesicle with calcium ions, alternatively light-sensitive caged calcium or thermosensitive microcapsules \\ \addlinespace
			\textbfsf{T3 (Ion channels)} & Vesicle with acid and passive transport via voltage-gated proton channel \cite{DeCoursey2018} & Vesicle with calcium ions and passive transport via calcium channel, different gating mechanisms:\newline
			\textbullet~Voltage-gated (e.g. Ca$_v$1.1 \cite{Catterall2005})\newline
			\textbullet~Ligand-gated (e.g. 5-HT-3A + serotonin \cite{Ronde1998})\newline
			\textbullet~Mechanosensitive (e.g. TRPV4 \cite{Ho2012}) \\ \addlinespace
			\textbfsf{T4/T5 (Ion pumps)} & Vesicle with acid and active transport via proton pump, different energy sources:\newline
			\textbullet~Light-driven (e.g. bacteriorhodopsin \cite{Arjmandi2016})\newline
			\textbullet~ATP-driven (e.g. V-ATPases, P-ATPases \cite{Anandakrishnan2017}) &
			Vesicle with calcium ions and active transport via calcium pump, different energy sources:\newline
			\textbullet~Light-driven (e.g. engineered pump from \cite{Bennett2002}) \newline
			\textbullet~ATP-driven (e.g. Ca$^{2+}$-ATPase \cite{Toyoshima2009}) \\ \addlinespace
			\midrule
			\textbfsf{R1 (Measuring instrument)} & pH-meter & Conductivity measuring instrument \\ \addlinespace
			\textbfsf{R2 (Dye)} & pH-sensitive dye/pH indicator & Calcium-sensitive fluorescent dye (e.g. Fura-2, Indol-1, Fluo-3, Fluo-4, Calcium Green-1 \cite{Bootman2013, Lee1999}) \\ \addlinespace
			\textbfsf{R3 (Fluorescence proteins)} & GFP (different variants available) \cite{Miesenbck1998, Ogoh2013} & Fusion constructs of GFP variants and calmodulin (e.g. GCaMP \cite{Nakai2001}) \\ \addlinespace
			\textbfsf{R4 (FRET)} & $\alpha$-helix stabilization due to protonation, fluorescence proteins at the termini forming a FRET pair  \cite{Finke2007} & Calcium indicators containing troponin C and a FRET pair (e.g. TN-XL (CFP+YFP) \cite{Mank2006}) \\
			\bottomrule
		\end{tabularx}
	\end{table*}

		\subsection{Transmitters}
		
		Transmitter \textbf{T1 (pipette)} as well as the simple vesicle-based transmitter \textbf{T2~(degenerating vesicle)}, which does not contain specific membrane proteins, may be adapted for calcium ions simply by replacing the acid with a solution containing a calcium salt such as CaCl$_2$. Compared to protons, there are some additional degenerating carriers triggered by light or temperature which may be used as an alternative\footnote{One type of such carriers may be so-called light-sensitive caged compounds \cite{EllisDavies1996, EllisDavies2007}, which are already commercially available. Like in the vesicle, the calcium ions are shielded at the beginning and thus biologically inactive, which is due to a bound photoswitchable molecule. Upon irradiation with light of a certain wavelength, the shielding agent gets cleaved and the calcium ions are released from their cage \cite{EllisDavies1996, EllisDavies2007}. The same general concept, but with heat instead of light as a trigger mechanism, could also be realized by the use of thermosensitive microcapsules as described in \cite{Nuyken1991}.}. 
		
		Biological transmitters T3-T5 can be constructed by replacing the described proton channels and proton pumps in Fig.~\ref{fig:proton_transmitters} by calcium-specific proteins with an analogous function. Some suggestions for how this could be realized in detail are given below.\\   
		\textbf{T3 (Ion channels):}  As proposed for protons, the outward transport of calcium ions may be accomplished with ion channels as well. There exist some voltage-gated channels for calcium ions such as Ca$_v$1.1 \cite{Catterall2005} (Fig.~\ref{fig:strukturabbildungen}a). Although a stimulation by electricity would be convenient in testbeds and for applications outside the human body, the medical use of MCSs may require a release of calcium ions by triggers which are less invasive and therefore more biocompatible. Fortunately, for calcium ions, there are some additional gating mechanisms available compared to protons.  
		
		 One particular interesting example are ligand-gated channels. These ion channels become only permeable if a ligand binds from the outside causing thereby a conformational change of the protein. One possible candidate for calcium ion receptors is the 5-HT-3A receptor \cite{Ronde1998} (Fig.~\ref{fig:strukturabbildungen}e) which opens in response to serotonin. So, serotonin could be administered in order to release calcium ions from the vesicle. Since serotonin is a NT (see Section~\ref{sec:neurotransmitter}) and thus a natural signaling particle of the human body, it might even be possible to couple the calcium release directly to the activation of a serotonergic\footnote{Neuron/synapse which produces serotonin or uses serotonin as an NT.} neuron.  This would allow to use the input from a nerve fiber for activation of the transmitter. This general principle can also be applied to other pairs of ligand-gated channels and their physiological ligand, of course. This allows for the direct coupling of a biological process and a synthetic MCS. 	
		
		Another interesting option are mechanosensitive calcium channels, which are opened in response to mechanical stress \cite{Martinac2012}. One possible approach could be the insertion of the vesicles between the fibers of the extracellular matrix at some location in the body. Upon mechanic shear or pressure on the respective tissue, a calcium release would be triggered. This principle could be useful in the context of targeted drug delivery, e.g. in order to facilitate the local administration of an anesthetic. One example for such a mechanosensitive calcium channel involved in nociception, i.e., the encoding and processing of pain stimuli in the human body, is the transient receptor potential vanilloid 4 (TRPV4) \cite{Ho2012} which is depicted in Fig.~\ref{fig:strukturabbildungen}f.\\
		\textbf{T4/T5 (Ion pumps):} Besides passive channels, similar to protons, calcium ions can actively be transported using either a light-driven pump, that has been artificially created \cite{Bennett2002}, or an ATP-driven Ca$^{2+}$-ATPase \cite{Toyoshima2009}. If two light-driven pumps are intended to be used in the same vesicle to allow for transmitter regeneration (reversibility), the second pump would need to be engineered to work at a different wavelength as has already been reported for some bacteriorhodopsin mutants \cite{Soppa1989}.

		\subsection{Receivers}
			Regarding possible receivers for calcium ions, building blocks which are similar to R1-R4 presented for protons can be employed (see Table~\ref{tab:proton_calcium_vergleich} for a comparison).\\
            %Measure electrical conductivity
			\textbf{R1 (Conductivity measurement):} In response to the release of calcium ions, the conductivity of the medium surrounding the transmitter would increase. Analogous to the usage of a pH-meter in case of protons, the easiest way to construct a receiver for calcium ions is thus an instrument, which can measure the conductivity of the solution in the channel. While this approach might be suitable for testbeds and applications outside the body, it is challenging for biological systems because the high background concentration of other ions in the environment requires the detection of rather small changes of the total ionic strength.\\
            %Measure fluorescence
			\textbf{R2 (Fluorescent dye):} Similar to a pH indicator, a calcium-sensitive fluorescent dye could be used to detect changes in the calcium ion concentration. When such a dye is illuminated with light of a certain wavelength, fluorescence occurs if calcium ions are present. Examples for such dyes are Fura-2, Indol-1, Fluo-3, Fluo-4, and Calcium Green-1 \cite{Bootman2013, Lee1999}, which all have a high specificity for calcium ions in common. \\
            %Another fluorescence
			\textbf{R3 (Fluorescence proteins)}: For the third receiver structure, we propose the use of calcium-sensitive fluorescent proteins. They are generally fusion constructs of GFP (Fig.~\ref{fig:strukturabbildungen}c) or one of its variants, and the calcium-binding protein calmodulin. One example is GCaMP \cite{Nakai2001} which shows only feeble activity if calcium ions are absent, but undergoes a conformational change upon calcium binding leading to a pronounced fluorescence. \\
            %Measure by some light excitement experiment
			\textbf{R4 (FRET):} Alternatively, as described for protons, calcium ions may be detected via receivers relying on the FRET mechanism \cite{Mank2006, Lindenburg2014}. There exist calcium ion indicators which are based on a FRET pair of two different fluorescent proteins, connected via the calcium-binding protein troponin C. One example, TN-XL \cite{Mank2006}, consists of CFP and YFP. If no calcium ions are present, the protein has an extended conformation where the two fluorescent subunits are distant from each other. If CFP is activated by illumination with a certain wavelength, only cyan fluorescence occurs. As soon as a calcium ion binds to troponin C, the conformation of the fusion protein changes, such that the two fluorescent building blocks get into mutual vicinity. Upon illumination and activation of CFP, the energy is partly transferred to YFP via FRET so that yellow fluorescence can be observed as well.

%%%%%%%%%%%%%%%%%%%%%%%%%%%%%%%%%%%%%%%%%%
\section{Neurotransmitters as Signaling Particles\label{sec:neurotransmitter}}
%%%%%%%%%%%%%%%%%%%%%%%%%%%%%%%%%%%%%%%%%%

\begin{figure*} 
	\centering
	\includegraphics[width=0.7\textwidth]{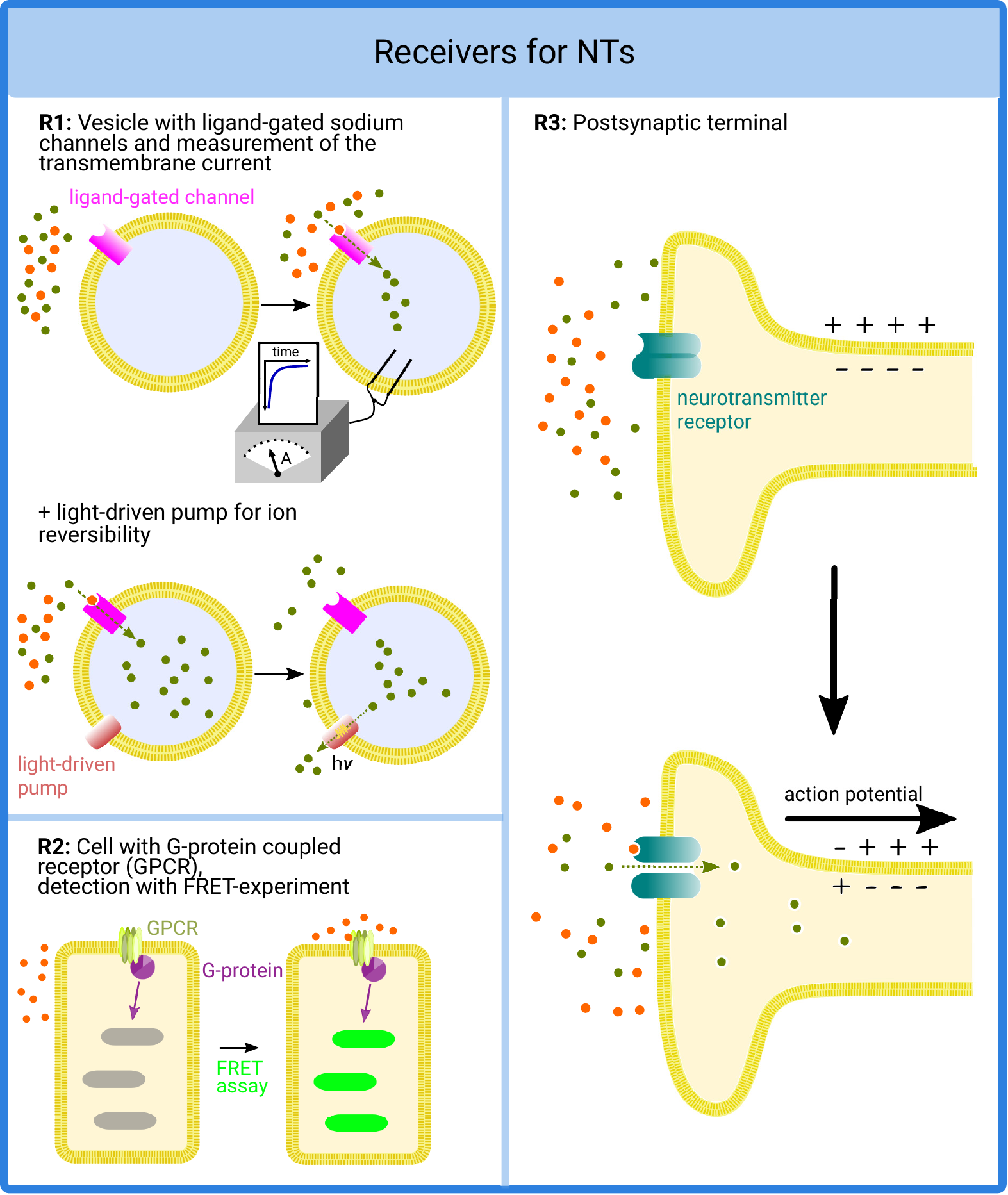}
	\caption{Receivers for NTs as signaling particles. The suggested building blocks for different receiver systems are presented in order of increasing complexity and include designed artificial systems and physiological receivers, which naturally occur in the human body. Neurotransmitters and sodium ions are indicated as orange and green dots, respectively.}
	\label{fig:neurotransmitter_receivers}
\end{figure*}

        %Motivation from nature for using neurotransmitters
Another important class of signaling particles well suited for the design of synthetic MCSs are NTs such as acetylcholine, dopamine, and serotonin. 
In the human body, NTs are used to transmit nerve signals (action potentials) at the chemical synapses between two neurons (e.g. dopamine, serotonin) or from a neuron 
to a muscle fiber (acetylcholine) \cite{Wecker2010}. In this section, we describe some general principles and building blocks for the design of synthetic transmitters and receivers for NTs. 
The general design strategies for the transmitter systems are depicted in Fig.~\ref{fig:neurotransmitter_transmitters}, those for the receiver systems are presented in Fig.~\ref{fig:neurotransmitter_receivers}. These general design principles can be applied to all three NTs discussed in this survey. Thus, the most suitable NT can be selected depending on the requirements imposed by the desired application. It is important to note that the design of a specific signaling pathway requires the choice of suitable protein components depending on the type of NT used. Candidate components 
for acetylcholine, dopamine, and serotonin are summarized in Table~\ref{tab:neurotransmitters}.
%%%%%%%%%%%%%%%%%%%%%%%%%%%%%%%%%%%%%%%%%%
\subsection{Transmitters}
\begin{figure*} 
	\centering
	\includegraphics[width=0.7\textwidth]{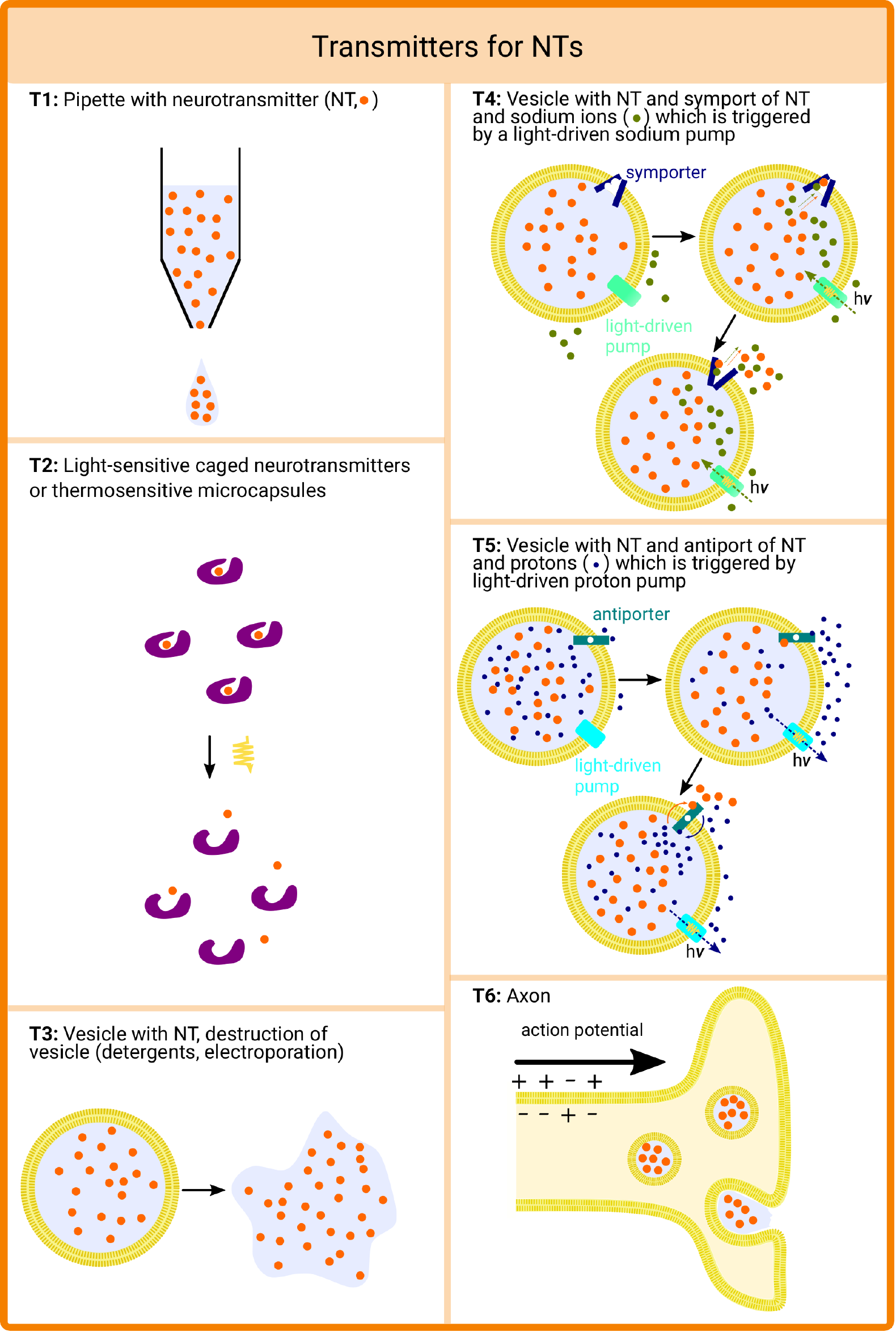}
	\caption{Transmitters for NTs as signaling particles. The suggested building blocks for different transmitter systems are presented in order of increasing complexity and include designed artificial systems and physiological emitters, which naturally occur in the human body.}
	\label{fig:neurotransmitter_transmitters}
\end{figure*}

%%%%%%%%%%%%%%%%%%%%%%%%%%%%%%%%%%%%%%%%%%
In the following, we consider six different possible types of transmitters for NTs. The first two transmitters (T1, T2) are simple and can be used as macroscale interfaces. Transmitters T3-T5 are vesicle based and employ different biological 
mechanisms for NT release. The final transmitter (T6) is the axon terminal of a nerve fiber. \\
            %Pipette or destroying vesicle
\textbf{T1 (Pipette):}  Analogous to cations (Fig.~\ref{fig:proton_transmitters}, Table~\ref{tab:proton_calcium_vergleich}), the simplest and least sophisticated transmitter is a pipette by which a solution containing the NTs can be released dropwise into the channel (Fig.~\ref{fig:neurotransmitter_transmitters}, T1). \\
% Destroying vesicle 
\textbf{T2 (Caged compounds):} As described for calcium ions, another interesting concept, which is based on shielding the NTs until they need to be released, are so called light-sensitive caged compounds (Fig.~\ref{fig:neurotransmitter_transmitters}, T2). In this approach, the NTs are enclosed and thus inactivated by a photoswitchable molecule. Upon irradiation with light of a certain wavelength, the photoswitchable molecule is either cleaved or changes its conformation and as a consequence, the NTs are released from their cage. Caged compounds have been developed for acetylcholine, dopamine, serotonin, and many other NTs \cite{Milburn1989, Araya2013, Cabrera2017}. Some of them are already commercially available. Recently, caged serotonin has been suggested to be used for targeted drug delivery in the context of neurodegenerative diseases \cite{Cabrera2017}. Alternatively, the NTs could be shielded using thermosensitive microcapsules \cite{Nuyken1991} which release their content upon an increase of temperature.\\
\textbf{T3 (Degenerating vesicles):} Vesicles containing a solution of particular NTs may be used as transmitter (Fig.~\ref{fig:neurotransmitter_transmitters}, T3). As a simple option, the vesicle can be destroyed to 
release its content as described previously. However, this has the disadvantage that only a one-time release is possible.  \\
            %Example of controlling neurotransmitter release
\textbf{T4 (Symporters):} If a vesicle-based approach is used as suggested in T3, a transmitter which pumps the NTs in a more controlled manner from the inside of the vesicle into the channel may be a better option than simply destroying the vesicle and releasing its entire content at once. To this end, special proteins for the outward transport of the NTs maybe inserted into the vesicle membrane. One possible outward transporter are NT sodium symporters (Fig.~\ref{fig:neurotransmitter_transmitters}, T4), which are driven by a sodium concentration gradient \cite{Shi2008}. The structure of sodium symporters for the target NT dopamine, i.e., dopamine transporters, is shown in Fig.~\ref{fig:strukturabbildungen}g. In particular, the underlying mechanism of sodium symporters is as follows: When both a sodium ion and an NT bind to the transporter simultaneously, they are carried across the vesicle membrane by a conformational change of the transporter. This means that the NT will be pumped outwards as soon as a sodium concentration gradient from the inside to the outside is established. Such a sodium gradient may be realized by a light-driven sodium pump which transports sodium inside when it is illuminated by light of a certain wavelength. Thus, the combination of an NT-sodium-symporter and a light-driven sodium-pump can be used for a finely controllable release of NTs from the vesicle.\\
            %Another example of controlling neurotransmitter relase
\textbf{T5 (Antiporters):} An alternative strategy with the same level of complexity and controllability as the previous transmitter structure (T4) is to use vesicles with NT antiporters instead of symporters \cite{Albers1999} (Fig.~\ref{fig:neurotransmitter_transmitters}, T5). In contrast to sodium symporters, in this case, the driving force for transportation of NTs across the vesicle is a proton gradient. In particular, if a proton binds to the antiporter from the outside of 
the vesicle and an NT simultaneously from the inside, a conformational change occurs by which the proton is transported inside and the NT is transported outside. To avoid a constitutive NT release, the inside of the vesicle has 
to be more acidic than the surrounding channel when the transmitter is inactive. Coupled with a light-driven proton pump such as bacteriorhodopsin, one can then remove protons from the vesicle upon illumination with a certain 
wavelength and thereby induce a proton gradient towards the inside which triggers a controlled release of the NTs. \\
            %Example of harnessing existing biological transmitter component
\textbf{T6 (Natural transmitters):} Besides the vesicle-based transmitters (T3-T5), a main advantage of NTs is the possibility to directly use a physiological transmitter, i.e., the axon terminal (presynaptic part) of a nerve fiber in the 
human body (Fig.~\ref{fig:neurotransmitter_transmitters}, T6). Upon excitation of a nerve fiber, an electrical signal (action potential) is generated, moves along the axon terminal, and triggers the release of the NTs, which are stored 
in vesicles \cite{Albers1999}. This natural biological transmitter is inexhaustible because the vesicles are regenerated by the neuron. Possible applications of such a direct interface to a neuron \cite{Nakano2012} are the bridging of nerve lesions, the release of drugs in certain conditions 
(e.g. an analgesic combined with a neuron involved in pain reception), and the movement of a prosthesis.

%%%%%%%%%%%%%%%%%%%%%%%%
\begin{table*} 
	\footnotesize
	\raggedleft
	\caption{Comparison of different NTs as signaling particles. Exemplary options for different components.}
	\label{tab:neurotransmitters}
	\begin{tabularx}{\linewidth}{p{0.85cm}C{3.4cm}ZZZ}
		\toprule
		\multicolumn{2}{l}{\textbf{\textsf{Component}}} & \textbf{\textsf{Acetylcholine}} & \textbf{\textsf{Dopamine}} & \textbf{\textsf{Serotonin}}\\
		\midrule
		\textbf{\textsf{T3-T5}} & \textbf{\textsf{Vesicle}} & yes & yes & yes \\ \addlinespace
		\textbf{\textsf{T4}} & \textbf{\textsf{Symporter (sodium-driven)}} & --- & dopamine transporter (DAT) \cite{Kruse2012} & serotonin transporter (SERT) \cite{DeFelice2016}\\ \addlinespace
		\textbf{\textsf{T5}} & \textbf{\textsf{Antiporter (proton-driven)}} & vesicular acetylcholine transporter (VAChT) \cite{Arvidsson1998}& vesicular monoamine transporter 2 (VMAT2) \cite{Wimalasena2010} & vesicular monoamine transporter 2 (VMAT2) \cite{Wimalasena2010} \\ \addlinespace
		\textbf{\textsf{T6}} & \textbf{\textsf{Physiological transmitter}} & axon terminal at neuromuscular junction \cite{Macintosh1952} & axon terminal in the central nervous system (CNS); regulation of executive functions, motor control, motivation, arousal, reinforcement, and reward \cite{Cools2010} & axon terminal in the CNS; regulation of mood, emotion, memory processing, sleep, cognition \cite{Cools2010}\\ \addlinespace 
		\midrule
		\textbf{\textsf{R1}} & \textbf{\textsf{Ligand-gated channel}} & nicotinic acetylcholine receptor (nAChR) \cite{Albuquerque2009} & --- & serotonin receptor subtype 5-HT$_3$ \cite{Thompson2006} \\ \addlinespace
		\textbf{\textsf{R2}} & \textbf{\textsf{G-protein coupled receptor (GPCR)}} & muscarinic acetylcholine receptors (mAChR) M$_{1-5}$ \cite{Haga2013} & dopamine receptors DRD1-DRD5 \cite{Beaulieu2011} & serotonin receptor subtypes 5-HT$_{1,2,4-7}$ \cite{Berumen2012}\\ \addlinespace
		\textbf{\textsf{R3}} & \textbf{\textsf{Physiological receiver}} & trigger contraction of muscle fiber \cite{Macintosh1952} & trigger nerve impulses in the CNS \cite{Cools2010} & trigger nerve impulses in the CNS \cite{Cools2010} \\ \addlinespace
		%					\textbf{\textsf{ISI}} & \textbf{\textsf{Signal removal (Enzyme)}} & acetylcholinesterase (AchE) & catechol-O-methyltransferase (COMT) or monoamine oxidase A/B (MAO-A/B) & monoamine oxidase A (MAO-A) \\ 
		\bottomrule	
	\end{tabularx}
\end{table*}

%%%%%%%%%%%%%%%%%%%%%%%%%% %%%%%%%%%%%%%%%%%%%%%%%%%%            
\subsection{Receivers}
%%%%%%%%%%%%%%%%%%%%%%%%%% %%%%%%%%%%%%%%%%%%%%%%%%%% 
 In this subsection, we discuss two synthetic receivers for NTs which employ NT receptors embedded into the membrane of a vesicle. Such NT receptors exist in the human body, e.g. in postsynaptic cells, and there are different types of 
NT receptors. Here, we consider ligand-gated sodium channels such as the serotonin 5-HT3A receptor (R1, Fig.~\ref{fig:strukturabbildungen}e) and G-protein coupled receptors (R2, Fig.~\ref{fig:strukturabbildungen}h). In addition, we also consider one natural receiver for NT which can serve as an interface for the control of biological systems (R3). \\           
%Receptor opening ion channel upon reception
\textbf{R1 (Ligand-gated ion channels):}  For ligand-gated sodium channels, upon binding of the target NT, a pore in the receptor is opened which allows sodium ions to pass through the membrane \cite{Lodish2000}. 
Based on the concentration gradient,  sodium ions move from the outside to the inside of the cell, which leads to the formation of a transmembrane current that can be measured by methods such as the two-electrode voltage clamp method 
(Fig.~\ref{fig:neurotransmitter_receivers}, R1, upper panel) \cite{Guan2013}. For proper functionality of the vesicle, the concentration of sodium ions inside the vesicle has to be lower than the concentration outside. 
To ensure that after detection the required sodium ion gradient is restored to enable future receptions, reversibility has to be integrated in the vesicle. This can be accomplished by further integrating light-driven 
sodium-pumps into the membrane of the vesicle (Fig.~\ref{fig:neurotransmitter_receivers}, R1, lower panel) \cite{Kato2016}. Then, the light-driven sodium pumps can operate e.g. in the time interval between two 
consecutive transmissions to pump out the sodium ions so that the receiver is replenished.\\
            %Receptors releasing inner-bound particle upon reception; can be visualized
\textbf{R2 (GPCRs):} Besides ligand-gated ion channels, G-protein coupled receptors (GPCRs) are another class of receptor for NTs \cite{Lodish2000}. The structure of one example GPCR, namely the M3 muscarinic receptor, is shown in Fig.~\ref{fig:strukturabbildungen}h. 
When an NT binds to a GPCR, this leads to a conformational change of the receptor. This conformational change is conveyed to the inside of the cell via an intracellular binding protein such as a G-protein or arrestin, where it may activate or inhibit a variety of second messenger molecules and thereby have an impact on the cell metabolism \cite{Gurevich2017} rendering it suitable for microscale applications. However, a GPCR could also be used as reception mechanism for a macroscale interface (Fig.~\ref{fig:neurotransmitter_receivers}, R2). Such a setup will most probably require an intact cell, which has such a receptor in its membrane, because it would be extremely difficult to synthetically reconstruct the corresponding complex signaling cascade in a vesicle. There are several commercially available kits which use FRET experiments allowing optical detection of the level of GPCR activation.\\	
            %Application: Stimulate muscle nerve
\textbf{R3 (Natural receivers):}  NTs can be used to directly interact with biological systems. For example, if the NTs are released in proximity of a postsynaptic terminal or a muscle fiber, they may stimulate a nerve or induce a muscle contraction 
(Fig.~\ref{fig:neurotransmitter_receivers}, R3). This stimulation occurs by binding of an NT to a receptor, e.g. a ligand-gated ion channel, on the membrane of the postsynaptic terminal. The resulting ion influx leads to a depolarization 
of the cell membrane which propagates then as a new action potential along the cell. This principle can be exploited in medical applications of MC \cite{Nakano2012} for bridging of nerve lesions or targeted intervention into a deregulated neuronal circuitry, 
e.g. in the context of neurodegenerative diseases.

%%%%%%%%%%%%%%%%%%%%%%%%%%%%%%%%%%%%%%%%%%%%%%%%%%%%%%%%%%%%%%%%%%%%%%%%%%%%%%%%%%%%%%%%%%%%%%%%
\section{Phosphopeptides as Signaling Particles\label{sec:peptides}}	
%%%%%%%%%%%%%%%%%%%%%%%%%%%%%%%%%%%%%%%%%%%%%%%%%%%%%%%%%%%%%%%%%%%%%%%%%%%%%%%%%%%%%%%%%%%%%%%%
%Peptide as carrier of peptide modification

As outlined in Section~\ref{subsec:signaling_particles}, protein modifications represent a widespread principle of signal transduction in nature. Phosphorylation, where a phosphoryl group is added to a peptide, represents one of the most frequent modifications in cellular signaling, and will be considered in the following in more detail. To exploit this principle for the design of MCSs, it appears advisable to reduce the size of the respective phosphoproteins to the vicinity of the phosphorylation sites. These smaller `phosphopeptides' have the advantage of faster diffusion due to their smaller size compared to intact proteins.

Phosphopeptides are complementary to the particles described in the previous sections as they have very different properties. The most important difference is that peptides do not function in isolation, but require attachment to a chemical functional group. This step is mediated by a kinase, a specific type of enzyme. It is important to note that the peptide unit represents more than a mere carrier molecule to transport the phosphoryl group from the transmitter to the receiver but also plays an important role for the specificity of the signal transduction process. The proteins discussed as receivers below generally do not only recognize the phosphoryl group itself, but also the physico-chemical properties of the peptide in its vicinity. This allows the design of various types of phosphopeptides with different signaling specificity. 

%%%%%%%%%%%%%%%%%%%%%%%%%%%%%%%%%%%%%%%%%%%%%%%%%%%%%%%%%%%%%%%%%%%%%%%%
\subsection{Transmitter}
%%%%%%%%%%%%%%%%%%%%%%%%%%%%%%%%%%%%%%%%%%%%%%%%%%%%%%%%%%%%%%%%%%%%%%%%
\begin{figure*} 
    \centering
    \includegraphics[width=0.7\textwidth]{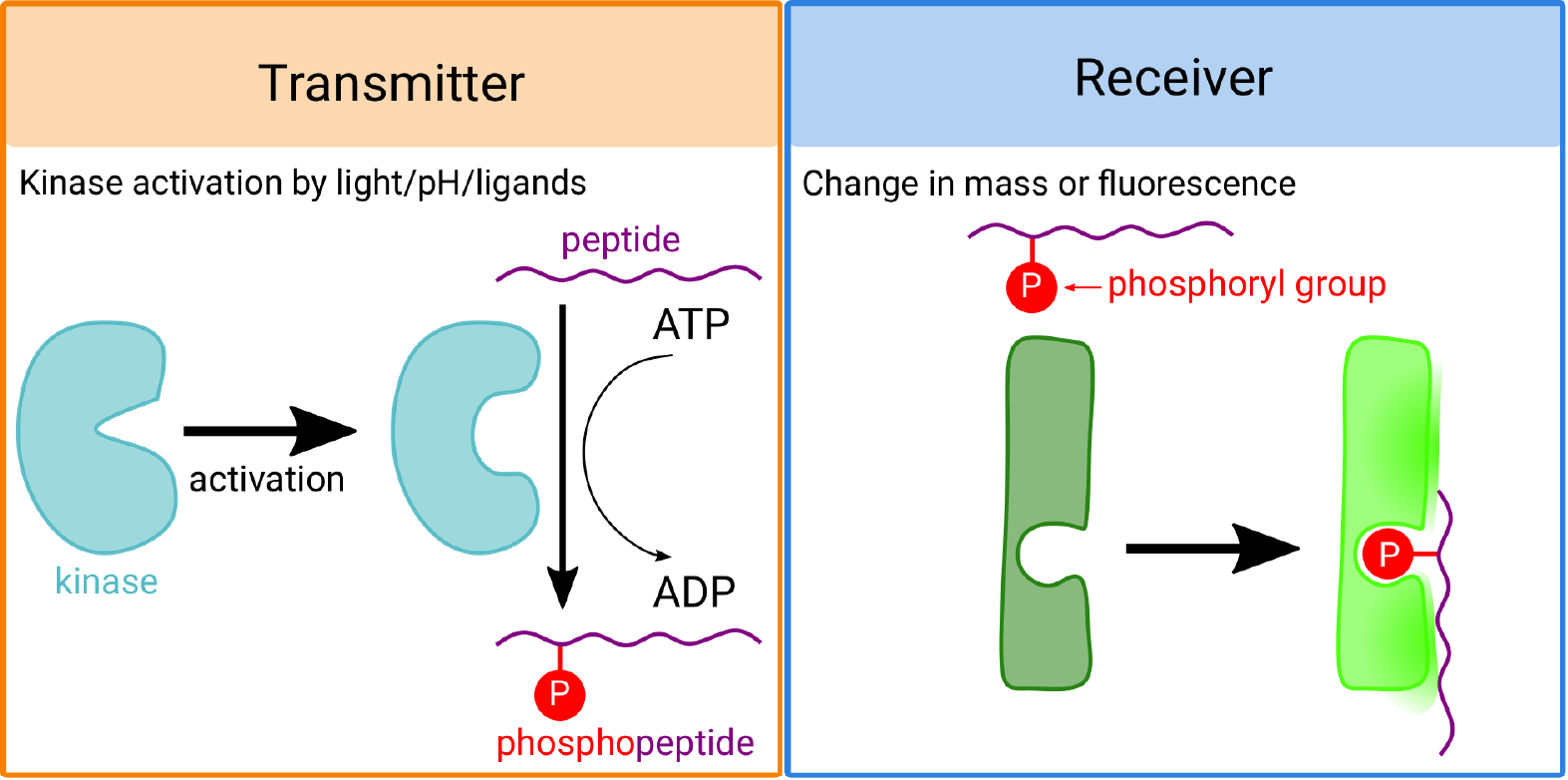}
    \caption{Phosphopeptides as signaling particles. General composition of the communication system, which uses switchable kinases (activation by light, ligands, or pH change) as transmitters. The receiver consists of a phosphopeptide-binding domain, which allows detection of the binding process via changes in tryptophan fluorescence or in the mass of the complex.}
    \label{fig:ptm_anordnung}
\end{figure*}
%
%Kinases generate peptide modifications upon stimulus
In contrast to the systems described in Sections~\ref{sec:protons}-\ref{sec:neurotransmitter}, in the case of peptide modifications, the signaling particles do not need to be stored at the transmitter but can be generated upon 
a stimulus (e.g. light or external ligand), see Fig.~\ref{fig:ptm_anordnung}. The stimulation is provided by kinases that transfer a phosphoryl group from the chemical energy carrier molecule ATP to 
a peptide, thereby creating a phosphopeptide, the signaling particle. In this process, ATP is hydrolyzed to adenosine diphosphate (ADP), which can be regenerated to ATP by other cellular processes.

%There are many peptide modifications
For phosphorylation, several different peptides and corresponding kinases are available. 
%Different types of transmitters can be designed using different kinases.
The human proteome contains at least 518 different protein kinases \cite{Manning2002}.  
Most protein kinases phosphorylate either the amino acids serine/threonine or tyrosine (specific types of amino acids).
However, there are also dual-specificity protein kinases that can phosphorylate both serine/threonine and tyrosine residues.
Within these groups, kinases additionally differ in their specificity, i.e., the amino acid sequence in the environment of the potential phosphorylation site can determine whether an amino acid becomes phosphorylated by a certain kinase or not.
Due to the large variability of the peptide sequences with the corresponding kinases, a large number of different signaling particles is available.

%Kinases can be switched on and off by a stimulus
An important prerequisite for the use of signaling particles in MCSs is the controllability of particle generation, i.e., in this case, the ability to switch the kinases on and off.
Because kinase activity has profound effects on cellular processes, protein kinases are generally highly regulated, i.e., there are many mechanisms for switching them on and off.	An overview of  physiological and engineered mechanisms for kinase regulation is given in Table~\ref{tab:kinase_regulation}.

	\begin{table*} 
	%\scriptsize
	\fontsize{8.5}{10.2}\selectfont
	\raggedleft
	\caption{Summary of the type of stimuli that can be used to control the activity of protein kinases. The table distinguishes between physiological ({\normalfont p}) principles of activation and those that were achieved by molecular engineering ({\normalfont e}). In addition to the target kinase and the type of stimulus, a brief description of the underlying molecular mechanism is given.}
	\label{tab:kinase_regulation}
	\begin{tabularx}{\linewidth}{p{3cm}C{1cm}C{2cm}Z}
	\toprule
	\textbf{\textsf{Type of stimulus}} & \textbf{\textsf{Origin}} & \textbf{\textsf{Target kinase}} & \textbf{\textsf{Molecular mechanism}} \\
	\midrule
	\textbf{\textsf{Phosphorylation}} & p & Lck & Lck contains two regulatory tyrosyl residues (Tyr394, Tyr505). Phosphorylation of these residues controls Lck activity in T cells \cite{Xu1995}. \\
	\midrule
	\textbf{\textsf{Ubiquitination}} & p & receptor tyrosine kinases (RTKs) & RTKs can become modified by ubiquitin, which causes their endocytosis from the plasma membrane and degradation \cite{Critchley2018}.\\
	\midrule
	\textbf{\textsf{pH change}} & p & egg cortex tyrosine kinase & This kinase shows significant changes of activity within the physiologically relevant pH range from 6.8 to 7.3 and may therefore be used as a pH sensitive transmitter \cite{Jiang1991}. \\
	\midrule
	\textbf{\textsf{Regulatory protein}} & p & cyclin dependent kinases (CDKs) & The activity of CDKs is modulated by the interaction with specific cyclins that act as regulatory partners \cite{Pines1995}. \\ 
	\midrule
	\textbf{\textsf{Allosteric ligand}} & e & Fyn, Src, Lyn, Yes, PAK1 & Through insertion of a modified FK506 binding domain, these kinases were engineered to allow activation by the allosteric ligand rapamycin \cite{Chu2014, Dagliyan2017}.\\
	\midrule
	\textbf{\textsf{Photoresponsive ligand}} & e & Protein kinase C (PKC) & When exposed to light, a photoresponsive small molecule becomes an active inhibitor of PKC. This turning on of enzyme inhibition with light allows to control enzyme function \cite{Wilson2017}.\\
	\midrule
	\textbf{\textsf{Light}} & e & receptor tyrosine kinases (RTKs) & RTKs were engineered to include light-oxygen-voltage (LOV)-sensing domains, resulting in kinases that can be activated by light \cite{Grusch2014}.\\
	\midrule
	\textbf{\textsf{Light}} & e & Tropomyosin-related kinase (Trk) & Trk was engineered to include the photolyase homology region of cryptochrome~2 (a blue-light photoreceptor) resulting in a light-controllable kinase \cite{Chang2014}. \\
	\midrule
	\textbf{\textsf{Light}} & e & Raf1, MEK1, MEK2, CDK5 & A photodissociable dimeric protein (Dronpa) was engineered that dissociates in cyan light and re-associates in violet light. Insertion of Dronpa into protein kinases allowed to create photo-switchable kinases \cite{Zhou2017}. \\
	\bottomrule	
\end{tabularx}
\end{table*}

%%%%%%%%%%%%%%%%%%%%%%%%%%%%%%%%%%%%%%%%%%%%%%%%%%%%%%%%%%%%%%%%%%%%%%%%
\subsection{Receiver}
%%%%%%%%%%%%%%%%%%%%%%%%%%%%%%%%%%%%%%%%%%%%%%%%%%%%%%%%%%%%%%%%%%%%%%%%
%Many reception binding mechanisms for peptide modifications
For the detection of phosphorylated peptides, there exists a large set of protein domains in nature that may be used as receivers in synthetic MCSs.
The binding of the phosphopeptide to such domains can for example be detected by a change in tryptophan fluorescence that occurs upon binding (Fig.~\ref{fig:ptm_anordnung}). Alternatively, the change in mass upon peptide binding may be detected via surface plasmon resonance \cite{Homola1999}. 

%Several examples of binding mechanisms
Similar to the versatility on the transmitter side, there exist many different adapter domains that can be used as specific receivers.
Serine/threonine phosphorylated peptides can be recognized by a large number of different domain types, including 14-3-3, BRCT, FF, WW, and FHA domains \cite{Seet2006}.
Tyrosine phosphorylated peptides can be recognized by SH2 or PTB domains \cite{Jin2012}.
The SH2 domain family represents the largest class of tyrosine phosphopeptide recognition modules and is found in 111 different human proteins \cite{Liu2011}.
In addition to the phosphorylated tyrosine residue (pTyr) itself, these domains also recognize peptide residues adjacent to the phosphorylation site.
For example, the SH2-domains of the SHP protein preferentially bind to a pTyr-X-X-Leu sequence stretch, i.e., they recognize a leucine (Leu), which is three amino acids apart from the phosphorylation site (``X'' denotes a variable amino acid).
In contrast, CRK SH2-domains recognize a pTyr-X-X-Pro sequence, which contains a proline (Pro) instead of leucine at the respective sequence position \cite{Tinti2013}.
This recognition of additional residues in the peptide ensures a high specificity at the receiver side and underscores that the peptide moiety of the signaling particle is more than a 
mere carrier, but instead plays an important role for the construction of specific transmitter-receiver pairs.

%\subsection{Signaling Particle Removal}
%
%%Peptide can be unmodified
%After the signal has been received, the phosphorylation can be removed by phosphatases to avoid interference with subsequent signals.
%One way to regulate such phosphatases are light-controlled inhibitors, which are reversibly activated by irradiation with UV light \cite{Wagner2015}.
%Alternatively, phosphatases with pH-dependent activity were developed \cite{Makde2006}. 
%A modular protein assembly approach was used to design a prototype that can detect tyrosine phosphorylation and immediately activate phosphatase without requiring further external stimuli \cite{Sun2017}.
%%%%%%%%%%%%%%%%%%%%%%%%%%%%%%%%%%%%%%%%%%%%%%%%%%%%%%%%%%%%%%%%%%%%%%%%
\subsection{More Complex Architectures for MC}\label{sec:phosphopeptides_complex}
%%%%%%%%%%%%%%%%%%%%%%%%%%%%%%%%%%%%%%%%%%%%%%%%%%%%%%%%%%%%%%%%%%%%%%%%

%Binding is a fundamental and powerful mechanism

\begin{figure*}
	\centering
	\includegraphics[width=0.7\textwidth]{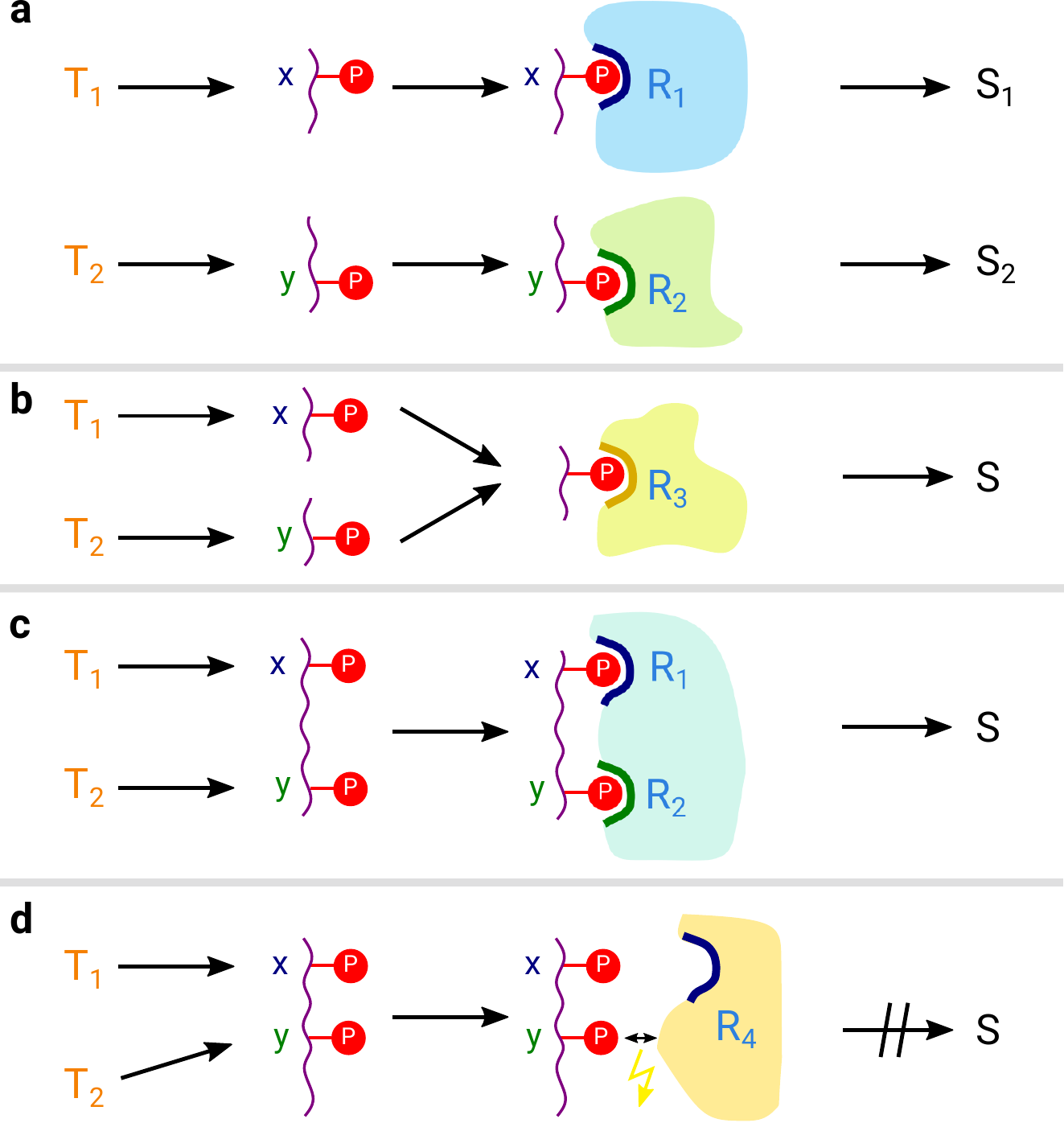}
	\caption{More complex architectures for MC based on phosphopeptides. (a) Simultaneous orthogonal transmission of two different signals (S1, S2). MCSs realizing (b) diversity, (c) coding, and (d) jamming. Transmitters (T) and 
		receivers (R) of different types are labelled with small subscript numbers. $x$ and $y$ denote two distinct phosphorylation sites either in two different peptides (a, b) or within the same peptide (c, d).}
	\label{fig:ptm_verschaltung}
\end{figure*}

Compared to cations and NTs, phosphopeptides have a more sophisticated structure which provides more degrees of freedom for system design. In particular, by combining several kinases with receiver domains of corresponding specificity, various communication concepts can be realized. Here, we discuss orthogonal channels, diversity, coding, and jamming, see Fig.~\ref{fig:ptm_verschaltung}.

\noindent
%Multiple peptide modifications can be employed in parallel
\textbf{P1 (Orthogonal channels):} Orthogonal channels can be realized by using two kinases, which differ in the type of their activation mechanism and the specificity of their phosphorylation (Fig.~\ref{fig:ptm_verschaltung}a). 
For example, transmitter 1 (T1) could be a light-activated kinase and T2 a pH-activated kinase, each combined with a specific receiver domain (R1 or R2). In this system, changes in irradiation and pH can then be 
detected in the same setup based on the signals S1 and S2. This setup allows the interference free multiplexing of signals. \\
%Example: OR processing
\textbf{P2 (Diversity):} 
By selecting suitable signal peptides and receiver domains, two signals cannot only be observed separately, but can also be processed jointly to produce a combined output signal.
One setup for such a processing is shown in Fig.~\ref{fig:ptm_verschaltung}b. Here, it is sufficient if one of the two stimuli is present to trigger the signal at the receiver.
The difference to the situation shown in Fig.~\ref{fig:ptm_verschaltung}a is that instead of two specific recognition domains, a receiver domain (R3) with low specificity is now used 
which can bind the phosphorylation sites $x$ and $y$ of both peptides. This can be interpreted as a form of diversity. For example, let's assume that both peptides convey the same information 
(e.g. both convey information bit ``1''). If we further assume that diffusion is the main transportation mechanism to bring the phosphorylation sites of the peptides into contact with the recognition domain at the receiver, 
then, due to the random nature of the diffusion process, one of the peptides may not arrive at the receiver. Alternatively, one of the peptides may not be phosphorylated at all, because the respective kinase was not activated by a stimulus. However, for the considered setup, it is sufficient if one of the peptides carrying one of the phosphorylation sites 
arrives at the receiver, which implies a diversity gain.\\
%Example: AND processing
\textbf{P3 (Coding):} For the architecture shown in Fig.~\ref{fig:ptm_verschaltung}c, both stimuli (e.g. light and pH change) must be present so that a signal can be detected at the receiver.
The carrier molecule used is a peptide that has two distinct phosphorylation sites for kinases T1 and T2. This requires a receiver with two recognition sites for phosphoryl groups, each of which on its own binds the phosphoryl groups too weakly to trigger the signal. The simultaneous binding of two phosphoryl groups results in a significantly stronger 
binding, which triggers a detectable signal at the receiver. This may be seen as a form of repetition coding as a signal is generated only if both phosphoryl groups are observed at the receiver. This principle is used in nature, 
for example, by the ZAP70 adapter protein, which has two SH2 domains. The simultaneous binding of both SH2-domains causes a $>$100-fold increase in affinity compared to the interaction of a single SH2 domain \cite{Bu1995}.\\
%Example: XOR processing
\textbf{P4 (Jamming):} For the architecture shown in Fig.~\ref{fig:ptm_verschaltung}d, the transmitter and the signal peptides are similar to those for the coding scheme shown in Fig.~\ref{fig:ptm_verschaltung}c. However, 
the receiver (R4) has different properties compared to R3. If a second phosphorylation is added at position $y$, this leads to a weakening of the binding due to unfavorable interactions with the receiver, such that no signal is detected.
In a communication context, this may be interpreted as a jamming of the signal. In particular, if the intended message is encoded via phosphorylation site $x$, adding the second phosphorylation $y$ jams the received signal.  
An example of such a receiver in nature is a 14-3-3 protein that specifically recognizes a Cdc25B signal protein phosphorylated at the serine 323 position. If a second phosphorylation is added at the adjacent serine 321, the interaction 
with the receiver is disrupted \cite{Astuti2010}.

%Application: Interact with intracellular signaling
The principles for switchable interactions described in Fig.~\ref{fig:ptm_verschaltung} are widely used in nature.
The ELM.switches database \cite{VanRoey2013} provides an overview of known switchable systems, which might be exploited for MCS design.
As a long-term goal, a quantitative understanding of these signaling processes may guide the design of signaling particles that interfere with cellular signal transduction processes in a desired fashion, e.g. by counter-balancing impaired signaling originating from disease. 

\begin{remark}
	We emphasize that cations and NTs also allow the realization of some of the above complex architectures albeit to a lesser degree. For instance, the number of NT-receptor pairs that allow realization of orthogonal channels is much smaller compared to what can be realized by phosphopeptides. In fact, the sophisticated structure of phosphopeptides allows the system designer to engineer multiple types of signaling molecules whose release and detection can be controlled either separately or jointly, depending on the desired application. Such high level of flexibility does not exist for cations and NTs.
\end{remark}

%%%%%%%%%%%%%%%%%%%%%%%%%%%%%%%%%%%%%%%%%%%%%%%%%%%%%%%%%%%%%%%%%%%%%%%%
\section{Comparison, Applications, and Practical Considerations}\label{sec:GC}
%%%%%%%%%%%%%%%%%%%%%%%%%%%%%%%%%%%%%%%%%%%%%%%%%%%%%%%%%%%%%%%%%%%%%%%%

In this section, we first compare the properties, advantages, and disadvantages of the signaling particle classes studied in this paper. Subsequently, we present several medical applications of the proposed biological MCSs and corresponding design options for the transmitter, receiver, and signaling particles. Furthermore, we discuss some practical communication-related considerations of the considered MCSs, namely ISI mitigation and the speed of communication.

\subsection{Comparison of the Considered Signaling Particles}
%Pros and Cons of the proposed particle types
In Sections~\ref{sec:protons}-\ref{sec:peptides}, we have described three different classes of candidate signaling particles for synthetic MCSs. Each of these classes has certain advantages and disadvantages for implementation. 
For example, ions exhibit a high particle stability and speed of diffusion, but different ions may generate the same signal at the receiver. This may cause considerable interference in  physiological environments where synthetic and natural MCSs employ different ions that interact with the same receiver. NTs provide a highly unique and specific signal, but the reversibility of their use still represents a bottleneck in the designs proposed here.
In contrast to the other classes, phosphopeptide-based communication does not require vesicles and is highly versatile\footnote{Phosphopeptides are versatile signaling particles because they can be used with large variations in the setup, i.e., different transmitters (different kinases) and receivers (different recognition domains).}, but the particle stability (susceptibility to proteases and phosphatases) may constitute a limitation.
The properties of the considered signaling particles are summarized in Table~\ref{tab:properties_advantages} and will be discussed more in detail in the following with respect to their medical applications, options for mitigation of ISI, and the speed of communication.

\begin{table*}[!htb]
	\footnotesize
	\raggedleft
	\caption{Properties and advantages of the different classes of signaling particles.}
	\label{tab:properties_advantages}
	\begin{tabularx}{\linewidth}{C{3.8cm}ZZZ}
		\toprule
		\textbf{\textsf{Properties}} & \textbf{\textsf{Ions}} & \textbf{\textsf{NTs}} & \textbf{\textsf{Phosphopeptides}}\\
		\midrule
		\textbf{\textsf{Transmitter control}} & light, voltage, mechanical stress, ligand & indirect via ion channel (light, voltage) & light, ligand \\ \addlinespace
		\textbf{\textsf{Membrane vesicles required}} & yes & yes & no \\ \addlinespace
		\textbf{\textsf{Regeneration possible (Reversibility)}} & yes, by returning signaling particles to the vesicles & difficult: in the presented settings only with pipette or axon terminal & yes, kinase is not destroyed (ATP is regenerated in living cells) \\ \addlinespace
		\textbf{\textsf{Energy source \newline (if reversible)}} & light by using light-driven pumps &	irreversible & chemical (ATP) \\ \addlinespace
		\textbf{\textsf{Receiver output signal}} & optical (fluorescence), 
		ion flow (conductivity), ligand release from protein & ion flow (conductivity),
		biological (GPCR signal) & optical (fluorescence), surface plasmon resonance\\ \addlinespace
		\textbf{\textsf{Signal removal}} & external addition of chemicals: 
		base (OH$^-$), chelation (EDTA) & enzymatic degradation:
		acetylcholinesterase (AChE), 
		monoamine oxidases (MAO) & enzymatic dephosphorylation with phosphatases \\ \addlinespace
		\textbf{\textsf{Underlying physiological communication type}} &	intercellular &	intercellular & intracellular \\ \addlinespace
		\textbf{\textsf{System complexity}} & moderate & high & moderate \\ \addlinespace
		\textbf{\textsf{Signaling particle stability}} & very high & high & moderate \\ \addlinespace		
		\textbf{\textsf{Speed (diffusion)}}	& very fast  & fast & moderate  \\ \addlinespace
		\textbf{\textsf{Uniqueness of signal}} & moderate & very good & good \\ \addlinespace
		\textbf{\textsf{Versatility}} & high & moderate & very high \\		
		\bottomrule	
	\end{tabularx}
\end{table*}

\subsection{Potential Medical Applications}

In the following, we present some practical examples for potential medical applications of the proposed building blocks of synthetic MCSs. The signaling particles presented in this article were chosen because changes in their concentration are either causative for diseases or because they may be used as important diagnostic markers to detect pathological conditions. Consequently, a combination of appropriate transmitters and receivers could allow for the construction of medical nanomachines that would be able to recognize specific regions of the body or abnormal tissue such as inflammations or tumors due to an altered concentration of certain signaling particles. 
%Ideally, it would then be possible to deliver drugs exactly to the locations where they are required in order to treat the respective disease. For example, cytostatics could be released specifically within a tumor to enhance their effectiveness while dramatically reducing side effects. 
An overview of possible medical applications covering a broad range of diseases such as bacterial or viral infection, cancer, inflammation, and neurodegeneration or circulatory diseases is provided in Table~\ref{tab:medical_applications}. This table lists the respective signaling particles and proposes building blocks which could be used to design interfering nanomachines.

\begin{table*} 
	%\footnotesize
	\fontsize{7.9}{9.48}\selectfont
	\raggedleft
	\caption{Examples for medical \textit{in vivo} applications of MCSs including the respective medical category and the involved signaling particles. Building blocks, which are suitable for the design of the corresponding nanomachines, may be realized using the transmitters and receivers proposed in Sections~\ref{sec:protons} to \ref{sec:peptides}.}
	\label{tab:medical_applications}%\scriptsize
	\begin{tabularx}{\linewidth}{C{2cm}p{8cm}Z}
				\toprule
		\textbf{\textsf{Medical category}} & \textbf{\textsf{Exemplary disease  mechanism and signaling particle}} & \textbf{\textsf{Proposed building blocks}} \\
		\midrule
		\textbf{\textsf{Bacterial infection}} & Some bacterial proteins interfere with physiological signaling pathways based on phosphopeptides or phosphoproteins \cite{zhang2015specific}. An example is SpvC from \textit{Salmonella} Typhimurium, which can  inactivate MAP kinases that are involved in the activation of the immune response \cite{Li2007}. & The depleted kinases could be replaced by a synthetic transmitter in order to restore the physiological immune response {(e.g., transmitters in Fig.~\ref{fig:ptm_anordnung} and Table~\ref{tab:kinase_regulation})}. \\
		\midrule
		\textbf{\textsf{Viral infection}} & Some viral proteins interfere with physiological signaling pathways based on signal peptides {\cite{davey2011viruses}}. One example is HIV Vpu \cite{Margottin1998} which contains phosphorylated residues that are detected by the receiver protein h-$\beta$TrCP within cells of the human immune system. This interaction leads to a degradation of the immune receptor CD4 which is required for pathogen recognition {\cite{devarajan2018pathogen}}. & One possibility to interfere with this pathological signaling pathway would be to design an artificial receiver that binds and sequesters Vpu in a more affine manner than h-$\beta$TrCP   (e.g., receiver in Fig.~\ref{fig:ptm_anordnung}).  %Alternatively, a phosphatase could be used to remove the phosphate groups from Vpu and therefore make it biologically inactive.
		\\
		\midrule
		\textbf{\textsf{Cancer or \newline  inflammation}} & Proton concentrations are elevated in inflamed tissue {\cite{cartwright2020adaptation}} or cancer { \cite{Tannock4373}}, \cite{Van1999} due to an enhanced/altered metabolism. & MCSs with receivers for protons could be used to detect affected tissue, which would be useful for targeted drug delivery {(e.g., R5 in Fig.~\ref{fig:proton_receivers})}.\\
		\midrule
		\textbf{\textsf{Depressive disorders or Parkinson's disease}} & Dysregulations within the dopaminergic system resulting in a lack of the NT dopamine play a role for the development of the hedonic deficits of patients suffering from depression \cite{Belujon2017}. A selective loss of dopaminergic neurons within the \textit{Substantia nigra} is causative for Parkinson's disease \cite{Dhall2016}. & Receivers which are activated by dopamine could be used to detect dopaminergic synapses. Then, transmitters releasing dopamine could be used to potentiate the signal {(e.g., T4 in Fig.~\ref{fig:neurotransmitter_transmitters} and R1 in Fig.~\ref{fig:neurotransmitter_receivers})}. \\
		\midrule
		\textbf{\textsf{Dysregulated signaling pathways}} & Many physiological signaling pathways are based on  phosphopeptides or phosphoproteins {\cite{jin2012modular,tinti2013sh2}}. Mutations of the respective phosphorylation sites which occur in various diseases can thus have deleterious effects {\cite{kadaveru2008viral}}. One example are several mutations in the protein c-Myc that lead to the loss of a phosphorylation site.  As a result, c-Myc is inefficiently degraded which may lead to tumor formation (Burkitt's lymphoma) \cite {Bahram2000}.& Synthetic receivers for mutant c-Myc could be used in order to bind and thus inactivate the signaling pathway that leads to tumor formation {(e.g., receiver in Fig.~\ref{fig:ptm_anordnung})}. \\
		\midrule
		\textbf{\textsf{Hypertension}} & One approach to treat hypertension is the administration of drugs which lead to a widening of blood vessels and can thereby reduce the vascular resistance \cite{Levenson1985}. An increased proton concentration results in ``acidic-metabolic'' vasodilatation \cite{Modin2001}. & Transmitters for protons could be used to increase the proton concentration in peripheral blood vessels in order to treat hypertension  {(e.g., T3 in Fig.~\ref{fig:proton_transmitters})}. \\
		\midrule
		\textbf{\textsf{Nerve lesions}} &  NTs such as acetylcholine, dopamine, or serotonin are missing because the physiological transmitter, i.e., the releasing neuron is damaged. Therefore, the transmission of nerve signals is interrupted {\cite{peer2008high}}. & The damaged nerve could be repaired using a bridging device consisting of two MCSs, cf. Fig.~\ref{fig:nerve_lesion} {(e.g., T4 in Fig.~\ref{fig:neurotransmitter_transmitters} and R1 in Fig.~\ref{fig:neurotransmitter_receivers})}.\\
		%\midrule
		%\newpage
		\bottomrule
		\end{tabularx}
\end{table*}

\begin{figure*} 
	\centering
	\includegraphics[width=.7\textwidth]{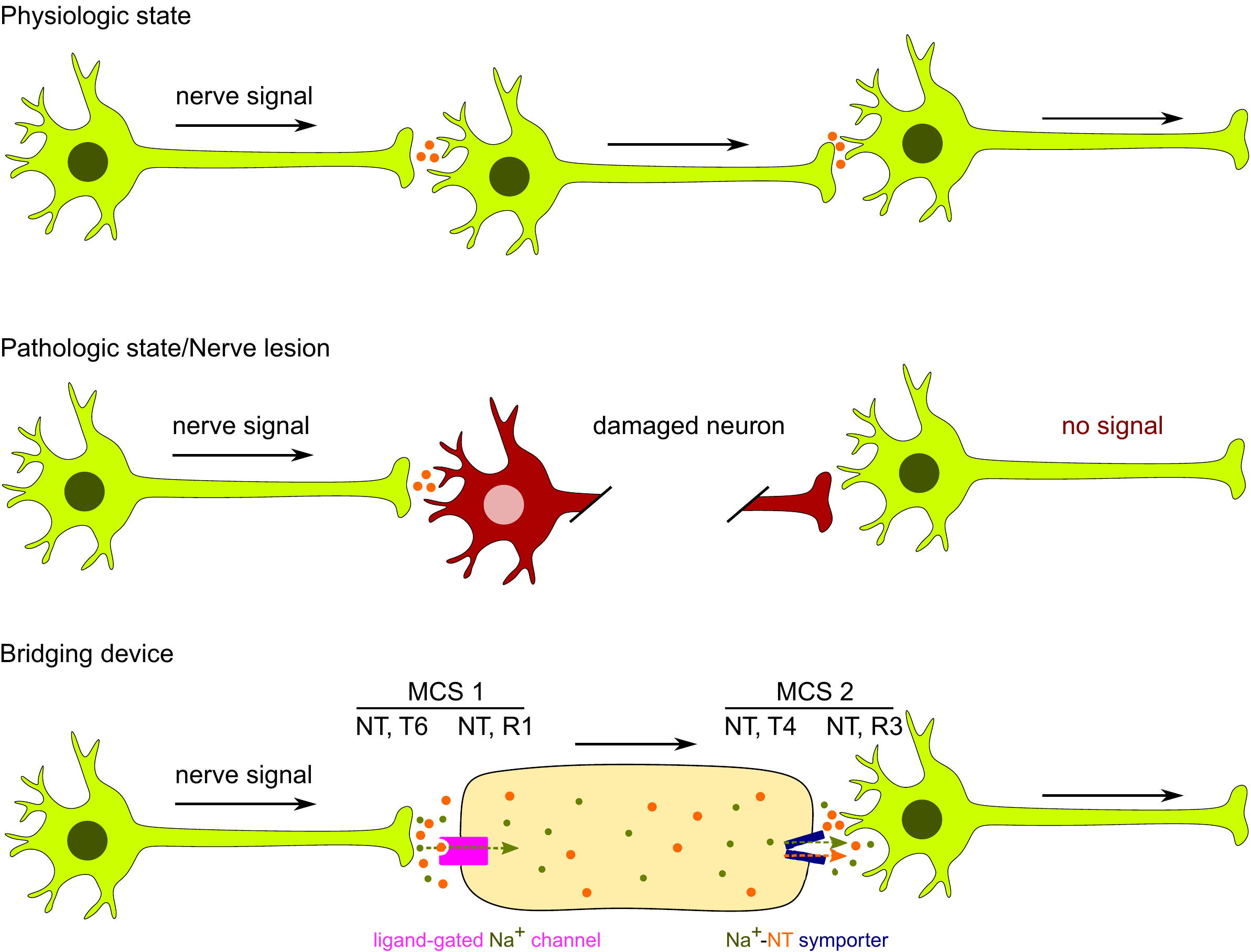}
	\caption{ Bridging device for nerve lesions as a potential medical application for MCSs. The upper panel shows the physiological state with three intact neurons. The middle panel illustrates a nerve lesion in which the second neuron is damaged so that the signal is not transmitted. In the lower panel, a potential vesicle- or cell-based bridging device is shown that could be used to repair the damaged nerve. The building blocks used in the MCS are described in Figs.~\ref{fig:neurotransmitter_transmitters} and \ref{fig:neurotransmitter_receivers}.}
	\label{fig:nerve_lesion}
\end{figure*}

 Which combination of transmitter and receiver is most appropriate critically depends on the  application of interest, of course. 
In the following, we focus on one of the medical applications listed in Table~\ref{tab:medical_applications}, namely the design of a  bridging device for a damaged neuron, cf. Fig.~\ref{fig:nerve_lesion}, in order to illustrate the design principles for choosing suitable biological building blocks for the desired MCS. In the physiologic state (Fig.~\ref{fig:nerve_lesion}, upper panel), a nerve signal is transduced consecutively from one neuron to the next via the release of NTs at the chemical synapse between adjacent neurons. If one of the neurons within this chain is damaged due to an injury or a neurodegenerative disease (Fig.~\ref{fig:nerve_lesion}, middle panel), signal transduction is interrupted. Depending on the type of nerve, this could result in a paralysis or a  loss of sensory perception. One conceivable option to repair this damage is the insertion of a bridging device in order to replace the damaged neuron (Fig.~\ref{fig:nerve_lesion}, lower panel). Such an artificial cell or micelle could be fully integrated within physiological signaling. Theoretically, it would consist of two MCSs, one between the healthy neuron on the left-hand side and the bridging device (MCS1) and one between the bridging device and the healthy neuron on the right-hand side (MCS2). In MCS1, the healthy neuron on the left side would serve as a physiological transmitter (building block NT, T6). As soon as a nerve signal occurs, NTs are released within the synaptic cleft. On the surface of the bridging device, ligand-gated Na\textsuperscript{+} channels (building block NT, R1) could be expressed, which upon binding of the NT would transport Na\textsuperscript{+} ions to the inside.  This would lead to an enhanced concentration of Na\textsuperscript{+} inside the vesicle. Thereby, MCS2 on the right-hand side of the bridging device would be activated. Na\textsuperscript{+}-NT symporters (building block NT, T4) would export NTs into the second synaptic cleft from where they would be detected by the healthy neuron on the right-hand side (building block NT, R3). Consequently, the interruption within the transmission of the nerve signal would be repaired. This exemplary MCS may be further optimized: Reversibility could be ensured by adding transporters for a reuptake of the NT to the left-hand side of the second synaptic cleft. Such a bridging device is well suited for \textit{in vivo} application as it is based on a biocompatible vesicle or cell and no external equipment such as light sources or devices for the measurement of membrane potentials are needed.

	\begin{remark}
		We note that for the exemplary application scenarios in Table~\ref{tab:medical_applications}, the end-to-end MCS may be  in general partially \textit{synthetic} and partially \textit{natural}. For example, in Fig.~\ref{fig:nerve_lesion}, the receiver in MCS~1 is a synthetic cell whereas the transmitter is a natural cell (the nerve cell on the left-hand side in Fig.~\ref{fig:nerve_lesion}). On the other hand, for MCS~2, the transmitter is a synthetic cell whereas the receiver is a natural cell (the nerve cell on the right-hand side in Fig.~\ref{fig:nerve_lesion}). More advanced applications of MCSs may necessitate the use of both partially and fully synthetic end-to-end MCSs. For instance,  in targeted drug delivery for cancer therapy, on the one hand, the drug-carrying nanomachines may need to communicate with each other in order to collaboratively find cancerous tissue \cite{Okonkwo2017,Mosayebi2019Cancer}. Here, the end-to-end MCS  is synthetic, i.e., transmitter and receiver are nanomachines and the signaling particles have to be chosen by the system designer by considering the specific characteristics of the application. On the other hand, after the targeted cancerous tissue has been localized, the drug-carrying nanomachines that are close to the tissue have to release their payload drug for cancer therapy \cite{Gao2014,cao2019diffusive}. This communication system can be  interpreted as a partially synthetic MCS whereby the transmitter is a synthetic nanomachine, the particles are the drug molecules, and the receiver is the cancerous tissue. 
	\end{remark}

\subsection{ISI Mitigation via Signal Removal} 
MCSs are impaired by ISI which is caused by the dispersive nature of the diffusion process. In particular, when the transmitter emits consecutive symbols, the signaling particles of previously transmitted symbols may be observed at the receiver
in the current symbol interval. Since ISI is a common problem also in conventional wireline and wireless communication systems, many approaches for ISI mitigation exist \cite{Proakis}. In conventional communication systems, equalization at the receiver is employed to combat ISI. Although, equalization has also been proposed as an option for MCSs \cite{Kilinc2013}, it may be challenging to implement even simple schemes, such as linear equalization, at nano-scale. Alternatively, in natural MCSs, ISI is usually mitigated by removing the signaling particles from the channel. This biological approach also seems to be a viable option for ISI mitigation in synthetic MCSs. In the following, we will discuss mechanisms for particle removal for each of the considered signaling particles. 

\begin{itemize}
	\item \textbf{Protons}: If protons are used as signaling particles, then adding a basic solution capable of reversing the pH change caused by the transmitter system is the simplest option for removal of the signaling particles. Like the acidic solution (Fig.~\ref{fig:proton_transmitters}, T1), this basic solution can be released into the channel e.g. by an electrically controlled pipette after symbol detection. In a physiological environment, where such a pipette is not available, raising the pH value can be 
	achieved for example by decarboxylation reactions of amino acids or urea, which consume protons \cite{Foster2004}. In the presence of the enzyme urease, urea is cleaved into CO$_2$ and NH$_3$. The latter acts as a base and can thus lead to a neutralization of the acidic liquid in the channel. This cleavage of urea occurs also in nature, e.g. as a mechanism to remove acidic carbohydrate fermentation products in human saliva \cite{Khramov1997}. Moreover, pathogenic \textit{Helicobacter pylori}, which evokes gastritis and gastric cancer in humans, uses urease to buffer the highly acidic gastric liquid \cite{Mobley1996}.
	\item \textbf{Calcium ions}: Calcium ions can be chemically shielded with a chelating agent such as ethylenediaminetetraacetate (EDTA) \cite{Ullmann2000} or ethylene glycol-bis($\beta$-aminoethyl ether)-N,N,N'\\,N'-tetraacetic acid (EGTA). In particular, these substances can bind to calcium ions and prevent them from further interaction 
	with their environment. By releasing an EDTA/EGTA solution into the channel as soon as a signal has been detected, the calcium ions released in the current symbol interval are neutralized and cannot cause interference in future symbol intervals. 
		Hence, ISI is mitigated. 
	\item \textbf{NTs}: For NTs, in nature, signal removal is either achieved by symporters, which pump the NTs into separated compartments \cite{Shi2008}, or by enzymes, which degrade the NTs \cite{Wecker2010}. The latter 
	principle has the advantage that it can be easier to implement in synthetic MCSs as fewer biological building blocks are needed. A theoretical study of NT removal via enzyme has been provided for acetylcholine in \cite{Noel2012}, which 
	can analogically to a physiological process at neuromuscular junctions be eliminated by the enzyme acetylcholinesterase (Fig.~\ref{fig:strukturabbildungen}i). In an analogous manner, dopamine and serotonin can be degraded by monoamine oxidase-A. 
	\item \textbf{Phosphopeptides}: For modified peptides as signaling particles, the phosphorylation can be removed by phosphatases to avoid interference with subsequent signals. One option for regulating such phosphatases are light-controlled 
	inhibitors, which are reversibly activated by irradiation with UV light \cite{Wagner2015}. Alternatively, phosphatases with pH-dependent activity were developed \cite{Makde2006}. A modular protein assembly approach was used to design a 
	prototype that can detect tyrosine phosphorylation and immediately activate phosphatase without requiring further external stimuli \cite{Sun2017}.     
\end{itemize}

\subsection{Speed of Communication}
%Look at book "Biology by the Numbers"!
The time scales at which the different processes needed for signal transmission occur at transmitter and receiver as well as in the channel are crucial for MCS design as they ultimately limit the achievable data rate.
The quantity, which is the most straightforward to measure, is the speed of diffusion (Fig.~\ref{fig:signal_particles}). However, in order to estimate the speed of the entire communication process, the speed of the proteins 
involved as well as the time for building up concentration gradients sufficient for detection need to be taken into account as well. A comprehensive review of these processes is beyond the scope of this article.
We will therefore only give a broad overview of the respective time scales; a comprehensive overview of time ranges of biological processes with many useful model calculations and the corresponding database are given in \cite{BioNumBook} and \cite{BioNum}, respectively. One approach that may be exploited to speed up synthetic MCSs is flow. In particular, flow is also used in nature as a biological process that facilitates the transportation of chemical species across the human body.   

%Single molecular actions are fast
\begin{itemize}
\item \textbf{Transport in channel:}
Individual molecular events, such as an ion flowing through an ion channel or an enzyme catalyzing a reaction, are typically very fast. For example, one bacteriorhodopsin molecule can pump about 188 protons per second \cite{BioNum}. 
However, depending on the density of bacteriorhodopsin in the vesicle, the proton gradient needed to activate the receiver, and the distance between transmitter and receiver, it may take considerably longer to achieve a desired 
macroscopic effect. Some model calculations for such macroscopic processes for bacteriorhodopsin have been performed in \cite{Arjmandi2016}. In an experimental study, \textit{E.~coli} cells were used to synthetize with blue-absorbing and green-absorbing proteorhodopsins, which are two different types of proton pumps closely related to bacteriorhodopsin. A change of $5-30~$nM in proton concentration of the cell suspension was achieved after about 60 s with an approximately linear evolution over time \cite{Wang2003}.

%Overall detectable output speed depends on exact composition
\item \textbf{Ion channels:}
For ion channels, a good estimate for the number of ions that can be transported through a single channel is on average about 10$^7$ per second \cite{BioNum} with a range from about 10$^3$~s$^{-1}$ to 10$^8$~s$^{-1}$. Some exemplary permeability rates (number of ions that flow through the channel in a certain time span) are summarized in Table~\ref{tab:channels}. This compilation includes a subset of the channels suggested in this article for which such information was available in the literature. Although the channels are rather fast on the level of a single molecule, the time needed to detect an effect at the cellular level (for example in \textit{Xenopus} oocytes) is in the range of a few seconds \cite{Carattino2007}. 
However, if artificial vesicles with ion channels are created, the required time span will be highly dependent on the density of the ion channels in the membrane. 

\begin{table*}[!t]
	\footnotesize
	\raggedleft
	\caption{Permeability rates for selected ion channels. Channel 1 is an example for a voltage-gated proton channel (Protons, T3), channel 2 is a voltage-gated calcium channel (Calcium ions, T3). Channel 3 can be used as a receiver for acetylcholine (NTs, R1). Channels 4, 5, 6, and 7 are additional examples demonstrating the typical range of ion transport rates.}
	\label{tab:channels}
	\begin{tabularx}{\linewidth}{p{0.2cm}C{6.8cm}ZZ}
		\toprule
		\multicolumn{2}{l}{\textbfsf{Ion channel}} & \textbfsf{Permeability rate [s}$^{-\text{\textbfsf{1}}}$\textbfsf{]} & \textbfsf{References} \\
		\midrule
		\textbfsf{1} & \textbfsf{Voltage-gated proton channel H$_\text{\textbfsf{v}}$1} & $6.0\cdot10^3$ & \cite{Musset2012, DeCoursey1993} \\ \addlinespace
		\textbfsf{2} & \textbfsf{Voltage-gated calcium channels (e.g. Ca$_\text{\textbfsf{v}}$1.1)} & $\approx10^6$ & \cite{Clapham2007} \\ \addlinespace
		\textbfsf{3} & \textbfsf{Nicotinic acetylcholine receptor} & $2.5\cdot10^7$ & \cite{Berg2002} \\ \addlinespace
		\textbfsf{4} & \textbfsf{Potassium channel KcsA} & $\approx10^8$ & \cite{MoraisCabral2001} \\ \addlinespace
		\textbfsf{5} & \textbfsf{Voltage-gated sodium channel Na$_\text{\textbfsf{v}}$1.4} & $\approx10^7$ & \cite{Bendahhou2012} \\ \addlinespace
		\textbfsf{6} & \textbfsf{Cation channel gramicidin} & $1.5\cdot10^7$ & \cite{Stein2015} \\ \addlinespace
		\textbfsf{7} & \textbfsf{Calcium channel CRAC} & $1.1\cdot10^4$ & \cite{Zweifach1993} \\ 
		\bottomrule
	\end{tabularx}
\end{table*}

\item \textbf{Transporters:}
Transporters are in most cases much slower than ion channels with a typical transport rate of $\approx$~100 molecules s$^{-1}$ \cite{BioNumBook}. Nevertheless,
if a cell has e.g. 10,000 transporters of a certain type in its membrane, 10$^6$ molecules can be transported in total per second.
For glucose transporters in certain cell types, it has been estimated that they make up about 2 \% of the total membrane surface \cite{BioNumBook}. How many molecules can be exported per time unit
will critically depend on how easily the corresponding membrane protein can be inserted into a vesicle membrane and what density can be achieved.

\item \textbf{Receptors:}
For GPCRs, ligand binding happens on a microsecond time scale \cite{Clark2017} and the conformational rearrangements required for receptor activation take place on a micro- to millisecond range \cite{Heyden2013}.
However, current FRET experiments to readout the receptor activation at the cellular level would then take multiple tens of seconds or even minutes \cite{Ayoub2015}.

%Example calculation of reaction speed
\item \textbf{Enzymatic reactions:}
For some of our proposed building blocks, such as the degradation of NTs and the addition and removal of peptide modifications, enzymes are required.
The speed at which an enzyme catalyzes its reaction is dependent on the concentration of the enzyme itself and the substrate (target molecule) the enzyme binds to. 
The general form of an enzymatic reaction is as follows: 
\begin{equation}
    %\ch[math-space=0pt]{ E + S <=>[ $k_1$ ][ $k_{-1}$ ] ES  <=>  [ $k_{\text{cat}}$ ] E + P }
    \text{E} + \text{S} \overset{k_1}{\underset{k_{-1}}{\rightleftharpoons}} \text{ES} \overset{k_{\text{cat}}}{\rightharpoonup}  \text{E} + \text{P}
    \label{eq:enzyme_1}
\end{equation}
where enzyme E forms with its substrate S an enzyme-substrate complex ES with a certain rate constant $k_1$.
This complex can either dissociate, with the reaction rate constant $k_{-1}$, and form E and S, or react, with the reaction rate constant $k_{\text{cat}}$, and form an E molecule and a product molecule P \cite{Koolman2003}.
Note that the enzyme itself is not consumed in this process. The speed $v$ of this reaction may be approximated (apart from additional factors such as cooperative binding that can affect $v$) by use of the 
Michaelis-Menten equations \cite{Koolman2003} as follows
\begin{equation}
    v=k_{\text{cat}} \cdot [\text{E}]_t \cdot \frac{[\text{S}]}{K_M+[\text{S}]}=v_{\text{max}} \cdot \frac{[\text{S}]}{K_M+[\text{S}]}
    \label{eq:enzyme_2}
\end{equation}
where square brackets denote the concentrations of the respective molecules, $[\text{E}]_t$ is the total concentration of the enzyme (free+substrate-bound), $v_{\text{max}}$ is the maximum reaction speed at the given enzyme concentration, 
and $K_M$ is the Michaelis constant. Thus, knowing $k_{\text{cat}}$, $K_M$, and the concentrations of enzyme and substrate, the speed of the reaction can be calculated. In particular, in \eqref{eq:enzyme_2}, $k_{\text{cat}}$ stands, 
in simple terms, for the number of reactions an enzyme makes per unit time and $K_M$ for the substrate concentration at which 50~\% of the maximum reaction speed is reached \cite{BioNumBook}.

\begin{table*}[!t]
		\footnotesize
		\raggedleft
		\caption{Kinetic constants for selected enzyme-substrate combinations. Enzymes 1 and 2 can be used for NT degradation, enzyme 3 is an exemplary enzyme for protein phosphorylation and enzyme 4 for protein dephosphorylation (Their indicated kinetic values were determined for a sample peptide.). Enzymes 5, 6, and 7 are additional examples demonstrating the activity range of different enzymes.}
		\label{tab:enzymes}
		\begin{tabularx}{\linewidth}{p{0.2cm}C{3.8cm}ZZZZ}
			\toprule
			\multicolumn{2}{l}{\textbfsf{Enzyme}} & \textbfsf{Substrate} & \textbfsf{k}$_{\text{\textbfsf{cat}}}$~[\textbfsf{s}$^{-\text{\textbfsf{1}}}$] & \textbfsf{K}$_\text{\textbfsf{M}}$~[\textbfsf{mM}] & \textbfsf{References} \\ 
			\midrule
			\textbfsf{1} & \textbfsf{Acetylcholinesterase} & acetylcholine & $3.0\cdot10^{4}$ & 0.1 & \cite{Rosenberry1975, Purich2010}\\  \addlinespace
			\multirow{2}{14mm}{\textbfsf{2}} & \multirow{2}{18mm}{\textbfsf{\hbox{Monoamine~oxidase-A} (MAO-A)}} & dopamine & 1.83 & 0.23 & \multirow{2}{14mm}{\cite{Ramsay2017}} \\
			& & serotonin & 1.80 & 0.40 & \\ \addlinespace
			\textbfsf{3} & \textbfsf{Protein kinase N1} & peptide & 0.7 & 0.013 & \cite{Falk2014} \\ \addlinespace
			\textbfsf{4} & \textbfsf{Protein phosphatase 2C\text{\boldmath$\alpha$} } & peptide & 0.3 & 0.09 & \cite{Fjeld1999} \\ \addlinespace
			\textbfsf{5} & \textbfsf{Catalase} & H$_2$O$_2$ & $4.0\cdot10^{7}$ & 1.6 & \cite{Ogura1955, Purich2010} \\ \addlinespace
			\textbfsf{6} & \textbfsf{Triosephosphate isomerase} & glyceraldehyde 3-phosphate & $4.2\cdot10^{3}$ & 0.5 & \cite{Putman1972, Purich2010} \\ \addlinespace
			\textbfsf{7} & \textbfsf{Chymotrypsin} & N-acetyl-Val-OMe &  $1.7\cdot10^{-1}$ & 88 & \cite{Voet2005, Purich2010} \\ 
			\bottomrule
		\end{tabularx}
\end{table*}
 
\begin{remark}
Typical values of $k_{\text{cat}}$ vary between 10$^{-2}$~s$^{-1}$ and 10$^{7}$~s$^{-1}$ with the median approximately at 14~s$^{-1}$. Moreover, the range of the values of $K_M$ varies approximately from 10$^{-5}$~mM up to 
10$^{3}$~mM with a median value of 0.130~mM \cite{BioNumBook, BarEven2011}. Please note that these values are not only enzyme-specific, but they also depend on the respective substrate and environmental conditions such as pH or ionic strength. For illustration, we provide concrete values of $k_{\text{cat}}$ and $K_M$ for some selected exemplary enzyme-substrate combinations in Table~\ref{tab:enzymes}. Specific values for many more enzymes and substrates are available in the BioNumbers data bank~\cite{BioNum}.
\end{remark} 
\end{itemize}    

\section{Future Research Directions\label{sec:futurework}}
\begin{figure*} 
	\centering
	\includegraphics[width=0.9\textwidth]{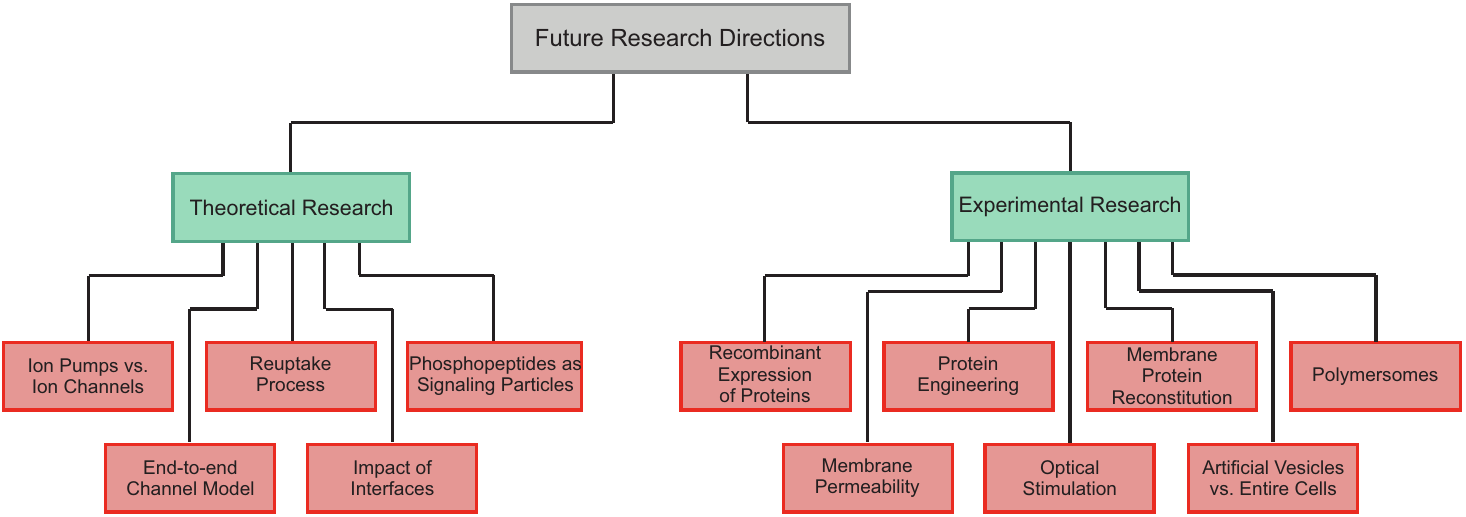}
	\caption{Summary of potential future research directions.}
	\label{fig:futurework}
\end{figure*}

In this section, we outline some future research directions including open research problems that should be tackled in Stage~2 of the roadmap in Fig.~\ref{fig:MCS_Roadmap}  by theoreticians for modeling and designing the  biological building blocks proposed in this paper as well as the challenges that should be addressed by experimentalists for implementation of these  building blocks in Stage~4 of the roadmap.  {A summary of the proposed future research directions are provided in Fig.~\ref{fig:futurework}.}

\subsection{Directions and Challenges for Theoretical Research}

The design of reliable and efficient synthetic MCSs crucially depends on the accuracy and simplicity of the corresponding transmitter, channel, and receiver models. Simplicity is desirable since complicated models typically do not allow for the derivation of insightful design guidelines which are required for efficient system design while a certain level of accuracy is also needed to ensure the relevance of the results and the corresponding design. Although there exists a rich literature in biology that analyzes the proposed biological building blocks for MCSs, the corresponding models are too complex to be readily applicable to communication system design. In fact, for many of the transmitter and receiver architectures proposed in this paper, communication-theoretical models have not been developed by the MC community, yet. In the following, we highlight some of the related open research problems:
\begin{itemize}
	\item \textbf{Ion pumps vs. ion channels:} While ion channels allow passive diffusion of the signaling particles across the membrane and rely on a concentration gradient, ion pumps enable the transport of the signaling particles across the membrane even against a concentration gradient at the cost of energy. Communication-theoretical channel models have been developed for ion pumps and ion channels in \cite{Arjmandi2016} and \cite{Arjmandi2016ion}, respectively. However, these results are preliminary and contain no comparative study of ion pumps and ion channels.  To quantitatively compare these options, the impact of the energy consumption of ion pumps and  the concentration gradient of ion channels have to be modeled and analyzed. This has not been fully addressed in the existing MC literature, yet.
	\item \textbf{Reuptake:} The particle reuptake mechanism enables a reduction of ISI and the possibility of ``harvesting" signaling particles for re-transmission. However, several aspects affect the performance of the reuptake process including the position of the reuptake pumps on the cell surface, their number/density, and the required energy consumption. Related preliminary results by the MC community can be found in \cite{Deng2017EE,Guo2018SMIET,lotter19}, where the authors proposed the transmission of signaling molecules by one node and harvesting them by inward reuptake  pumps  by the same or another node.
	\item \textbf{Phosphopeptides:} As discussed in Section~\ref{sec:phosphopeptides_complex}, phosphopeptides are quite flexible which allows the construction of complex system architectures. In addition, unlike other types of signaling particles, peptide molecules can be always present in the channel and become phosphorylated to create the signaling molecules, namely phosphopeptides, upon an external stimulus (e.g. light or external ligand).  This interesting class of signaling particles has not been investigated in the MC literature, yet. Hence, there exist many open research problems ranging from the modeling of the synthesis and reception of phosphopeptides to the design of new network architectures addressing the needs of different applications.
	\item \textbf{End-to-end channel model:} In communication theory, we are often interested in an end-to-end channel model which relates the received signal to the transmit signal. In the context of MCs, the number of signaling molecules released by the transmitter is typically considered as the transmit signal and  the number of signaling molecules counted at the receiver constitute the received signal. However,  the release and reception mechanisms in MCSs may include multiple stages which  should be carefully taken into account for end-to-end channel modeling. For instance, for the NT receiver R1 in Fig.~\ref{fig:neurotransmitter_receivers}, the NTs activate ligand-gated sodium channels which allow the sodium ions to enter the receiving cell and potentially trigger another action. Therefore, the number of sodium ions entering the receiver should be considered as the actual received signal rather than the number of activated sodium channels. Although these two quantities are related, the latter accounts for the available concentration of sodium ions outside the receiver, which may considerably affect the performance. For example, although signaling particles (i.e., NTs) may activate many ligand-gated sodium channels, a strong signal cannot be generated inside the receiver if the concentration of sodium ions outside the receiver is low.  
	\item \textbf{Impact of interfaces:} We presented various micro-to-macroscale interfaces to facilitate the experimental evaluation of the proposed biological building blocks for MCSs. For the experimental verification in Stage~4 of the proposed development roadmap in Fig.~\ref{fig:MCS_Roadmap}, the relative impact of the microscale MCSs and the interface on the received signal have to be carefully investigated. For instance, the impairments (noises and nonlinearities) that diffusion, reactions, and interference from natural cells cause to the received signal are different from those that a light source, a pH meter, or a current meter  cause. These two different categories of impairments have to be separately  and quantitatively analyzed. 
\end{itemize}

\subsection{Implementation Challenges\label{subsec:feasibility}}

%Proposed mechanisms are feasible 
In order to realize the biological building blocks proposed in this article, some technical challenges have to be met, which are discussed in the following. 
\begin{itemize}
	\item \textbf{Recombinant expression of proteins:} To generate sufficient amounts of the required proteins (e.g. those shown in Fig.~\ref{fig:strukturabbildungen}), they have to be expressed recombinantly in cells from either prokaryotic (e.g. \textit{E. coli}) or eukaryotic  origin (e.g. yeast, insect or mammalian cells). Therefore, a DNA coding for the respective protein has to be inserted, such that the native cellular machinery of protein biosynthesis can be utilized \cite{Kubicek2014}. The fact that crystal structures for most of the proteins shown in Fig.~\ref{fig:strukturabbildungen} exist indicates that protocols for the recombinant expression and subsequent purification of larger amounts of these proteins are already available.
	
	\item \textbf{Additional engineering:} The vast majority of the proteins proposed as biological transmitter and receiver components were directly adapted from functional biological systems and do not require further modification, i.e., they can be readily obtained by recombinant expression or by using them directly in the context of the native biological system. 	However, protein engineering \cite{Lluis2012, Lin2016} offers the possibility to modify protein function in order to make them more suitable for application in MCSs. The simplest strategy is the introduction of point mutations, i.e., the exchange of single amino acids within the protein. This strategy has for example been applied to engineer a light-driven calcium pump \cite{Bennett2002}, bacteriorhodopsin mutants with activity at different wavelengths \cite{Soppa1989}, and GFP variants with different pH optima \cite{Miesenbck1998}. This strategy may also be used to design light-driven calcium pumps which operate at different wavelengths (prerequisite for reversibility, Table~\ref{tab:proton_calcium_vergleich}, T5). Such components have not yet been reported, but can presumably be constructed in a similar manner as 
	the already available bacteriorhodopsin variants (Fig.~\ref{fig:proton_transmitters}, T5) by introducing point mutations. Besides point mutations, another strategy for protein engineering is the fusion to unrelated protein domains of synergistic function\footnote{Cooperative function of two or more agents which allows for faster achievement of a common goal.}, which can be considered as part of a modular toolbox \cite{Lin2016}. For instance, a fusion construct of GFP (Fig.~\ref{fig:strukturabbildungen}c)  and the calcium-binding protein calmodulin has led to the development of the calcium sensor GCaMP which shows a pronounced fluorescence only upon the presence of calcium ions \cite{Nakai2001}. Moreover, light-inducible protein kinases have been developed using a similar approach \cite{Grusch2014, Chang2014, Zhou2017}. Another interesting application would be a fusion to domains that allow for a fixation at certain carrier materials (e.g. magnetic nanoparticles) \cite{Horn2015, Banerjee2018, Bruun2018} to facilitate the design of MCSs with a defined geometry.  
	
	Additional engineering will also be required to implement some of the more complex systems proposed in this survey. For example, membrane proteins have to be forced into the correct orientation when they are inserted into vesicles (see below).

	\item \textbf{Membrane protein reconstitution:} Some of the proposed systems require the reconstitution of membrane proteins into vesicles, which is a challenging task. However, the artificial construction of such proteoliposomes has been widely reported 
	\cite{Rigaud2003, Beales2017, Garni2017, Skrzypek2018, Kuhn2017, Jorgensen2016}, and due to the fast progress in the field of synthetic biology, there is a continuous development of corresponding new methods \cite{Rigaud2003}.
	As vesicles containing only physiological types of lipids are sometimes not sufficiently stable, so-called polymersomes which have membranes consisting of 
	non-lipid polymers, or hybrid vesicles containing both polymers and lipids may be used as an alternative \cite{Beales2017}.

	\item \textbf{Polymersomes:}
	The significantly higher mechanical stability is not the only advantage that polymersomes have over liposomes in the context of MC. The higher stability goes hand in hand with a lower intrinsic permeability, since both aspects correlate with membrane thickness \cite{discher2002polymer}. While lipid membranes are usually 3 – 5 nm thick \cite{le2011recent}, polymersomes with up to 40 nm thick membranes have been reported \cite{chen2009mechanical}. This leads to permeability coefficients that are several orders of magnitude lower than the corresponding values determined for liposomes \cite{poschenrieder2017determination}. As a consequence, the uncontrolled passive diffusion of substances over the compartment boundaries and, thus, the background noise in MCS is much lower. The excellent stability and encapsulant retention are also reasons why polymersomes have attracted significant attention as vehicles for (targeted) drug delivery \cite{palivan2016bioinspired}. In addition, the chemical versatility of the amphiphilic polymer monomers from which the polymersomes are formed enables the membrane properties to be tailored to the intended application. For some polymers, such as the ABA-type triblock co-polymer poly(2-methyloxazoline)-poly(dimethylsiloxane)-poly(2-methyloxazoline) (PMOXA-PDMS-PMOXA) it has been shown that they do not cause cytotoxic effects or inflammatory responses in biological systems \cite{tanner2011can,brovz2005cell}. PMOXA-PDMS-PMOXA also allows the functional integration of diverse transmembrane proteins, as reviewed recently \cite{gaitzsch2016engineering}. With regard to the communication speed, the number of proteins per vesicle is decisive. In polymersomes with a mean diameter of 100 nm, a maximum number of 120 transmembrane proteins per vesicle has been inserted so far \cite{schwarzer2018membrane}. This number corresponds to a coverage of 3\% of the available surface area. However, due to the rapid advances in the field of preparative purification of membrane proteins and their functional reconstitution in artificial membranes, it is expected that this value will increase  in the future \cite{beales2017durable}. A systematic study of membrane protein reconstitution in liposomes and polymersomes revealed that similar amounts of proteins can be incorporated in both types of vesicle, with a tendency to slightly higher values in liposomes. Interestingly, the specific activity of reconstituted proton pumps was higher in polymer membranes, which points to the possibility that the same biological effect can be achieved with a smaller number of membrane proteins \cite{goers2018optimized}.

	\item \textbf{Membrane permeability:} The permeability coefficients of lipid membranes for charged molecules are generally low. As a result, pure lipid membranes are  often referred to as ``impermeable'' for ions if there is no protein-mediated transport. However, whether or not a certain membrane can be considered (im)permeable also depends on the considered time frame, since any molecule will permeate at some point due to passive diffusion if a concentration gradient across the membrane is applied. This leads to an uncontrolled mass transport over the compartment boundaries of vesicles, which increases the background noise in MCS. This aspect is of special importance for proton-based transmitters involving lipid vesicles (T2-T5, Fig.~\ref{fig:proton_transmitters}) since the permeability coefficients of protons ($10^{-4}$ cm s$^{-1}$) are several orders of magnitude higher than the corresponding values for other ions, such as calcium ($2.5$ x $10^{-11}$ cm s$^{-1}$) \cite{rossignol1985relationship}. As a consequence, the intrinsic membrane permeability may not be negligible for certain MCSs.

	\item \textbf{Optical stimulation:} Some of the suggested building blocks require external optical stimulation. Although it would be preferable to use internal stimuli for \textit{in vivo} applications because they are less invasive, light is still considered as a unique stimulus since it is easy to control and does not interfere with other physiological stimuli. For research purposes, it is already nowadays the state of the art to use optical systems in animals (e.g, living rats) in order to stimulate light-dependent proteins \cite{Abe2017}. For instance, optogenetics employs light to stimulate genetically engineered neurons, which express light-driven ion channels \cite{Balasubramaniam2018Optogenetics,Wirdatmadja2017brain,Li2014tutorial,Pashaie2014Optogenetic}. A further miniaturization of the respective equipment could eventually allow for a medical use in humans. The example of deep brain stimulation, which is widely applied for therapy of Parkinson's disease, dystonia, and obsessive compulsive disorders \cite{Hammond2008}, shows that such interfaces between technical devices and the human body are feasible.
	
	\item \textbf{Artificial vesicles vs. entire cells:} Instead of employing artificial vesicles, most of the presented building blocks could be realized using entire cells in which the respective membrane proteins are recombinantly expressed. While this may be easier from a technical point of view and facilitate the implementation of reversibility, it would also require higher concentrations of signaling particles due to the much bigger size of the transmitter/receiver systems. Moreover, it would most certainly lead to a high level of noise because of the high complexity and the plethora of natural transmembrane proteins and signaling cascades present in cells which could interfere with the desired signaling processes. Finally, in the context of applications within the human body (e.g. targeted drug delivery), the usage of entire cells may be problematic because of the high amount of surface proteins which could trigger an immune reaction.
\end{itemize}

\section{Conclusions}\label{sec:conclusions}
In this paper, we provided a comprehensive survey of the biological components that can serve as building blocks for the transmitter and the receiver and as signaling particles for synthetic MCSs. Adopting a signaling particle
centric presentation, we argue that cations, NTs, and phosphopeptides represent promising candidate signaling particles. They can be used to interact with natural physiological processes and as information carriers in synthetic MCS employing engineered protein systems as transmitters and receivers. The engineered transmitter/receiver systems considered in this survey mainly rely on physiological protein components of well-characterized functionality. However, the engineering of functional transmitter/receiver systems based on these isolated components still remains a challenging task, particularly when insertion of proteins into vesicle membranes with defined inside-outside geometry is required. For each of the presented synthetic MCSs architectures, the ensuing advantages and limitations are outlined and references to the relevant literature in synthetic biology are provided. This survey will help both theoreticians and experimentalists to develop a better understanding of the options available for the design and implementation of biological synthetic MCSs and interfaces for natural MCSs. Some of the related open research problems for communication-theoretical modeling/design and technical implementation of the proposed MCS building blocks were outlined.  

\appendices
\section{Glossary of Biological Concepts and Terms}
\label{Sec;Glossary}
For convenience, in Table~\ref{tab:glossary}, we explain the most important biological concepts and terms appearing in this article.

%	\section*{Funding information}
%		
%		This work was supported by the Emerging Fields Initiative from the Friedrich-Alexander-Universität Erlangen-Nürnberg (FAU), project ``Molecular communication'' to AB, RS, and HS. We acknowledge support by a grant from the Deutsche Forschungsgemeinschaft (Bu894/5-1 to AB, XXXX to RS, and Sti155/6-1 to HS). xxx should we also mention the Staedtler-Stiftung here? 
	
	\bibliographystyle{IEEEtran}
	
    %\clearpage
	\bibliography{efi_paper} % .bib ending has to be dropped
	%\bibliography{main.bbl}
	%\begin{thebibliography}{1}	
	%\end{thebibliography}

	\clearpage
	\onecolumn
	{\footnotesize \centering
		\LTcapwidth=\textwidth
		\begin{table*}%[!h]
			\caption{Explanation of relevant biological and chemical terms. For more in-depth discussion, we recommend references \cite{Turner2005, Branden1999, Berg2015}.}
		\end{table*} \vspace{-50pt}
		\addtocounter{table}{-1}
		\begin{longtable}[c]{>{\raggedright}p{.27\textwidth}>{\raggedright\arraybackslash}p{.685\textwidth}}
			\label{tab:glossary}
			\endfirsthead
			\toprule 
			\textbfsf{Term} & \textbfsf{Description} \\
			\midrule
			\multicolumn{2}{c}{\itshape From amino acids to proteins} \\
			\textbfsf{Amino acid} & Organic compound containing an amine (NH$_2$) group, a carboxyl (COOH) group, and a specific side chain. Building block of a protein. \\
			\textbfsf{Peptide} & Short chain of amino acids (smaller than a protein) linked by peptide bonds. \\
			\textbfsf{Protein} & Large biomolecule, consisting of one or more long chains of amino acids.  \\
			\textbfsf{Crystal structure} & Three-dimensional model of the structure of a molecule (e.g. a protein) resolved by X-ray crystallography. \\
			\textbfsf{Residue} & Single amino acid within a peptide or protein.  \\ 
			\textbfsf{Proteome} & Entire set of proteins expressed by a certain organism, tissue or cell. \\
			\midrule
			\multicolumn{2}{c}{\itshape Protein folding and modifications} \\
			\textbfsf{Chaperone} & Protein that assists the folding process of other proteins or the assembly of macromolecular structures.\\
			\textbfsf{Molecular maturation} & Collective term for additional modifications taking place after protein biosynthesis, such as cleavage, posttranslational modifications and oligomerization.  \\
			\textbfsf{Posttranslational modification of proteins/peptides} & Covalent modification of a protein or a peptide after protein biosynthesis.\\ \addlinespace
			\textbfsf{Methylation/Acetylation/ \linebreak Ubiquitination} & Examples for posttranslational modifications: addition of a methyl (CH$_3$) / acetyl (CH$_3$CO) / ubiquitin (a small regulatory protein) group.\\
			\midrule
			\multicolumn{2}{c}{\itshape Different oligomerization states} \\
			\textbfsf{Monomer} & Single molecule which can undergo oligomerization and thereby contribute a constitutional unit to a macromolecule. \\
			\textbfsf{Dimer} & Chemical structure formed from two subunits.\\
			\textbfsf{Oligomer} & Macromolecule formed from a few subunits.\\
			\textbfsf{Homo-/Hetero-} & Consisting of identical/different sub-molecules.\\  
			\midrule 
			\midrule
			\multicolumn{2}{c}{\itshape Protein domains} \\
			\textbfsf{Protein domain} & Part of a protein which can evolve, fold, and function independently from the rest of the protein chain.\\
			\textbfsf{Adapter domain} & Domain mediating specific interactions with a binding partner. \\
			\textbfsf{Homologous domain} & Domain which is similar to another one due to a common evolutionary origin.\\
			\midrule 
			\multicolumn{2}{c}{\itshape Enzymes -- Biological catalysts} \\
			\textbfsf{Kinase} & Transfers a phosphate group from ATP to a protein.\\
			\textbfsf{ATPase} & Catalyzes cleavage of ATP to ADP and a phosphate group, uses released energy to drive a chemical reaction that would not occur otherwise.\\
			\textbfsf{Protease} & Cleaves a protein by hydrolysis of peptide bonds.\\
			\textbfsf{Phosphatase} & Removes a phosphate group from a phosphorylated protein.\\
			\textbfsf{Allosteric regulation} & Regulation of an enzyme by binding an effector molecule at a site other than the enzyme's active site.\\
			%Cognate specificity & Specificity of the same or similar nature, generically alike. & \href{https://www.merriam-webster.com/dictionary/cognate}{[L23]} \\
			\midrule
			%\pagebreak
			%\midrule
			\multicolumn{2}{c}{\itshape Overview of chemical terms} \\
			\textbfsf{Moiety} & Chemical group (e.g. methyl moiety, acetyl moiety, peptide moiety)\\
			\textbfsf{Acidic conditions} & Environmental conditions characterized by a low pH value, i.e., a high proton concentration.\\
			\textbfsf{Protonation} & Covalent addition of a proton to a molecule, formation of the conjugate acid.\\
			\textbfsf{Hydrolysis} & Cleavage of a biomolecule accompanied by the consumption of a water molecule.\\
			\textbfsf{Chelating agent} & Chemical agent that binds certain metal ions by forming two or more coordinate bonds. May be used to remove or chemically shield the metal ions.\\
			\midrule
			\multicolumn{2}{c}{\itshape Miscellaneous} \\
			\textbfsf{Surface-plasmon resonance} & Biophysical method to detect biomolecular interactions.\\
			\textbfsf{Recombinant DNA/protein} & DNA/protein sequence engineered by laboratory methods of genetic recombination such as molecular cloning.\\
			\bottomrule
		\end{longtable}
	}
	\clearpage
	\twocolumn
	
		\begin{IEEEbiography}[{\includegraphics[width=1in,height=1.5in,clip,keepaspectratio]{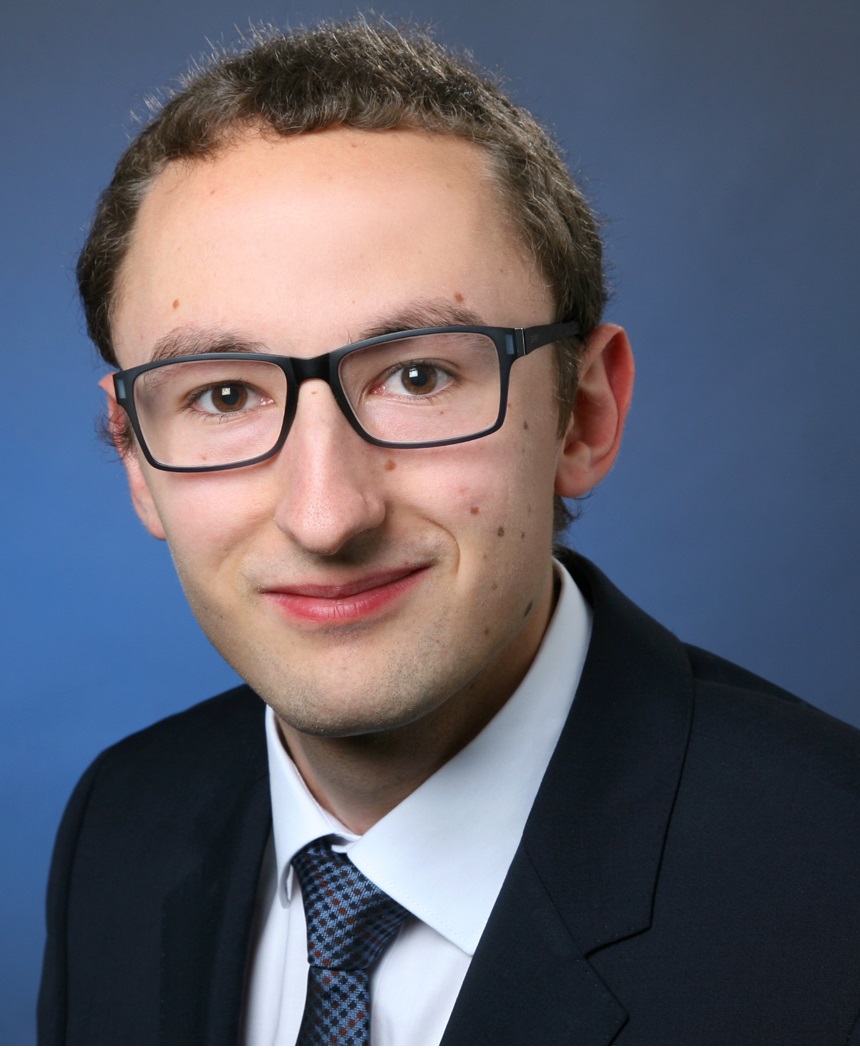}}]{Christian A. Söldner} 
			 received the B. Sc. and M. Sc. degree in molecular medicine from the Friedrich Alexander University Erlangen-Nürnberg, Germany, in 2014 and 2016, respectively. During his Ph.D. in the subject of bioinformatics, he focused his research interest on the analysis of receptor-ligand interactions. He studied GPCRs and the macrophage surface receptor Mincle by means of molecular dynamics simulations and developed a metadynamics-based protocol for the determination of ligand binding modes. After his Ph.D. degree, which he received in 2020, he joined Siemens Healthineers and is currently working as an application specialist and sequence developer for magnetic resonance imaging.
	\end{IEEEbiography}

	\begin{IEEEbiography}
		[{\includegraphics[width=1in,height=1.25in,clip,keepaspectratio]{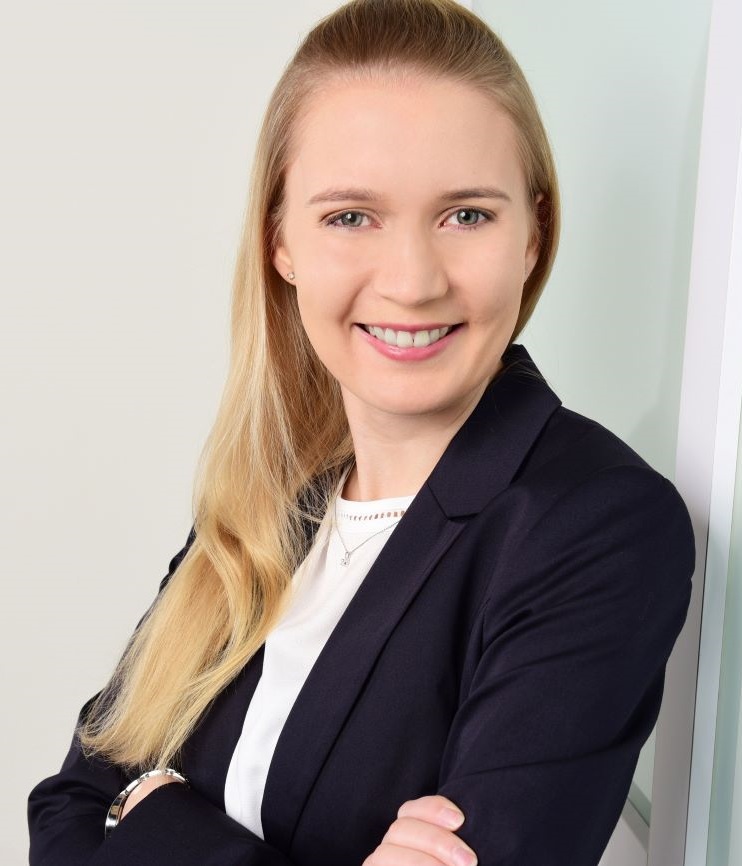}}]
		{Eileen Socher} received the B.Sc. and M.Sc. degrees in Molecular Medicine (with a focus on pathology and bioinformatics) from Friedrich-Alexander-University Erlangen-Nürnberg (FAU), Erlangen, Germany, in 2011 and 2012, respectively. In 2017, she received the doctoral degree in the field of bioinformatics from the Faculty of Sciences, Friedrich-Alexander-University Erlangen-Nürnberg (FAU). Currently, she is a postdoctoral researcher at the FAU Faculty of Medicine in Erlangen, Germany. Her research interests include structural bioinformatics in general and the molecular modelling of proteins as well as their investigation by using molecular dynamics simulations in particular. Within this field, she works, for instance, on the computer-based analysis of pH-induced effects on protein structure or protein-protein interactions between viral and human proteins.
	\end{IEEEbiography}
	
	\begin{IEEEbiography}[{\includegraphics[width=1in,height=1.25in,clip,keepaspectratio]{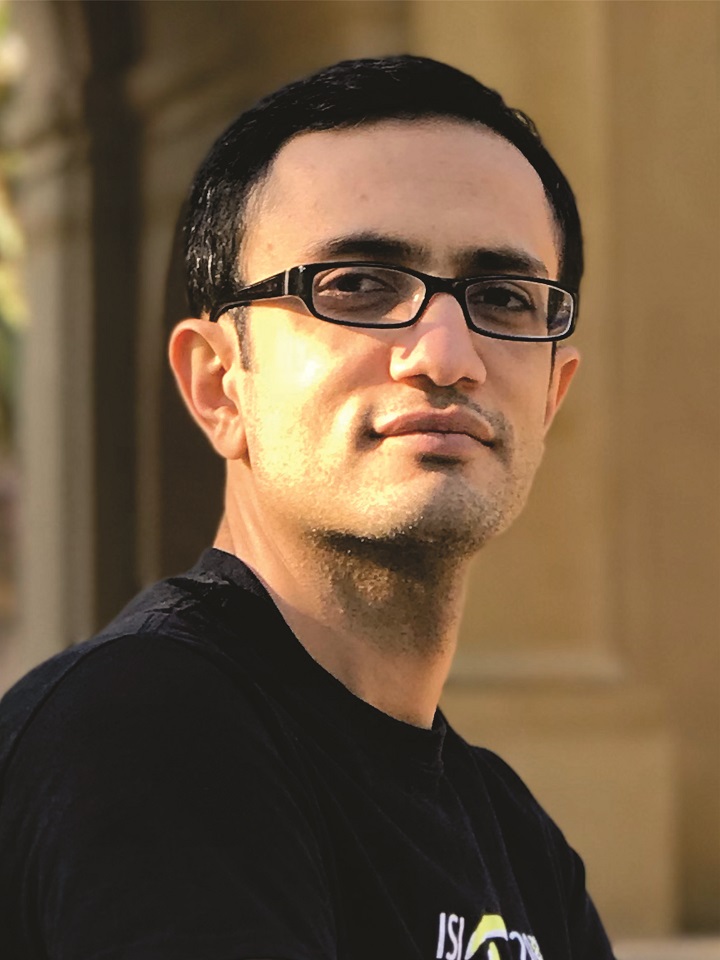}}]{Vahid Jamali} (S'12, M'20) received the B.Sc. and M.Sc. degrees (honors) in electrical engineering from the
		K. N. Toosi University of Technology, Tehran,
		Iran, in 2010 and 2012, respectively, and
		the Ph.D. degree (with distinctions) from
		the Friedrich-Alexander-University (FAU) of
		Erlangen-N\"urnberg, Erlangen, Germany,
		in 2019. In 2017, he was a Visiting Research Scholar with Stanford
		University, CA, USA. He is currently a Postdoctoral Researcher
		with the Institute for Digital Communication, FAU. His 
		research interests include wireless and molecular communications,
		Bayesian inference and learning, and multiuser information~theory.
		
		Dr. Jamali has served as a member of the Technical Program
		Committee for several IEEE conferences and he is currently an Associate Editor of the \textsc{IEEE Communications Letters} and \textsc{IEEE Open Journal of the Communications Society}. He received several
		awards for his publications and research work including the Best Paper Award from
		the IEEE International Conference on Communications in 2016, 
		the Doctoral Research Grant from the German Academic Exchange
		Service (DAAD) in 2017, the Goldener Igel Publication Award from
		the Telecommunications Laboratory (LNT), FAU, in 2018, 
		 the Best Ph.D. Thesis
		Presentation Award from the IEEE Wireless Communications and Networking Conference in 2018, the Best Paper Award from the ACM International Conference on Nanoscale Computing and Communication in 2019, and the Postdoctoral Research Fellowship by the German Research Foundation (DFG) in 2020. 
	\end{IEEEbiography}

\begin{IEEEbiography}
	[{\includegraphics[width=1in,height=1.25in,clip,keepaspectratio]{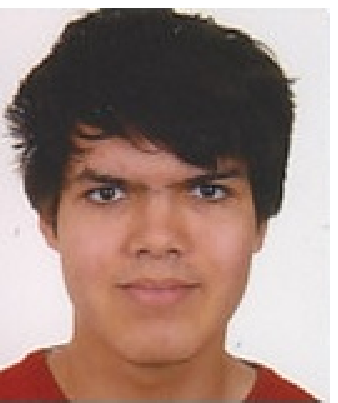}}]
	{Wayan Wicke} (S'17) was born in Nuremberg, Germany, in 1991.
	He received the B.Sc.\ and  M.Sc.\ degrees in electrical engineering from the Friedrich-Alexander University Erlangen-Nürnberg (FAU), Erlangen, Germany, in 2014 and 2017, respectively, where he is currently pursuing the Ph.D.\ degree.
	His research interests include statistical signal processing and digital communications with a focus on molecular communication.
\end{IEEEbiography}

\begin{IEEEbiography}[{\includegraphics[width=1in,height=1.25in,clip,keepaspectratio]{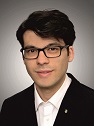}}]{Arman Ahmadzadeh} (S'14) received the B.Sc. degree in electrical engineering from the Ferdowsi University of Mashhad, Mashhad, Iran, in 2010, and the M.Sc. degree in communications and multimedia engineering from the Friedrich-Alexander-Universität (FAU) Erlangen-Nürnberg, Erlangen, Germany, in 2013, where he is currently working toward the Ph.D. degree in electrical engineering at the Institute for Digital Communications. His current research interests include physical layer molecular communications. Mr. Ahmadzadeh has served as a member of the Technical Program Committee of the Communication Theory Symposium for the IEEE International Conference on Communications (ICC) 2017-2020. He received several awards including the Best Paper Award from the IEEE ICC in 2016, IEEE ICC in 2020, and the Student Travel Grants for attending the Global Communications Conference (GLOBECOM) in 2017. He was recognized as an Exemplary Reviewer of \textsc{IEEE Communications Letters} in 2016.    
\end{IEEEbiography}

\begin{IEEEbiography}
	[{\includegraphics[width=1in,height=1.25in,clip,keepaspectratio]{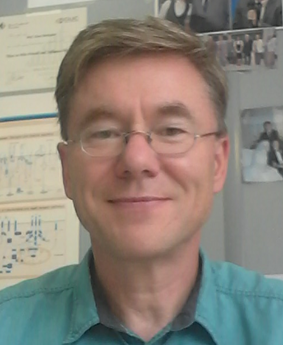}}]
	{Hans-Georg Breitinger}
	received the Ph.D. in Organic Chemistry from Heinrich Heine University (HHU), Düsseldorf, Germany, in 1993. He was a postdoctoral researcher at Cornell University from 1993-1997, and completed the Habilitation (University teaching certificate) for Biochemistry at Friedtrich-Alexander University, Erlangen, in 2003. In 2004 he joined the German University in Cairo as Full Professor and Head of the Department of Biochemistry. His research focuses on neuronal ion channels and viroporins, the biochemistry of membrane receptors, antimicrobial peptides and isolation of new pharmaceuticals from natural resources. His postdoctoral stay at Cornell University was supported by a fellowship of the Swiss Federation of Sciences. 
\end{IEEEbiography}

\begin{IEEEbiography}
	[{\includegraphics[width=1in,height=1.25in,clip,keepaspectratio]{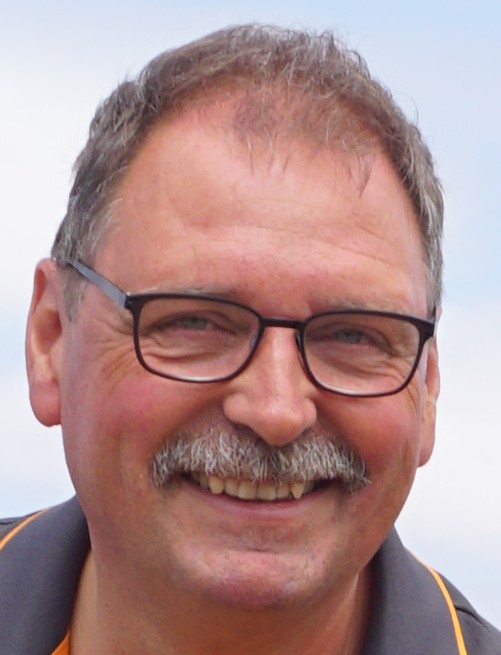}}]
	{Andreas Burkovski}
	 received his diploma in Biology in 1989 and his doctoral degree in 1993, both at the University of Osnabrück. After postdoc positions at the University of Osnabrück and the Research Center Jülich, he became group leader at the University of Cologne in 1997. In 2005 he received a professorship in Microbiology at the University of Erlangen-Nuremberg. His research interests focus on the analysis of host-pathogen-interactions, regulatory networks and synthetic biology.
\end{IEEEbiography}

\begin{IEEEbiography}
	[{\includegraphics[width=1in,height=1.25in,clip,keepaspectratio]{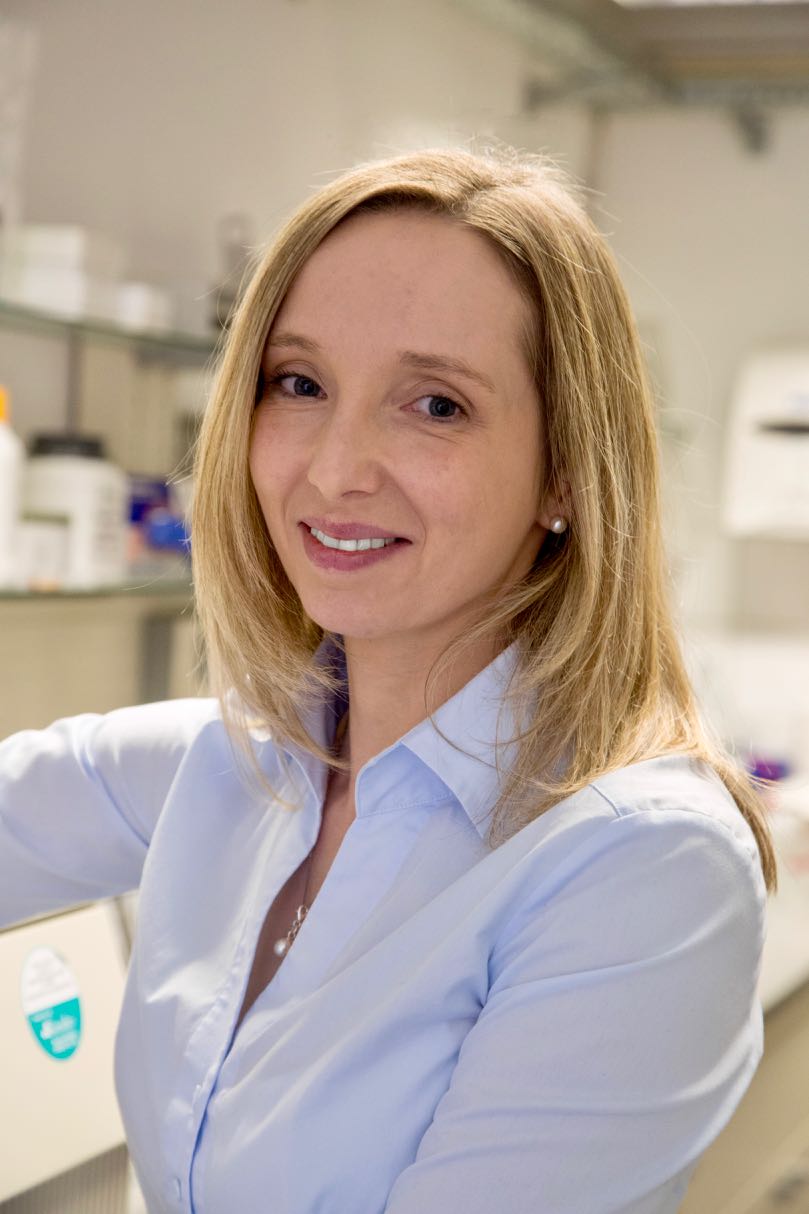}}]
	{Kathrin Castiglione}
	 received the B.Sc. and M.Sc. degrees in Molecular Biotechnology from the Technical University of Munich (TUM). After her PhD at the Institute of Biochemical Engineering of TUM in 2009, she moved to the Toyama Prefectural University in Japan as a postdoctoral fellow of the Japanese Society of the Promotion of Science. At the end of 2010, she returned to TUM and became head of the biocatalysis group at the Institute of Biochemical Engineering. Since 2012 she has been the leader of an independent
	junior research group working on artificial reaction compartments based on functionalized polymer vesicles. Since 2018 Kathrin Castiglione has headed the Institute of Bioprocess Engineering of the Friedrich-Alexander University Erlangen-Nürnberg (FAU). Among other things, she works on the design of functionalized polymersomes for molecular communication. 
\end{IEEEbiography}

\begin{IEEEbiography}
	[{\includegraphics[width=1in,height=1.25in,clip,keepaspectratio]{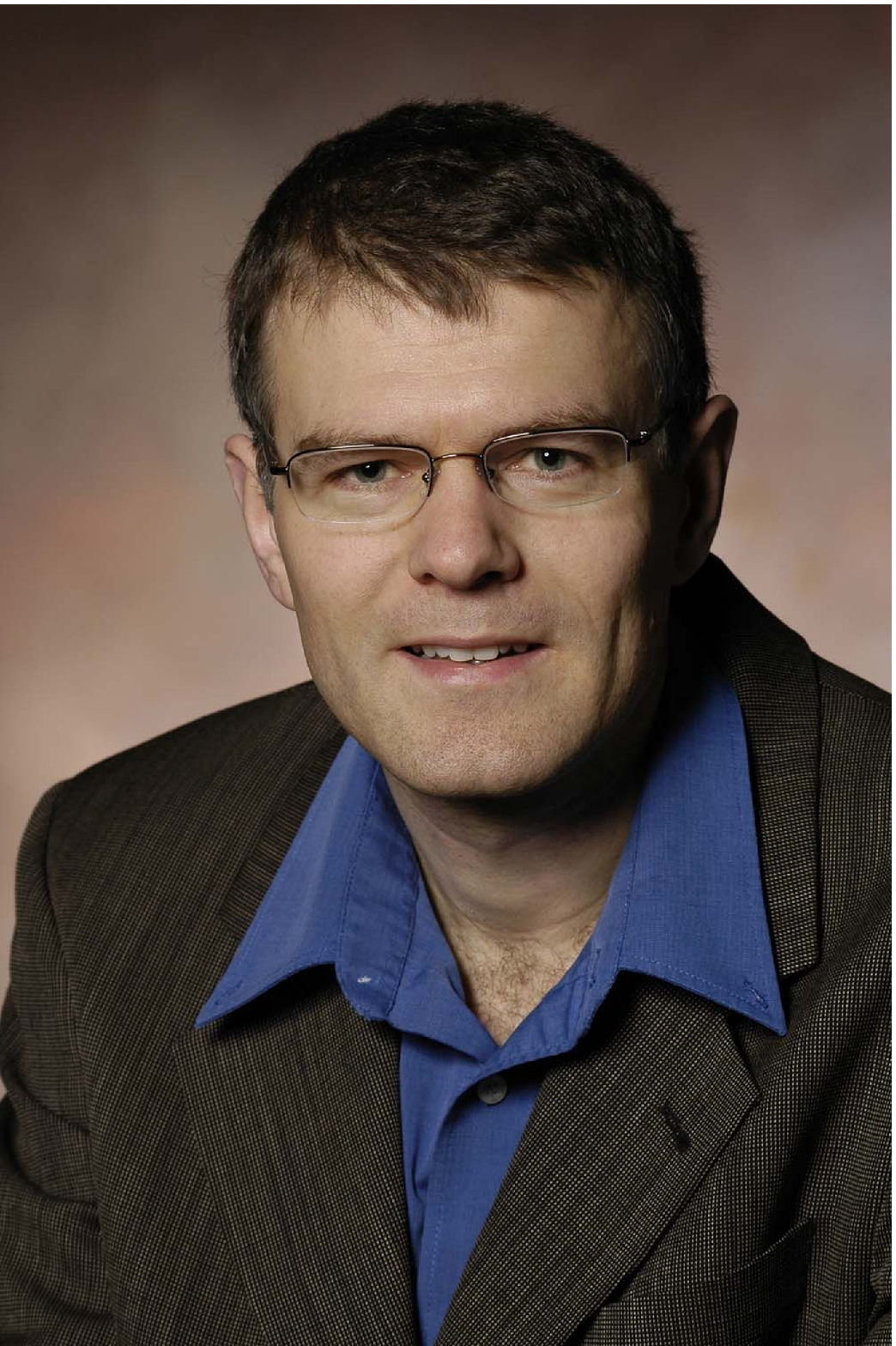}}]
	{Robert Schober}
	(S'98, M'01, SM'08, F'10) received the Diplom (Univ.) and the Ph.D.\ degrees in electrical engineering from the Friedrich-Alexander University (FAU) Erlangen-N\"urnberg, Erlangen, Germany, in 1997 and 2000, respectively.
	From 2002 to 2011, he was a Professor and Canada Research Chair at the University of British Columbia (UBC), Vancouver, Canada.
	Since January 2012, he is an Alexander von Humboldt Professor and the Chair for Digital Communication at FAU.
	His research interests fall into the broad areas of Communication Theory, Wireless Communications, and Statistical Signal Processing.
	
	Robert received several awards for his work including the 2002 Heinz Maier-Leibnitz Award of the German Science Foundation (DFG), the 2004 Innovations Award of the Vodafone Foundation for Research in Mobile Communications, a 2006 UBC Killam Research Prize, a 2007 Wilhelm Friedrich Bessel Research Award of the Alexander von Humboldt Foundation, the 2008 Charles McDowell Award for Excellence in Research from UBC, a 2011 Alexander von Humboldt Professorship, a 2012 NSERC E.W.R.\ Stacie Fellowship, and a 2017 Wireless Communications Recognition Award by the IEEE Wireless Communications Technical Committee.
	He is listed as a 2017 Highly Cited Researcher by the Web of Science and a Distinguished Lecturer of the IEEE Communications Society (ComSoc).
	Robert is a Fellow of the Canadian Academy of Engineering and a Fellow of the Engineering Institute of Canada.
	From 2012 to 2015, he served as Editor-in-Chief of the \textsc{IEEE Transactions on Communications}.
	Currently, he is the Chair of the Steering Committee of the \textsc{IEEE Transactions on Molecular, Biological and Multi-Scale Communications}, a Member of the Editorial Board of the Proceedings of the IEEE, a Member at Large of the Board of Governors of ComSoc, and the ComSoc Director of Journals.
\end{IEEEbiography}

\begin{IEEEbiography}
	[{\includegraphics[width=1in,height=1.25in,clip,keepaspectratio]{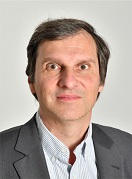}}]
	{Heinrich Sticht}
	 was born in 1968. He received the Diploma degree in biochemistry from the University of Bayreuth in 1993, the Ph.D. degree in 1995, and the Dr.habil. degree and Venia legendi from the University of Bayreuth in 1999. From 1996 to 1997, he was a Post-Doctoral Fellow with Oxford University, U.K. In 1997, he returned to the University of Bayreuth as a Research Assistant. From 1999 to 2002, he was a Group Leader with the University of Bayreuth and, in 2002, he became a Professor of Bioinformatics with Friedrich-Alexander University. His major research interests include molecular interactions and molecular communication processes, which are studied by a combination of computational tools (e.g., sequence data analysis, molecular docking, and molecular dynamics). He has authored or co-authored about 250 peer-reviewed journal papers, reviews, or book chapters.
\end{IEEEbiography}

\end{document}